\documentclass[trackchanges]{aastex701}
\usepackage{amsmath}

\begin{document}

\title{Measured energy exchange in coronal hole solar wind from its solar origins to the heliosphere}

\author[orcid=0000-0002-8748-2123, gname=Yeimy,sname=Rivera]{Yeimy J. Rivera}
\affiliation{Center for Astrophysics, Harvard \& Smithsonian, Cambridge, MA 02138, USA}
\email{yeimy.rivera@cfa.harvard.edu}
\correspondingauthor{Yeimy J. Rivera}

\author[orcid=0000-0002-3491-1983,gname=Alin,sname=Paraschiv]{Alin R. Paraschiv}
\affiliation{National Solar Observatory, Boulder, CO 80301, USA}
\email{arparaschiv@nso.edu}

\author[orcid=0000-0002-6145-436X, gname=Samuel,sname=Badman]{Samuel T. Badman}
\affiliation{Center for Astrophysics, Harvard \& Smithsonian, Cambridge, MA 02138, USA}
\email{samuel.badman@cfa.harvard.edu}

\author[orcid=0000-0003-3401-6409, gname=Alberto,sname=Vasquez]{Alberto M. Vasquez}
\affiliation{Instituto de Astronom\'ia y F\'isica del Espacio, CONICET--UBA, Ciudad de Buenos Aires, C1428EGA, Argentina}
\email{alberto.vasquez@conicet.gov.ar}

\author[orcid=0000-0003-2355-5853,gname=Federico,sname=Nuevo]{Federico A. Nuevo}
\affiliation{Instituto de Astronom\'ia y F\'isica del Espacio, CONICET--UBA, Ciudad de Buenos Aires, C1428EGA, Argentina}
\email{fnuevo@iafe.uba.ar}

\author[orcid=0000-0001-9726-0738, gname=Ryan,sname=French]{Ryan J. French}
\affiliation{Laboratory for Atmospheric and Space Physics, University of Colorado Boulder, Boulder, CO 80303, USA}
\email{ryan.french@lasp.colorado.edu}

\author[orcid=0000-0003-0583-0516, gname=Momchil,sname=Molnar]{Momchil Molnar}
\affiliation{Southwest Research Institute, Boulder, CO 80302, USA}
\email{momchil.molnar@swri.org}

\author[orcid=0000-0002-1017-7163, gname=Roberto,sname=Susino]{Roberto Susino}
\affiliation{Astrophysical Observatory of Turin, INAF, Pino Torinese, Turin, 101025, Italy}
\email{roberto.susino@inaf.it}

\author[orcid=0000-0002-4973-0018,gname=Zihao,sname=Yang]{Zihao Yang}
\affiliation{High Altitude Observatory, NCAR, Boulder, CO 80301, USA}
\email{zihao@ucar.edu}

\author[orcid=0000-0001-9921-1198, gname=Marco,sname=Romoli]{Marco Romoli}
\affiliation{Department of Physics and Astronomy, University of Florence, Sesto Fiorentino (FI), 50019, Italy}
\email{marco.romoli@unifi.it}

\author[orcid=0000-0002-0281-7677, gname=Richard,sname=Frazin]{Richard A. Frazin}
\affiliation{Climate and Space Sciences and Engineering, University of Michigan, Ann Arbor, MI 48109, USA}
\email{rfrazin@umich.edu}

\author[orcid=0000-0002-7728-0085, gname=Michael,sname=Stevens]{Michael L. Stevens}
\affiliation{Center for Astrophysics, Harvard \& Smithsonian, Cambridge, MA 02138, USA}
\email{michael.stevens@cfa.harvard.edu}

\author[orcid=0000-0002-1989-3596, gname=Stuart,sname=Bale]{Stuart D. Bale}
\affiliation{Physics Department, University of California, Berkeley, CA 94720-7300, USA}
\affiliation{Space Sciences Laboratory, University of California, Berkeley, 7 Gauss Way, Berkeley, CA 94720, USA}
\affiliation{The Blackett Laboratory, Imperial College London, London, SW7 2AZ, UK}
\email{bale@ssl.berkeley.edu}

\author[orcid=0000-0002-0665-2355, gname=Deborah,sname=Baker]{Deborah Baker}
\affiliation{Mullard Space Science Laboratory, University College London, Dorking, Surrey, RH5 6NT, UK}
\email{deborah.baker@ucl.ac.uk}

\author[orcid=0000-0001-5258-6128,gname=Jasper,sname=Halekas]{Jasper Halekas}
\affiliation{Department of Physics and Astronomy, University of Iowa, Iowa City, IA 52245, USA}
\email{jasper-halekas@uiowa.edu}

\author[0000-0002-7451-9804,gname="Thomas",sname="Schad"]{Thomas A. Schad}
\email{tschad@nso.edu}
\affiliation{National Solar Observatory, 22 \textquotesingle Ohi\textquotesingle a K\={u} Street, Pukalani, HI 96768, USA}

\author[orcid=0000-0002-5982-4667, gname=Christopher,sname=Owen]{Christopher J. Owen}
\affiliation{Mullard Space Science Laboratory, University College London, Dorking, Surrey, RH5 6NT, UK}
\email{christopher.owen@ucl.ac.uk}

\author[gname=Philippe,sname=Louarn]{Philippe Louarn}
\affiliation{Institut de Recherche en Astrophysique et Plan\'etologie, CNRS, Universit\'e de Toulouse, CNES, Toulouse, 31028 CEDEX 4, France}
\email{philippe.louarn@irap.omp.eu}

\author[orcid=0000-0002-4498-8706,gname=Jenna,sname=Samra]{Jenna Samra}
\affiliation{Center for Astrophysics, Harvard \& Smithsonian, Cambridge, MA 02138, USA}
\email{jenna.samra@cfa.harvard.edu}

\author[orcid=0000-0002-8439-6166, gname=Giuliana,sname='de Toma']{Giuliana de Toma}
\affiliation{High Altitude Observatory, NCAR, Boulder, CO 80301, USA}
\email{detoma@ucar.edu}

\author[orcid=0000-0003-2123-6605,gname=Daniela,sname=Lacatus]{Daniela A. Lacatus}
\affiliation{High Altitude Observatory, NCAR, Boulder, CO 80301, USA}
\email{dlacatus@ucar.edu}

\author[orcid= 0000-0002-4125-0204, gname=Giulio,sname='Del Zanna']{Giulio Del Zanna}
\affiliation{University of Cambridge and University of Leicester, Cambridge/Leicester, UK}
\email{gd232@cam.ac.uk}

\author[orcid=0000-0001-8775-913X,gname=Chad,sname=Madsen]{Chad A. Madsen}
\affiliation{Center for Astrophysics, Harvard \& Smithsonian, Cambridge, MA 02138, USA}
\email{chad.madsen@cfa.harvard.edu}

\author[orcid=0000-0002-9325-9884, gname=Enrico,sname=Landi]{Enrico Landi}
\affiliation{Climate and Space Sciences and Engineering, University of Michigan, Ann Arbor, MI 48109, USA}
\email{elandi@umich.edu}

\author[orcid=0000-0001-6913-1330,gname=Manolis,sname=Georgoulis]{Manolis K. Georgoulis}
\affiliation{Space Exploration Sector, Johns Hopkins Applied Physics Laboratory, Laurel, MD 20723, USA}
\email{manolis.georgoulis@jhuapl.edu}

\author[0000-0001-8702-8273,gname=Amir,sname=Caspi]{Amir Caspi}
\affiliation{Southwest Research Institute, Boulder, CO 80302, USA}
\email{amir.caspi@swri.org}

\author[orcid=0000-0003-4290-1897,gname=\'Eric,sname=Buchlin]{\'Eric Buchlin}
\affiliation{Institut d'Astrophysique Spatiale, Universit\'e Paris-Saclay, CNRS, Orsay, 91405, France}
\email{eric.buchlin@ias.u-psud.fr}

\author[orcid=0000-0003-2409-3742, gname=Nour,sname=Raouafi]{Nour E. Raouafi}
\affiliation{Johns Hopkins Applied Physics Laboratory, Laurel, MD 20723, USA}
\email{nour.raouafi@jhuapl.edu}

\author[orcid=0000-0002-7419-0527, gname=Luca,sname=Franci]{Luca Franci}
\affiliation{School of Engineering, Physics and Mathematics, Northumbria University, Newcastle upon Tyne, NE1 8ST, UK}
\email{luca.franci@northumbria.ac.uk}

\author[orcid=0000-0002-6203-5239, gname=Miho,sname=Janvier]{Miho Janvier}
\affiliation{ESTEC, ESA, Noordwijk, 2200 AG, Netherlands}
\email{miho.janvier@esa.int}

\author[orcid=0000-0003-1759-4354, gname=Cooper,sname=Downs]{Cooper Downs}
\affiliation{Predictive Science Inc., San Diego, CA 92120, USA}
\email{cdowns@predsci.com}

\author[orcid=0000-0002-0396-0547, gname=Roberto,sname=Livi]{Roberto Livi}
\affiliation{Space Sciences Laboratory, University of California, Berkeley, 7 Gauss Way, Berkeley, CA 94720, USA}
\email{rlivi@ssl.berkeley.edu}

\author[orcid=0000-0002-7287-5098, gname=Phyllis,sname=Whittlesey]{Phyllis Whittlesey}
\affiliation{Space Sciences Laboratory, University of California, Berkeley, 7 Gauss Way, Berkeley, CA 94720, USA}
\email{phyllisw@ssl.berkeley.edu}

\author[orcid=0000-0003-0519-6498, gname=Ali,sname=Rahmati]{Ali Rahmati}
\affiliation{Space Sciences Laboratory, University of California, Berkeley, 7 Gauss Way, Berkeley, CA 94720, USA}
\email{rahmati@berkeley.edu}

\author[orcid=0000-0001-5030-6030, gname=Davin,sname=Larson]{Davin Larson}
\affiliation{Space Sciences Laboratory, University of California, Berkeley, 7 Gauss Way, Berkeley, CA 94720, USA}
\email{davin@ssl.berkeley.edu}

\author[orcid=0000-0002-1128-9685, gname=Nikos,sname=Sioulas]{
Nikos Sioulas}
\affiliation{Imperial College London, South Kensington Campus, London SW7 2AZ, UK}
\email{nsioulas@berkeley.edu}

\begin{abstract}

Coordinated ground- and space-based remote observations are used to provide a comprehensive, multi-perspective view of the solar atmosphere that we connect with \emph{in situ} measurements in the heliosphere centered around the April 8 2024 Total Solar Eclipse observed across North America. These near-contemporaneous, multi-wavelength datasets are used to directly derive the plasma parameters and compute an energy budget of a solar wind stream at its low-coronal birthplace \textbf{(at an altitude of 1.05 $R_\odot$)}. The energy budget is also independently computed at and beyond the Alfv\'en surface from the \emph{in situ} observations of the same stream. The total wave energy flux dominates the low-coronal budget (11 W m$^{-2}$ Alfv\'en, 7 to 23 W m$^{-2}$ in compressive MHD modes), while enthalpy (13 W m$^{-2}$) and work done against gravity from 1.0 $R_\odot$ (2.5 W m$^{-2}$) make up the remainder. The total coronal energy budget is consistent with the total measured in situ by Parker Solar Probe if and only if compressive energy flux is included and attributed to the fast magnetosonic mode. Nearly all this thermal and wave energy is converted to ion heating, acceleration and gravitational potential energy within the Alfv\'en surface. The stream’s evolution is well-described as a thermal-pressure driven wind with additional energy deposited by MHD fluctuations. The observed evolution of the Alfv\'enic component of these fluctuations follows a near-dissipation-free curve below the Alfv\'en surface while deviating from it more strongly above this critical surface.

\end{abstract}


\section{Introduction}\label{sec1_intro}

Since the 1950s \citep{Parker1958}, the solar wind has been understood as the supersonic expansion of the Sun’s multi–million-Kelvin corona. Early observations confirmed this theory where measurements taken near the Earth indicated that solar wind reached velocities of 400--800~km~s$^{-1}$. However, hydrodynamic and magnetohydrodynamic (MHD) models struggled to self-consistently reproduce the velocity of the fastest solar wind \citep{Hundhausen1972, Leer1979, Holzer1980}. A study from the Mariner 5 mission found large-amplitude Alfv\'enic fluctuations permeated faster speed streams and were thought to be ``undamped remnants of waves generated at or near the sun'' \citep{Belcher1971}. For decades following that study, it was conjectured that Alfv\'en waves, which were routinely observed \emph{in situ}, were the critical energy source that gave rise to solar wind with the highest speeds \citep{Marsch1997, DePontieu2007} that originated from coronal holes \citep{McComas2000}. 

Contemporary missions, such as Parker Solar Probe \citep[\emph{Parker}; ][]{Fox2016}, have revealed that the fractional Alfv\'en wave energy flux increases significantly closer to the Sun, peaking within the highest speed streams. Recent studies \citep{Telloni2023, Halekas2023, Rivera2024, Rivera2025} present measurements of the inner-heliosphere which are consistent with a solar wind that behaves as a hydrodynamic fluid with added momentum and energy transfer from these Alfv\'enic fluctuations \citep{Jacques1977, Perez2021}. This has led to the overall conclusion that this Alfv\'enic energy flux is the primary additional energy term needed to provide the persistent additional heating and acceleration observed near and beyond the Alfv\'en surface required to explain the fast solar wind.

Connecting the solar wind from its source in the corona to its heliospheric evolution is crucial for extending the applicability of these results closer in to the Sun than current in situ constraints, understanding the mechanisms driving this outflow and constraining physical models of the Sun-Heliosphere system in state-of-the-art MHD models. A key constraint is the energy budget of the solar wind, which identifies the principal components heating and accelerating the plasma. Beyond the solar context, the accuracy of simulated magnetized winds is critical to other star systems as it determine the Alfv\'en critical surface location, the stellar wind speed that can be achieved, and inform on long-term stellar evolution properties such as angular momentum loss and spindown time \citep{Garraffo2018}. Constraining stellar wind evolution is necessary to determining its dynamic pressure and therefore understanding extra-solar space weather experienced by exoplanets in close proximity to their star \citep{Cohen2014, Garraffo2016, Garraffo2017, McCann2019}.

Detailed solar wind energy budget studies have traditionally been carried out with \emph{in situ} heliospheric data. While surveys at solar distances between 13 and $>$100~$R_{\odot}$ highlight the increasing role of Alfv\'en wave energy flux with solar wind speed \citep{Halekas2023, Rivera2024, Rivera2025}, few studies probe the critical region below 10~$R_{\odot}$. \citet{Telloni2023} computed the energy flux between 6--13~$R_{\odot}$ using \emph{Parker} \emph{in situ} data and Solar Orbiter’s \citep{Muller2020} Metis white light coronagraph \citep{Romoli2021_Metis} in quadrature, revealing comparable contributions from proton enthalpy, Alfv\'en waves, and kinetic energy at 6~$R_{\odot}$. However, Metis primarily provides density and speed diagnostics, and a comprehensive budget requires additional constraints on the ion and electron temperatures, magnetic field strength, and fluctuations quantities for a full assessment.

This work makes significant advances on these prior works by obtaining plasma parameters and the resulting energy fluxes from direct measurements in the low corona in conjunction with \emph{in situ} data. We presents a unique, comprehensive dataset from a coordinated campaign around the 8~April 2024 Total Solar Eclipse (TSE), capturing a single solar wind stream across critical evolutionary points: from sub-sonic/sub-Alfv\'enic states near the Sun to its super-Alfv\'enic state in the heliosphere. These coordinated remote and \emph{in situ} observations enable empirical estimates of the full kinetic, enthalpy, gravitational, and Alfv\'en wave energy fluxes at each distance, providing novel constraints on the energy flow all the way back to the stream's origin.


\begin{figure}[h]
\centering
\includegraphics[width=\textwidth]{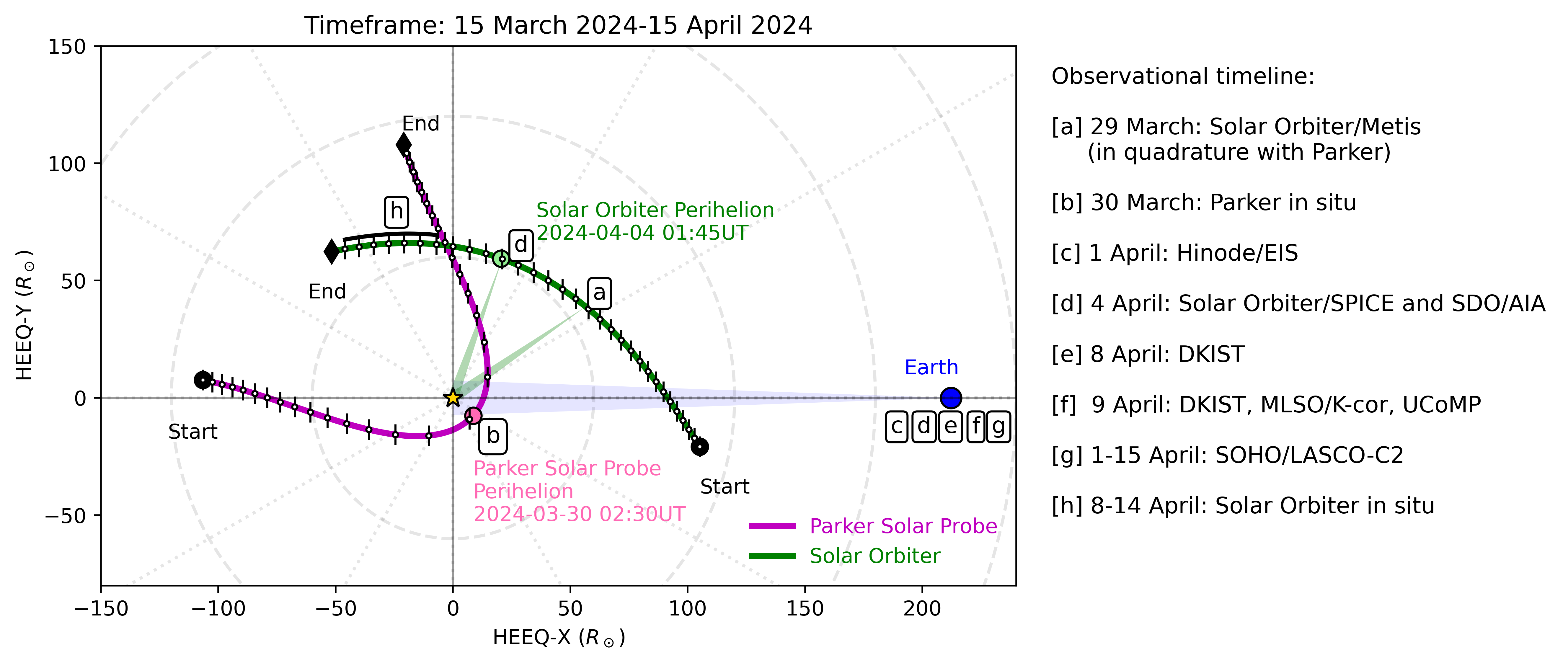}
\caption{Timeline of coordinated remote and \emph{in situ} observations taken between 15~March to 15~April summarizing all observations utilized. The figure shows the spacecraft trajectories of \emph{Parker} and Solar Orbiter as well as the Earth in a Heliocentric-Earth-Equatorial (Earth-fixed) frame. Events are listed in chronological order and are described in Section \ref{sec:event-obs}. The pink and lime dots indicate the location of \emph{Parker} and Solar Orbiter perihelion. The white dots and vertical markers along each trajectory are spaced 1~day apart starting on 15~March 2024. The shaded FOV cones originating from Earth (blue) and Solar Orbiter (green) represent periods where remote observations were taken by different telescopes from those perspectives. }
\label{fig0_5_context}
\end{figure}

\section{Observational Overview and Methodology}
\subsection{Event and Observations}\label{sec:event-obs}
To compute the solar wind energy budget, we obtained plasma density, velocity, temperature, and magnetic field measurements (including fluctuations) from a single solar wind stream at its coronal origin and at multiple heliospheric locations. While all these quantities are readily available simultaneously from the \emph{in situ} measurements, coronal measurements required a ``patchwork'' of coordinated remote sensing diagnostics \citep{Rivera2025_IAU}. These observations targeted an open magnetic field region connected to the stream encountered by the spacecraft. This section details the campaign surrounding the 8~April 2024 TSE, the selected coronal hole stream, and the combined remote and \emph{in situ} datasets. More complete methodological details and approach are provided in Appendix Section~\ref{sec:methods}.

\subsection{Coordination Campaign}\label{subsec:event}

In Figure~\ref{fig0_5_context}, we provide an overview of the series of measurements which resulted from the coordination campaign organized around the TSE. The plot shows measurement events [a--h] (labeled chronologically) with their locations presented in the Heliocentric-Earth-Equatorial (Earth-fixed) reference frame. Magenta and green lines indicate the motion of \emph{Parker} and Solar Orbiter which took \emph{in situ} measurements of the stream (events [a] and [h]). Other instruments on Solar Orbiter (events [a] and [d]) along with numerous Earth-based instruments (events [c--g]) took remote sensing data of the corona. 

This complex set of observations all targeted and captured the plasma properties of a single solar wind stream emanating from a narrow, negative magnetic polarity coronal hole. This source and its stream was observed:
\begin{itemize}
    \item On-disk from Earth, crossed \emph{in situ} by \emph{Parker} near its perihelion (11.60~$R_\odot$), and off limb by Solar Orbiter (while in quadrature with Earth) in events [a--c].
    \item On-disk from Solar Orbiter and from Earth on the West limb around the date of the solar eclipse, during events [d--g].
    \item Crossed \emph{in situ} by Solar Orbiter around 72.06~$R_\odot$ (shortly after its perihelion, during event [h]).
\end{itemize}

\begin{figure}[h]
\centering
\includegraphics[width=0.6\textwidth]{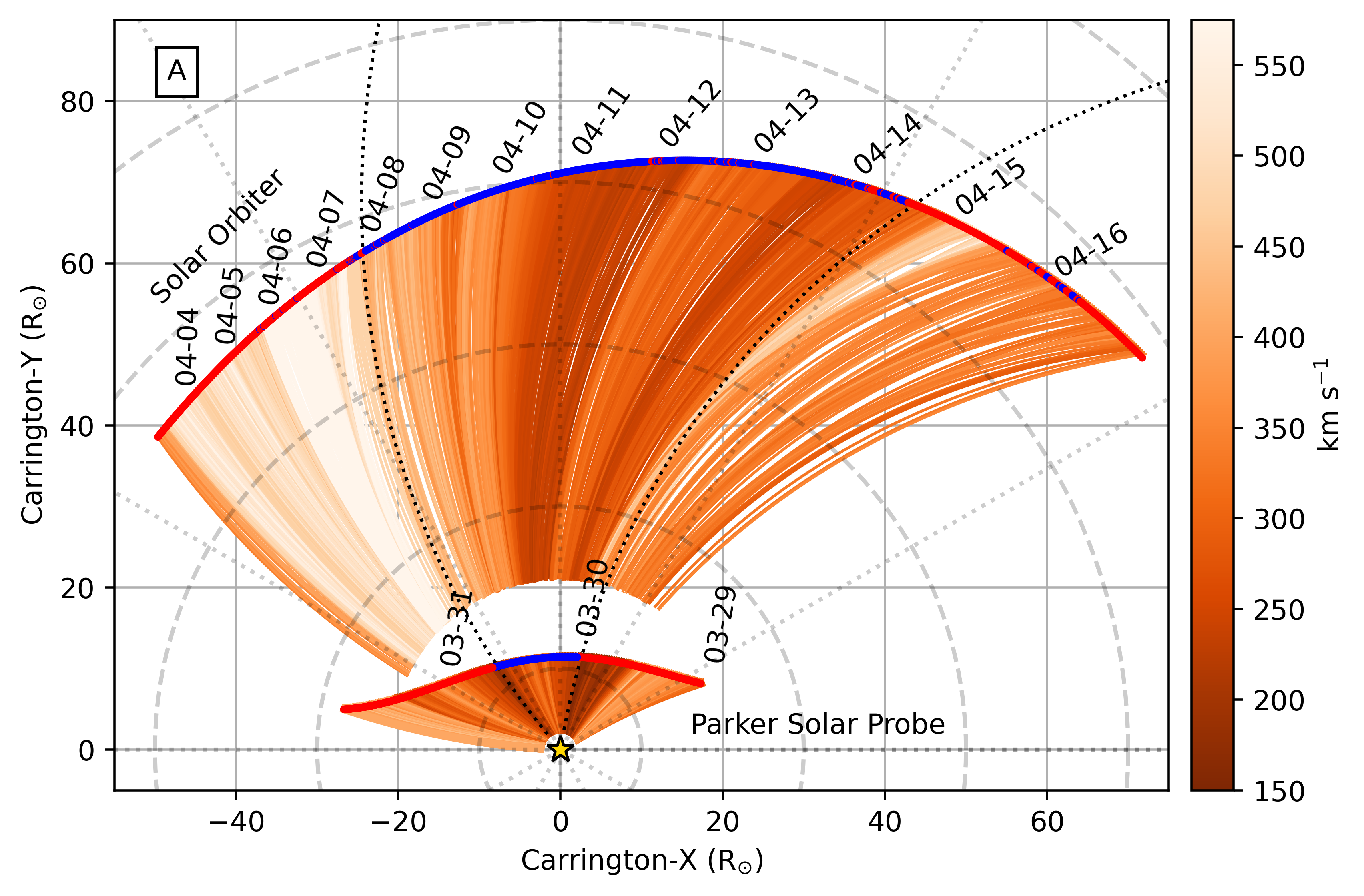}
\includegraphics[width=0.6\textwidth]{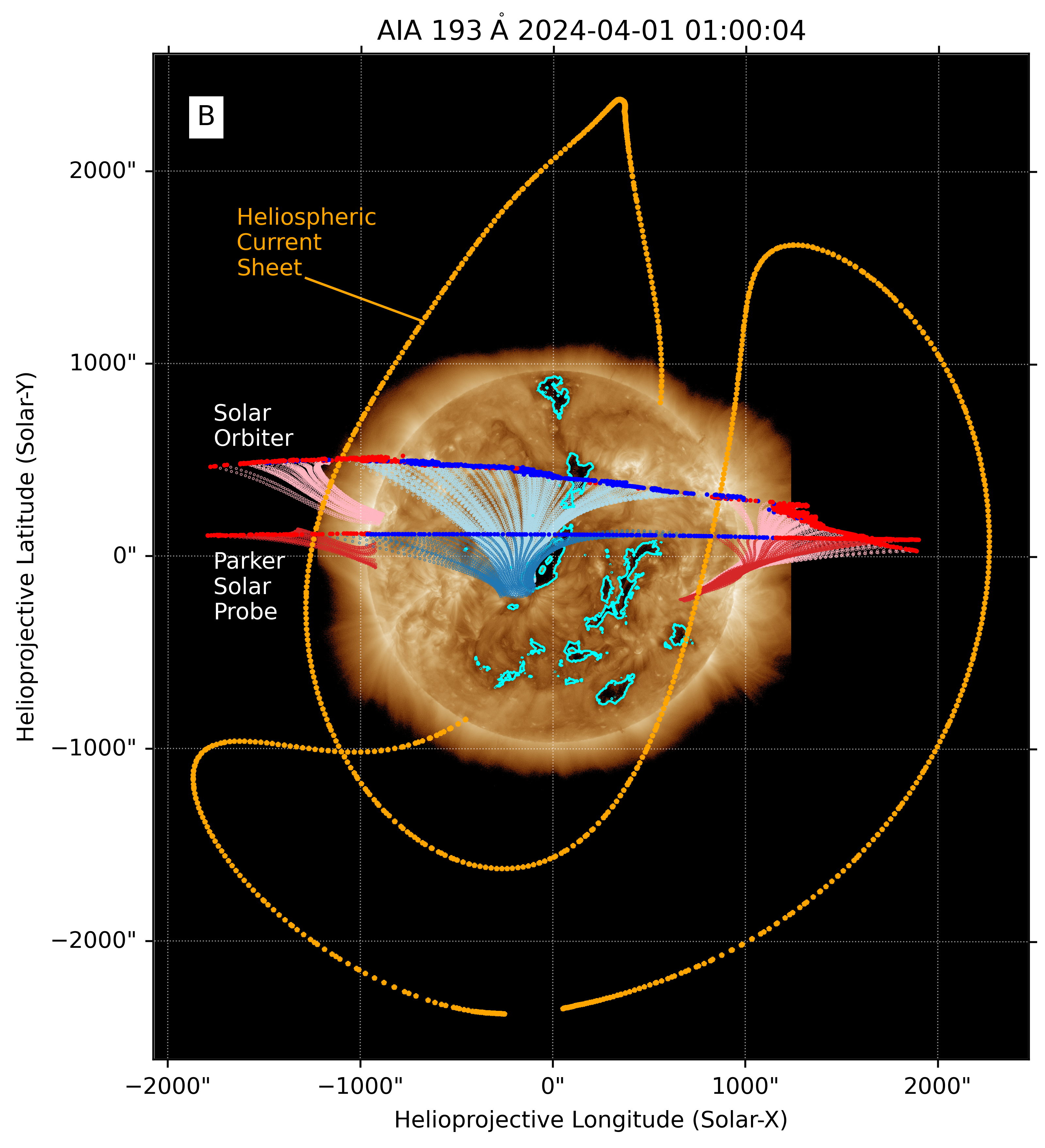}
\caption{Solar wind heliospheric mapping and field line tracing to the solar wind's coronal source. Panel~A shows a top-down view of the \emph{Parker} and Solar Orbiter trajectories in the Carrington (solar co-rotating) coordinate system within the period used in the analysis. The flow lines associated with each trajectory are colored by the local bulk speed measured at each spacecraft. The dashed black lines indicate the approximate location of the CSCs. Panel~B shows the projected flow line footpoints from each spacecraft from panel~A at the source surface, 2.5~$R_{\odot}$, plotted on an EUV image of the Sun on 1~April 2024. Coronal holes on the Sun are outline by cyan. The spacecraft trajectories are colored by the interplanetary magnetic field polarity where blue is negative and red is positive. The trajectories are mapped to the solar surface by tracing the associated PFSS magnetic field extrapolation to the photosphere where the heliospheric current sheet is plotted in orange. }
\label{fig1_orbit_mapping}
\end{figure}

\subsection{Solar-Heliospheric Connection} \label{subsec:connection}

Figure~\ref{fig1_orbit_mapping} shows the tracked coronal hole source, its resulting solar wind stream and the geometry and timing of the \emph{in situ} measurements in traversing it. Figure~\ref{fig1_orbit_mapping}A shows a top-down view of the \emph{Parker} and Solar Orbiter trajectories in the Carrington (solar co-rotating) coordinate system within the period used in the analysis. Both trajectories are colored by the polarity of the local interplanetary magnetic field measurement, with red and blue indicating positive and negative polarities, respectively. \emph{Parker} spirals colorized by the measured velocity show the ballistically-propagated trajectories, assuming a constant speed \citep{Nolte1973,Badman2020}, corresponding to both sets of measurements, see Appendix Section \ref{sec:methods:PFSS_helio_mapping}. Together, both spacecraft are observed to cross through an aligned negative polarity stream with \textit{Parker} crossing it at 11.60~$R_\odot$ on 31~March 2024 and Solar Orbiter at 72.06~$R_\odot$ from 8~April 2024 to 14~April 2024. The current sheets bounding this interval serve as robust alignment markers which independently identify the matched streams.

Potential-field Source-surface (PFSS; \cite{altschuler1969_pfss, Schatten1969_pfss}) modeling is used to trace the coronal magnetic field lines from the source surface (2.5~$R_{\odot}$) spacecraft locations to their photospheric origin, see previous work implementing this technique \citep{Zhao2017, Stansby2019, Stansby2021, Badman2020, Baker2023, Ervin2024, Ervin2024b, Rivera2025, Silwal2025}. The source surface endpoints of the ballistically mapped trajectories shown in Fig.~\ref{fig1_orbit_mapping}A are plotted and colorized by measured polarity at each spacecraft, over an extreme ultraviolet (EUV) 193~{\AA} image of the Sun as seen from Earth on 1~April 2024 in Figure~\ref{fig1_orbit_mapping}B. Comparing the source locations to the EUV image, and the location of the model's predictions for magnetic polarity inversions (orange curve) to the measured locations of these polarity changes, we identify robustly that these streams measured in the young solar wind originate from a narrow, negative polarity coronal hole, visible near disk center in the EUV image in Figure~\ref{fig1_orbit_mapping}B.

\begin{figure}[h!]
\centering
\includegraphics[width=0.37\textwidth]{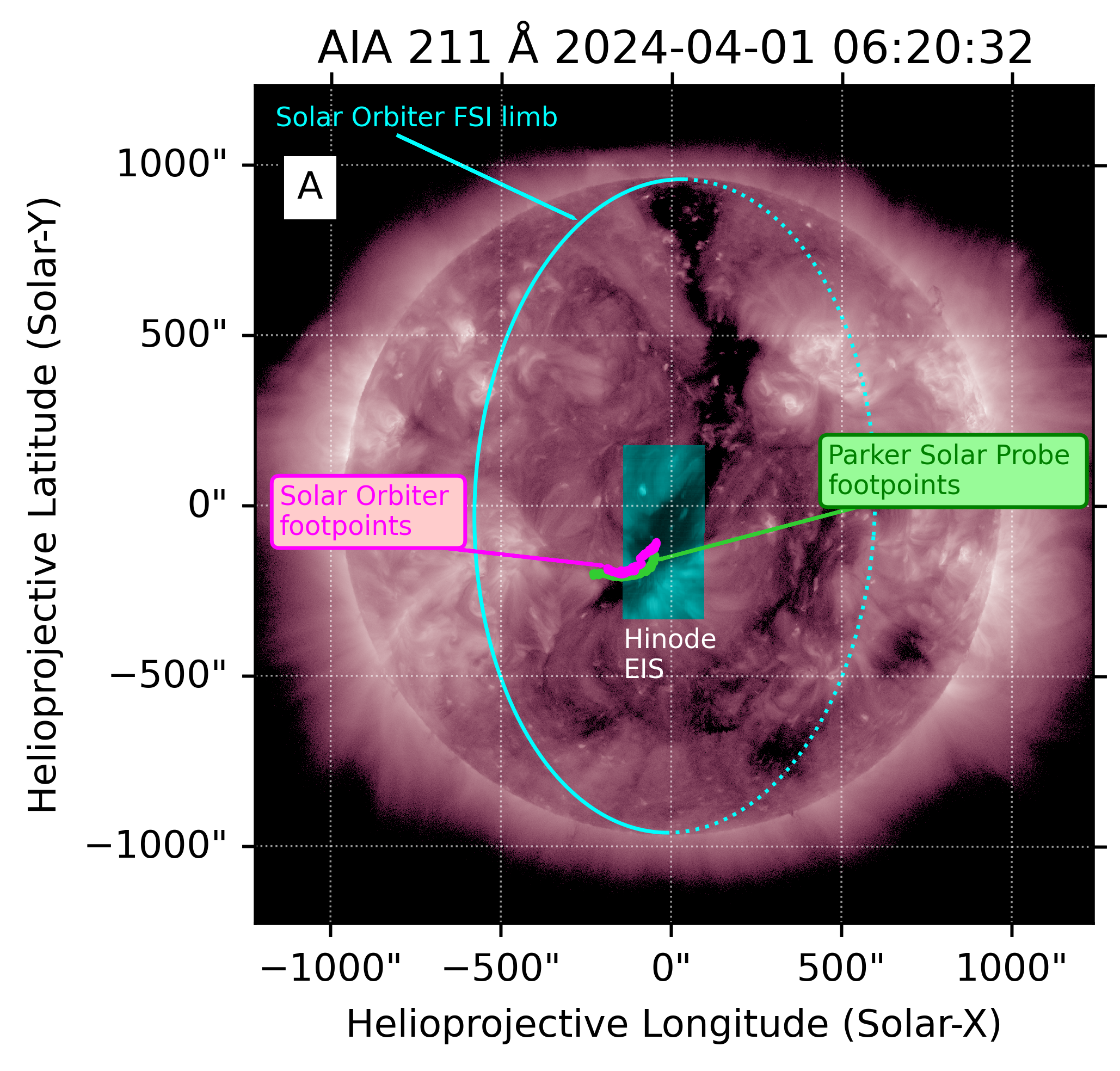}
\includegraphics[width=0.37\textwidth]{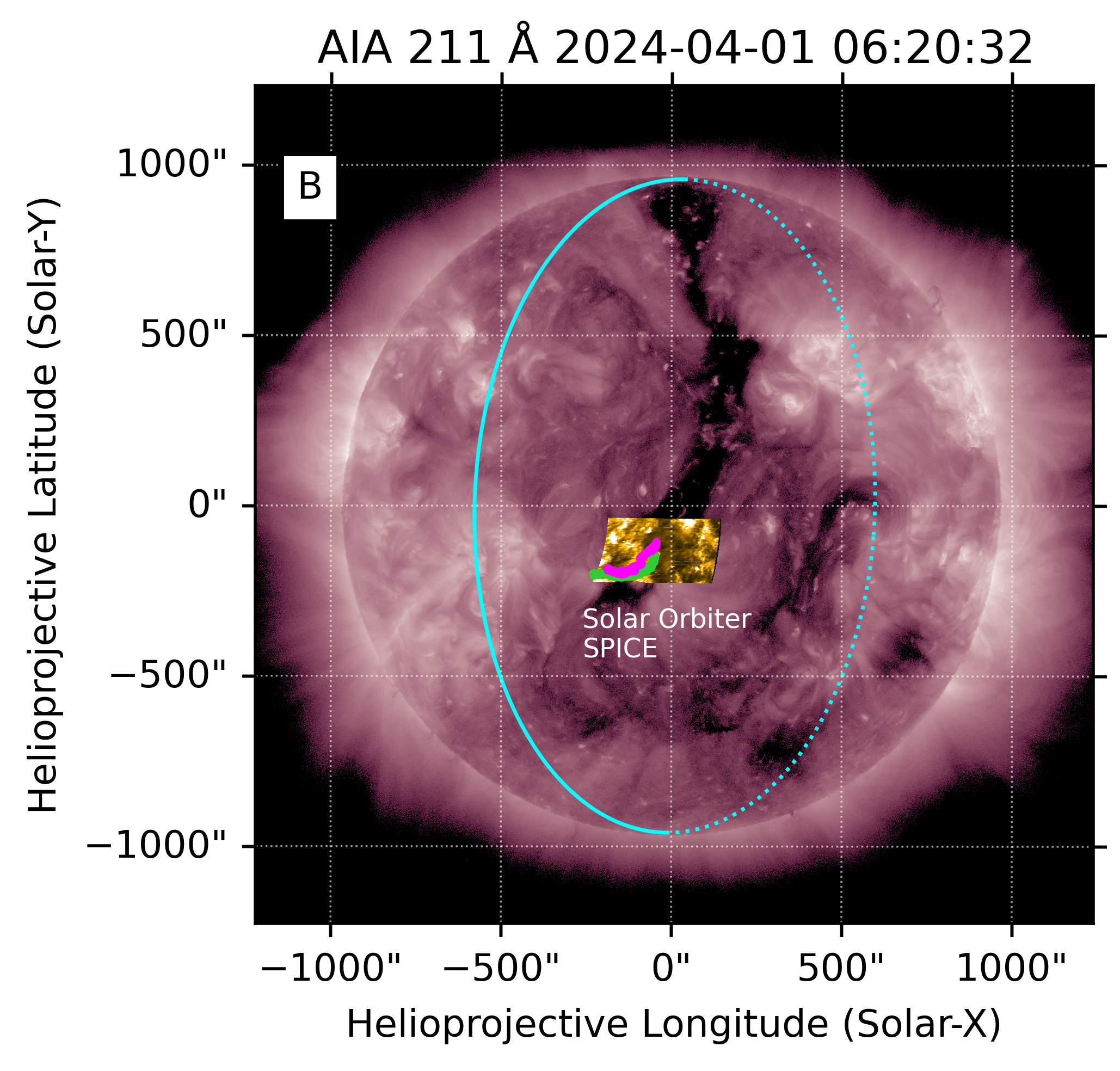}
\includegraphics[width=0.45\textwidth]{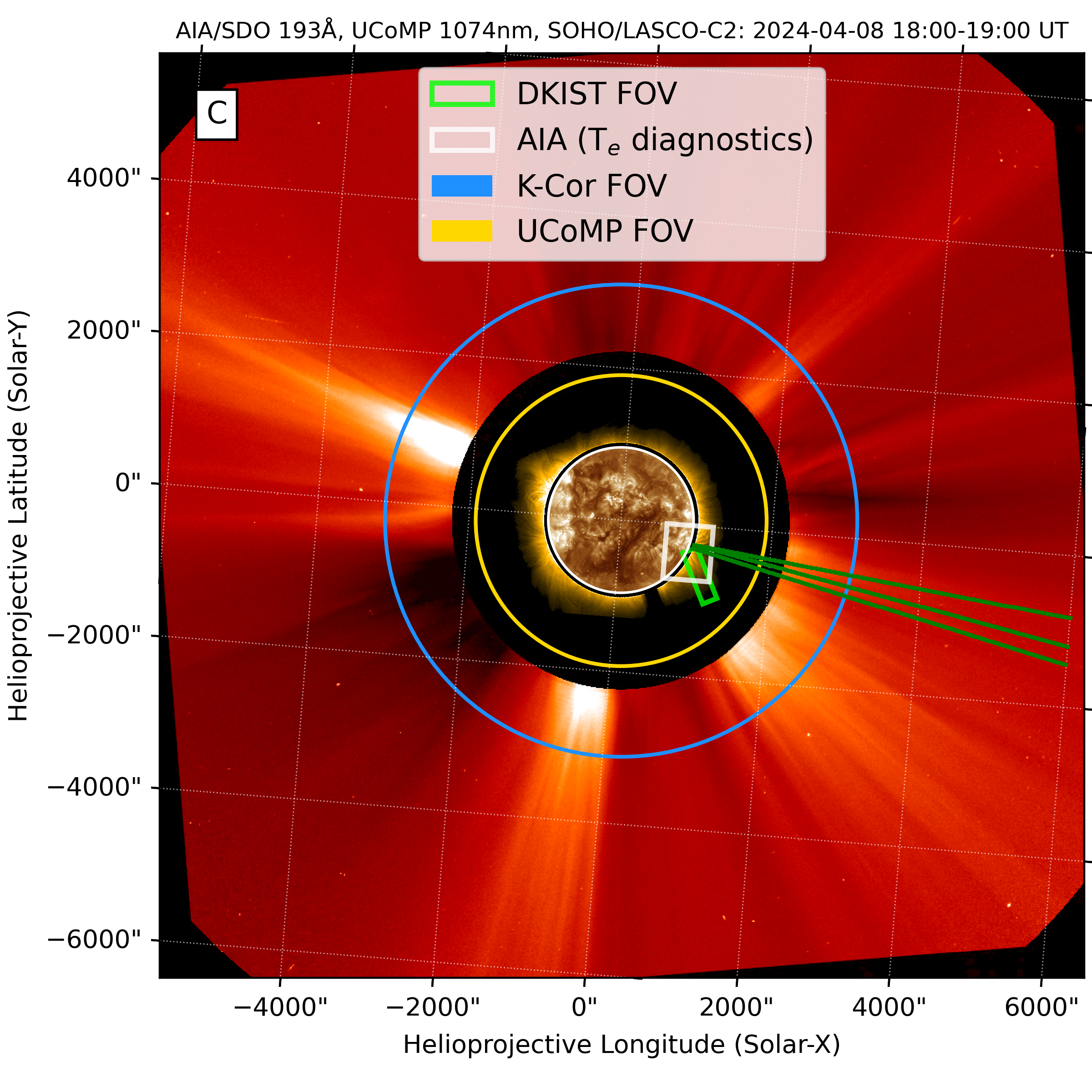}
\caption{Summary of the FOV for all instrumentation on 1~April and during the TSE. Panel~A, B show an EUV image from SDO/AIA on the Sun-Earth line along with the Solar Orbiter and Parker Solar Probe photospheric footpoints and the Solar Orbiter FSI limb during this date. Panel~A and B include the FOV of Hinode/EIS and Solar Orbiter/SPICE on 1~April. Panel~C shows the FOV of SDO/AIA, MLSO UCoMP, and SOHO/LASCO-C2 instruments during 8 and 9~April where the coronal hole is observed on the west limb. The green lines are the locations where the coronal hole was sampled along the radial direction. }\label{fig2_coordination}
\end{figure}

\subsection{Coronal Observations}\label{subsec:remote}

We note that due to the nature of the numerous and heterogeneous  remote sensing data sets included in this work (see Table~\ref{tab:remote_obs_diag_1}), each requiring a bespoke processing methodology described previously in the literature, we provide a fully referenced accounting of the methodology in the Appendix~\ref{sec:methods} as referred to throughout this section, while restricting the main text here to the more salient details for brevity and clarity.  

Having identified the coronal source prior to the TSE, various remote sensing instruments targeted this region, as illustrated in Figure~\ref{fig2_coordination}. Figure~\ref{fig2_coordination}A, B shows the coronal hole source region again near disk-center as viewed from Earth in the 211~{\AA} waveband on 1~April 2024 along with the spacecraft photospheric footpoints and the field of view of the Solar Orbiter/SPICE and Hinode/EIS instruments, which we use to determine the outflow speed and source electron temperature in the high chromosphere and at the base of the corona, respectively (see Appendix~\ref{sec:methods}). 

Figure~\ref{fig2_coordination}C illustrate the field of view of instruments which took observations on 8 and 9~April 2024, during and immediately following the TSE. By this time, the coronal hole source had rotated to the western limb from the Sun-Earth line perspective and therefore the properties of its outflow can be identified in the low corona off the limb, as evidenced by the approximate location of the coronal hole observed in quadrature (Appendix Figure~\ref{fig:EDF7_fsi_ch_outline}) reduced \ion{Fe}{13} emission and co-located open magnetic field lines (Appendix Figure~\ref{fig:EDF0_UCoMP_CH_onlimb}). In Figure~\ref{fig2_coordination}C, field of view annotations are placed on top of an image blending an EUV image of the solar disk, \ion{Fe}{13} emission from the UCoMP coronagraph and a SOHO/LASCO-C2 white light coronagraph image of scattered white light intensity out to a few solar radii. Electron temperature inversions were also obtained from SDO/AIA. Tomographic inversions were performed using SOHO/LASCO-C2 data to obtain density measurements and velocity estimates. The co-observing instruments include DKIST Cryo-NIRSP which provided velocity fluctuation measurements via non-thermal line broadening, K-Cor coronagraph which provided density estimates via polarized brightness measurements, and lastly UCoMP which provided further independent density estimates and, crucially, magnetic field estimates via a recently implemented coronal seismology technique \citep{Yang2020b}.

As shown in Figure~\ref{fig2_coordination}, some observations provide global coverage, while others only capture sub-regions of the Sun. For observations taken on-disk prior to the TSE, we utilize the footpoint mapping locations to select sub-regions from which to extract physical parameters, e.g., Appendix Figures~\ref{fig:EDF3_spice_aia_temp}, \ref{fig:EDF5_EIS_SPICE_doppshifts}. For observations taken during and just after the TSE (8 \& 9~April, with the source on the west limb), we sample the properties of the coronal hole associated with the green radial lines shown in Figure~\ref{fig2_coordination}C  (see also Appendix Figures~\ref{fig:EDF1_UCoMP_den_b}, \ref{fig:EDF2_k_cor_dens}, \ref{fig:EDF4_metis_dens_vels}) which coincide with the location in Appendix Figure~\ref{fig:EDF0_UCoMP_CH_onlimb}.  

A full accounting of all the observations and diagnostics utilized in the study are explained in Appendix Section~\ref{sec:methods} and shown in Table~\ref{tab:remote_obs_diag_1}. The table includes the instruments used, observation timeframes, spatial coverage, and the methodology utilized for the indicated derived diagnostic. The associated observations and derived diagnostics are shown in Figures~\ref{fig:EDF5_EIS_SPICE_doppshifts}, \ref{fig:EDF1_UCoMP_den_b}, \ref{fig:EDF2_k_cor_dens}, \ref{fig:EDF4_metis_dens_vels} and \ref{fig:EDF3_spice_aia_temp}. This broad array of observations, alignments, pointing limitations, and processing techniques emphasizes the challenging nature of collecting all quantities needed (density, velocity, temperature and magnetic field, fluctuation quantities) to fully characterize the stream's initial conditions.

\subsection{In Situ Solar Wind Observations}\label{subsec:in-situ}

\begin{figure}[h]
\centering
\includegraphics[width=0.73\textwidth]{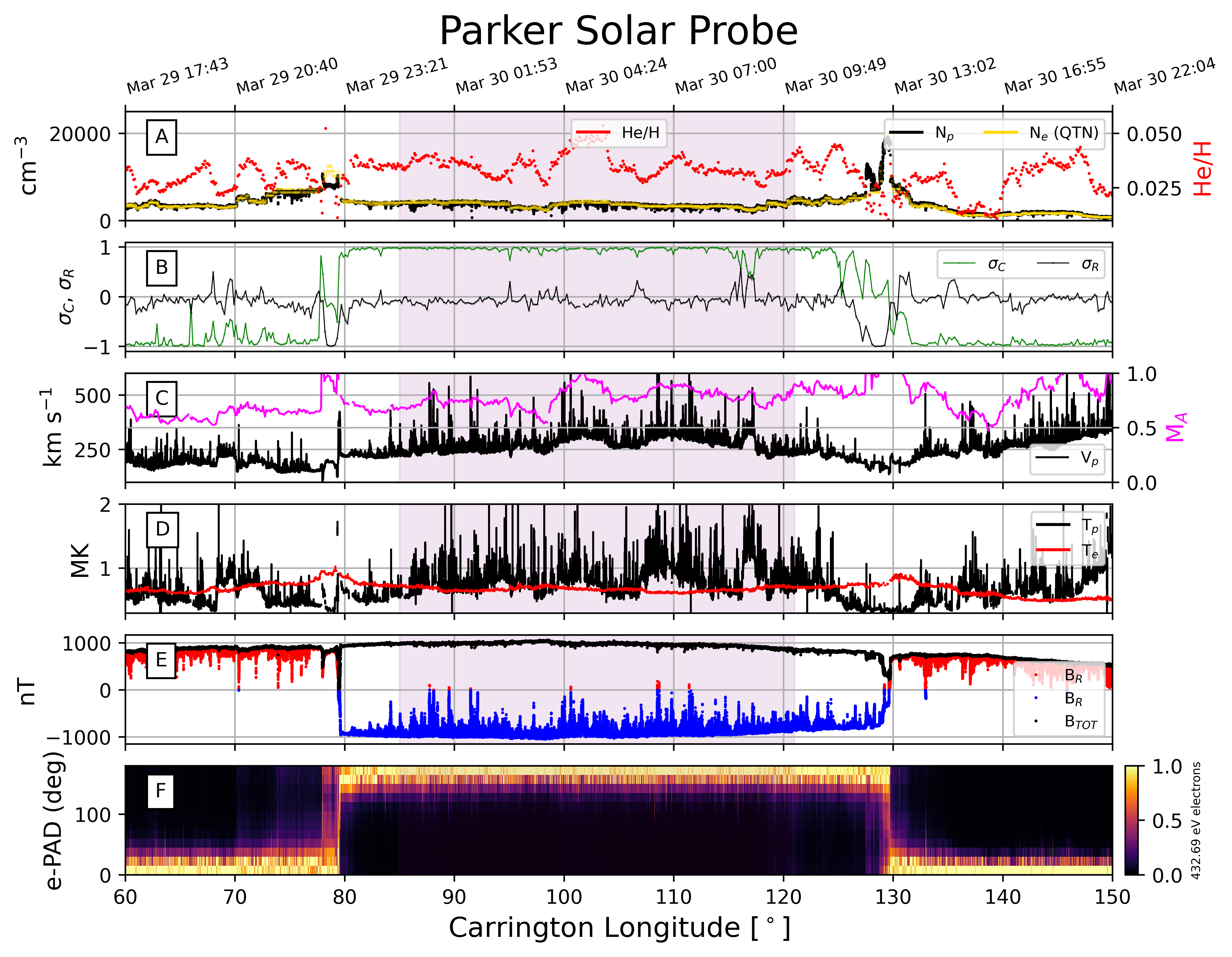}
\includegraphics[width=0.73\textwidth]{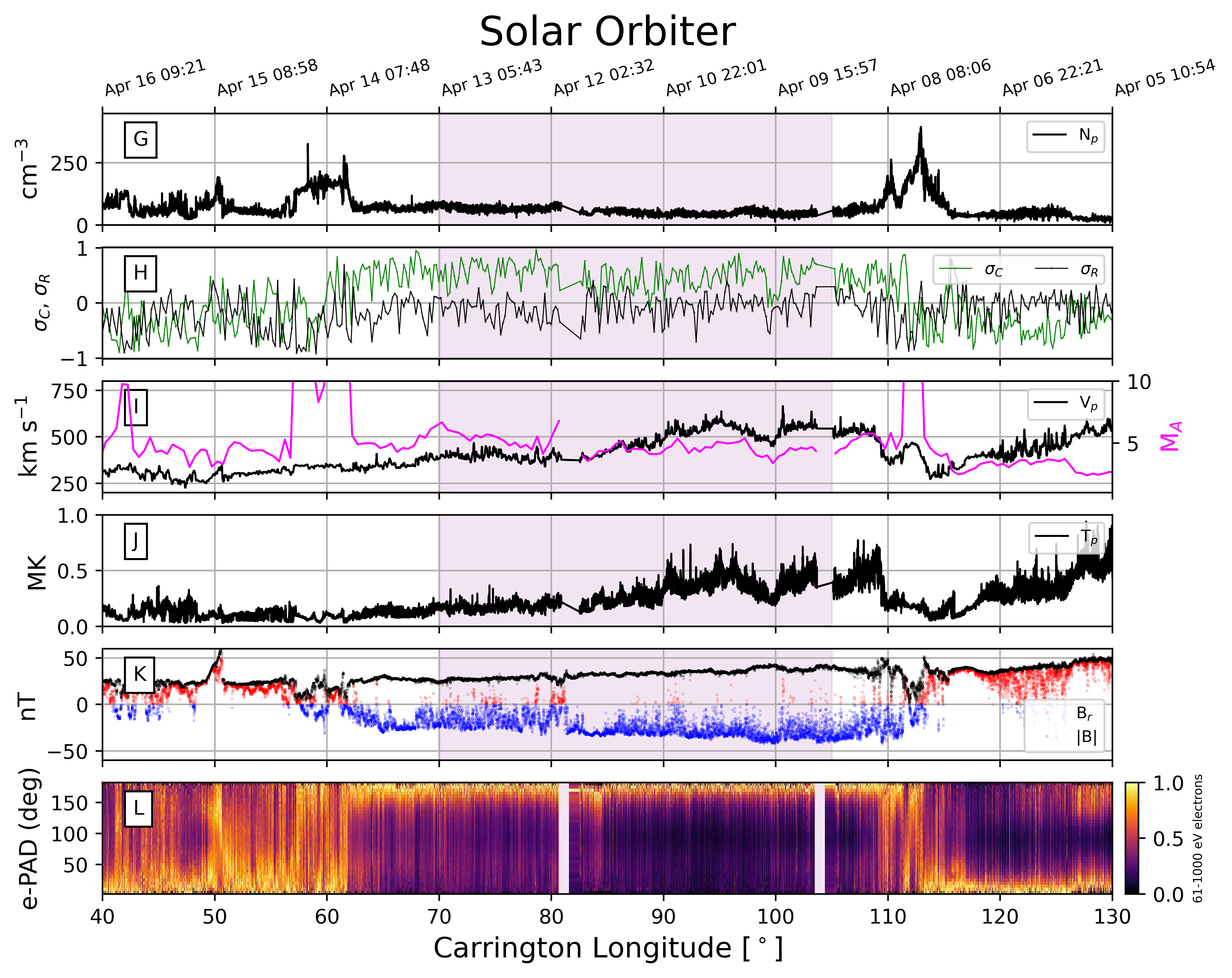}
\caption{\emph{In situ} observations of \emph{Parker} and Solar Orbiter plotted against spacecraft Carrington longitude as shown in the bottom panels and observations time, as indicated with the top panel. The purple shaded region indicates the period or longitude range at each spacecraft used in the study. (A)~shows the proton density (black), electron density (gold), and helium abundances (red). (B)~shows residual energy, $\sigma_{R}$ (black) and cross-helicity, $\sigma_{C}$ (green). (C)~the solar wind's radial velocity (black) and the Mach number (magenta), (D)~proton (black) and electron (red) temperature, (E)~magnetic field magnitude (black) and radial component (red/blue). (F~ shows the pitch angle distribution of the electron strahl. (G, H, I, J)~show the proton density, $\sigma_{C}$/$\sigma_{R}$, radial speed/Mach number, and proton temperature, respectively. (K)~shows the magnetic field magnitude and radial component (red/blue), and (L)~pitch angle distribution of the electron strahl. } \label{fig3_insitu}
\end{figure}

Figure~\ref{fig3_insitu}A--L shows the \emph{in situ} observations associated with the coronal hole stream. All of the quantities are plotted against each spacecraft's Carrington longitude (bottom $x$-axes) and Universal Time (UT) (top $x$-axes, note that \emph{Parker}'s time axis is inverted due to its angular motion being faster than the Sun's at this time). The coronal hole wind was sampled \emph{in situ} throughout \emph{Parker}'s 19th perihelion pass centered on 30~March 2024, roughly one week prior to the TSE. Figure~\ref{fig3_insitu} shows observations of \emph{Parker} and Solar Orbiter corresponding to the timeframe shown in Figure~\ref{fig1_orbit_mapping}A. From top to bottom, Panels A--F show at \emph{Parker}, proton density (black) and helium abundance (He/H, red), cross-helicity ($\sigma_{C}$, green) and residual energy ($\sigma_R$, black), radial speed (black) and Alfv\'en Mach number ($M_A=$V$_{r}$/V$_{A}$ where the Alfv\'en speed is V$_{A}=\frac{B}{\sqrt{\mu_{0}n m_p}}$), the proton (black) and electron (red) temperature, magnetic field magnitude and radial component, B$_{R}$, and the column normalized electron pitch angle distribution (e-PAD) of suprathermal electrons across 0--180$^{\circ}$ for the passband of 432.69~eV. 

The coronal hole stream was also intercepted by Solar Orbiter later in time (covering 7--14~April) and connected to the corresponding \emph{Parker} measurements through a comparison of their source surface longitude footpoints and the identification of corresponding current sheet crossings (CSCs). Figures~\ref{fig1_orbit_mapping}A  and \ref{fig3_insitu}G--L show the Solar Orbiter trajectory and corresponding \emph{in situ} observations for this period across Carrington longitude. As with \emph{Parker}, from top to bottom we plot the proton density, $\sigma_C$/$\sigma_R$, speed, Alfv\'en Mach number, proton temperature, magnetic field radial component and field strength, and the column normalized electron pitch angle distribution (e-PAD) of suprathermal electrons across 0--180$^{\circ}$ in the energy range of 61--1000~eV.

As mentioned above, the coronal hole solar wind stream is independently demarcated as a negative magnetic polarity interval between two CSCs observed \emph{in situ}, indicated by panels E, F at \emph{Parker} and K, L at Solar Orbiter, and which correspond to the blue region of the trajectory shown in Figure~\ref{fig1_orbit_mapping}. Accordingly, panels E, K show the polarity changes from red to blue to red, corresponding to positive, negative, positive observed at both spacecraft with average locations of 11.60 and 72.06~R$_{\odot}$, respectively. Panels F, K show corresponding changes in the field aligned and anti-aligned outward directed electron strahl, where sharp and large-scale changes in intensity between 0 and 180~degrees is indicative of a CSC. These clear boundaries observed both in \emph{Parker} and Solar Orbiter, along with the alignment and mapping presented in Appendix Section~\ref{subsec:connection}, robustly shows that this interval of measurements corresponds to the stream emerging from the targeted coronal hole and minimize reliance on ballistic mapping techniques. 

To obtain representative bulk properties of the stream at these two measurement points in the solar wind, the relevant quantities are averaged over longitude, as shaded in Figure~\ref{fig3_insitu}. As can be determined from the different quantities plotted, each set of \emph{in situ} data provides the necessary quantities for this study. Further details of the \emph{in situ} properties observed or derived are discussed in Appendix Section~\ref{sec:methods:in_situ_connection}.   

\section{Results}
\subsection{Radial Profiles of Coronal Hole Stream}\label{subsec:profiles}

\begin{figure}[h]
\centering
\includegraphics[width=0.37\linewidth]{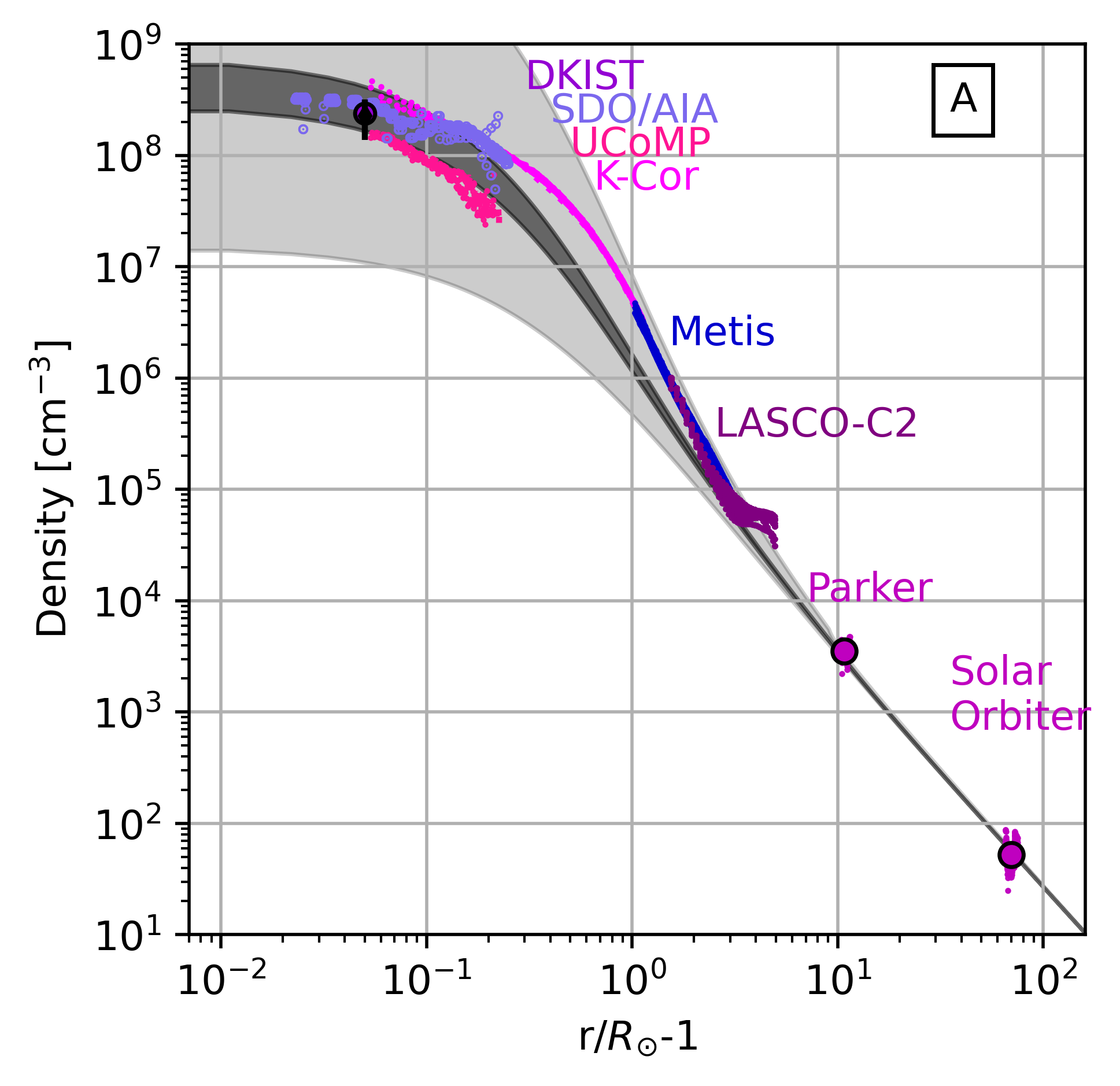}
\includegraphics[width=0.37\linewidth]{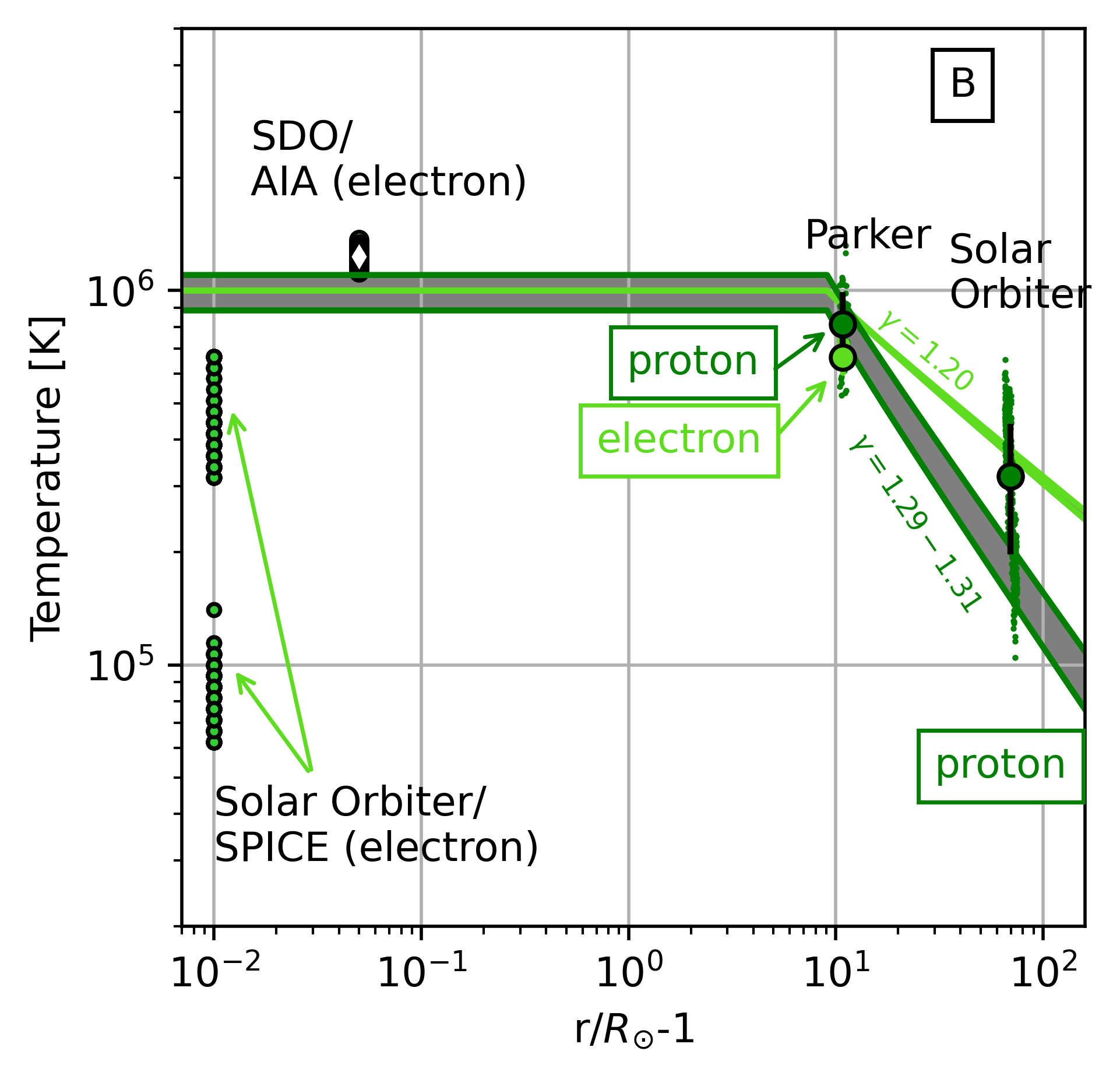}
\includegraphics[width=0.37\linewidth]{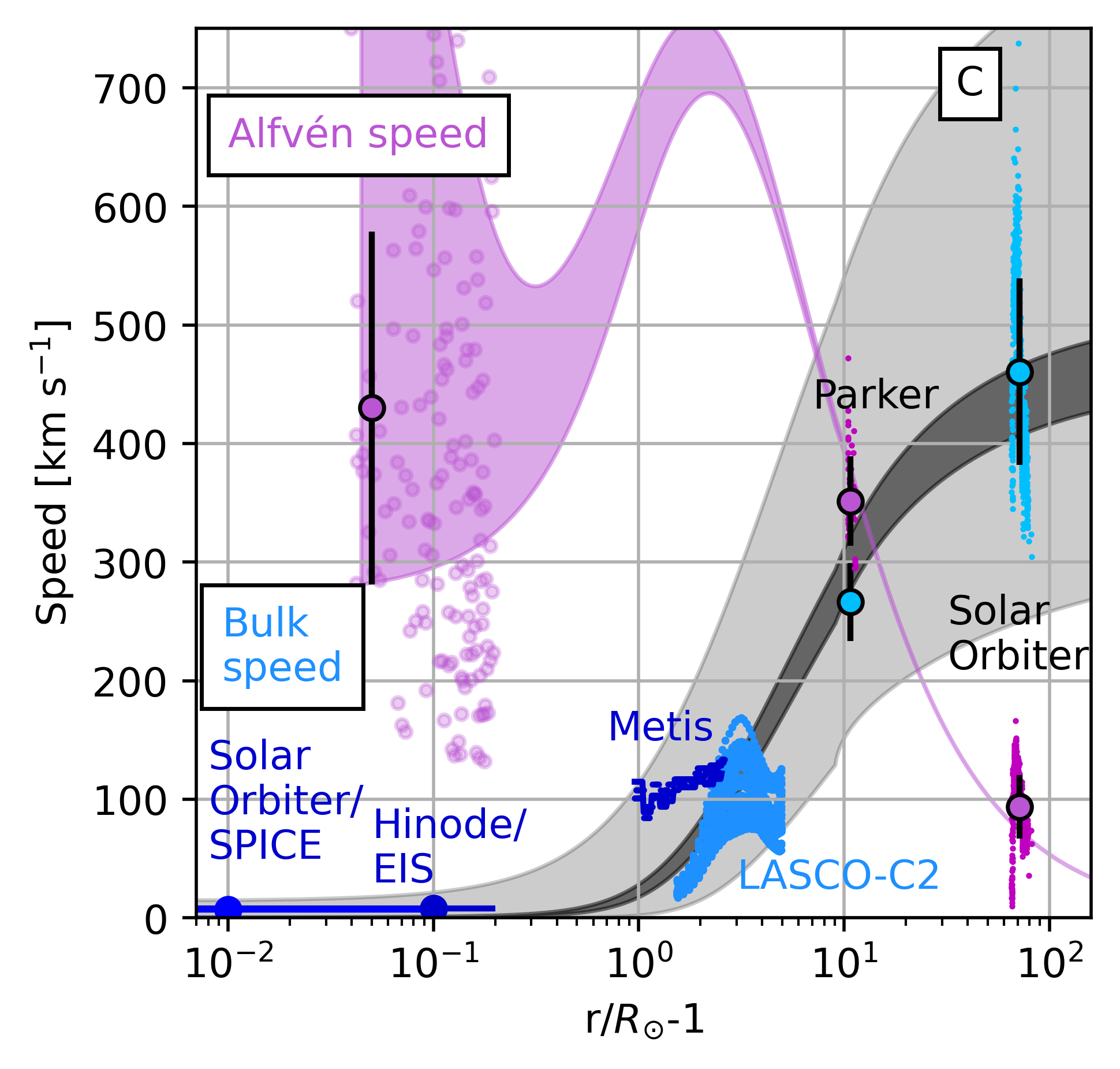}
\includegraphics[width=0.37\linewidth]{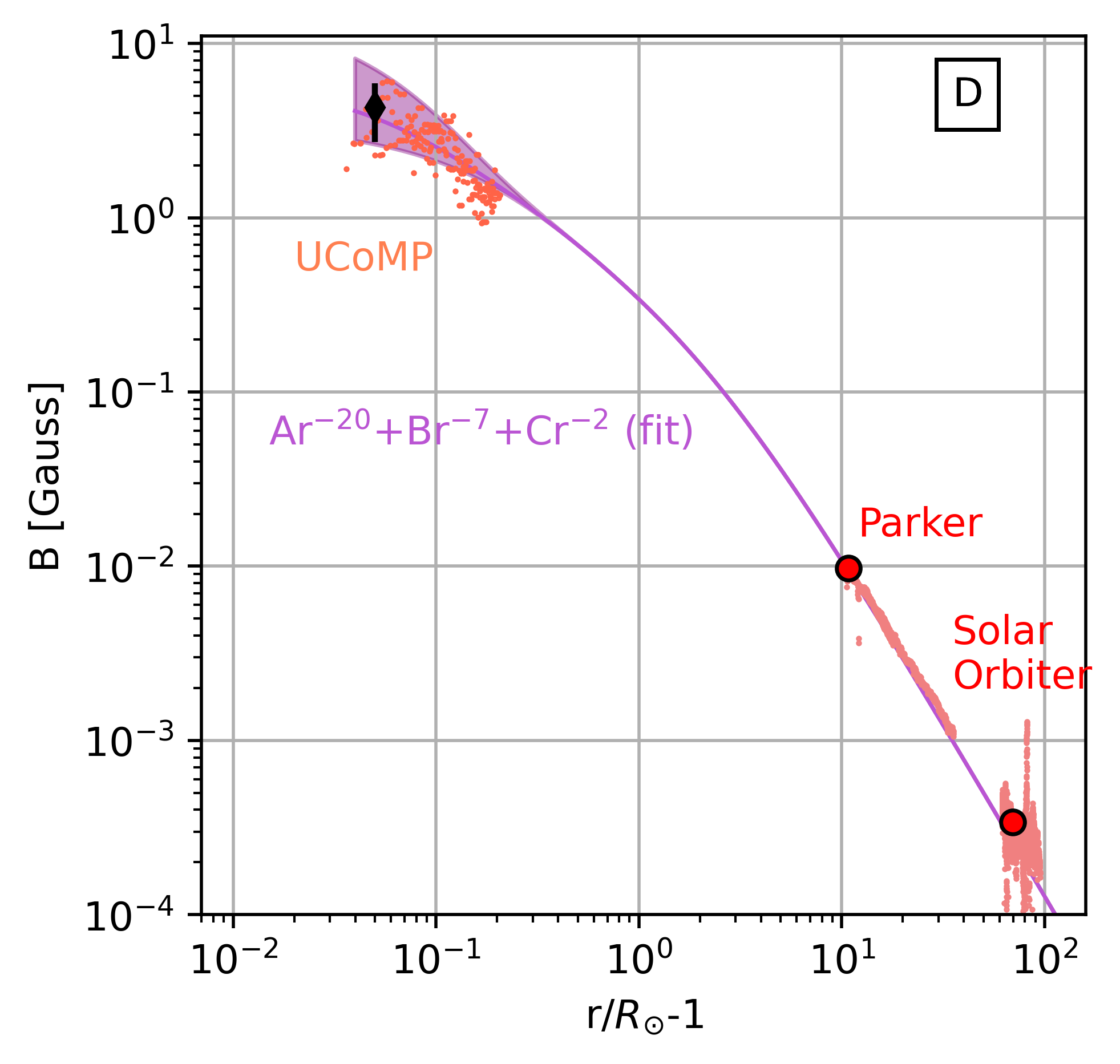}
\includegraphics[width=0.37\linewidth]{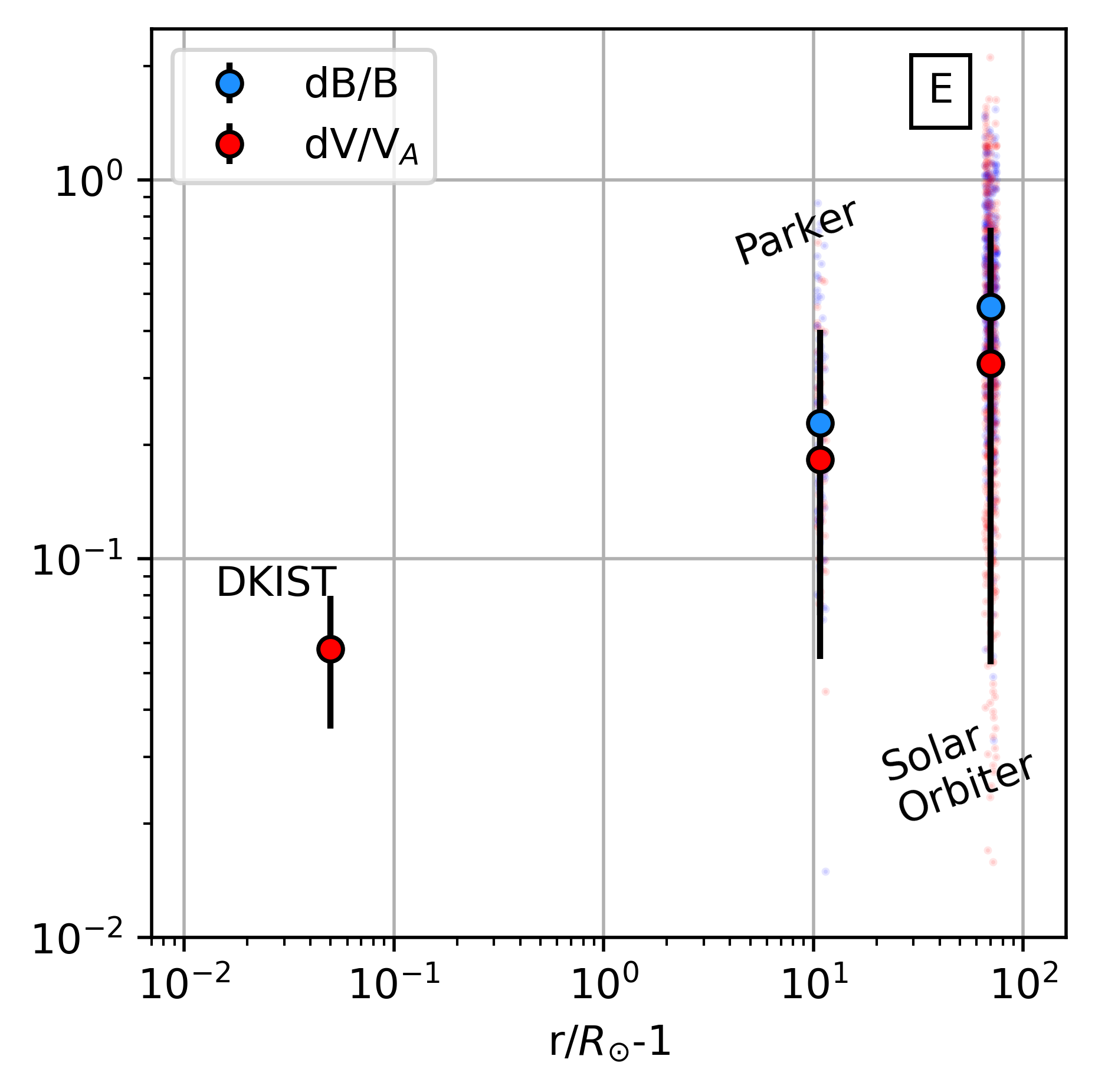}
\includegraphics[width=0.37\linewidth]{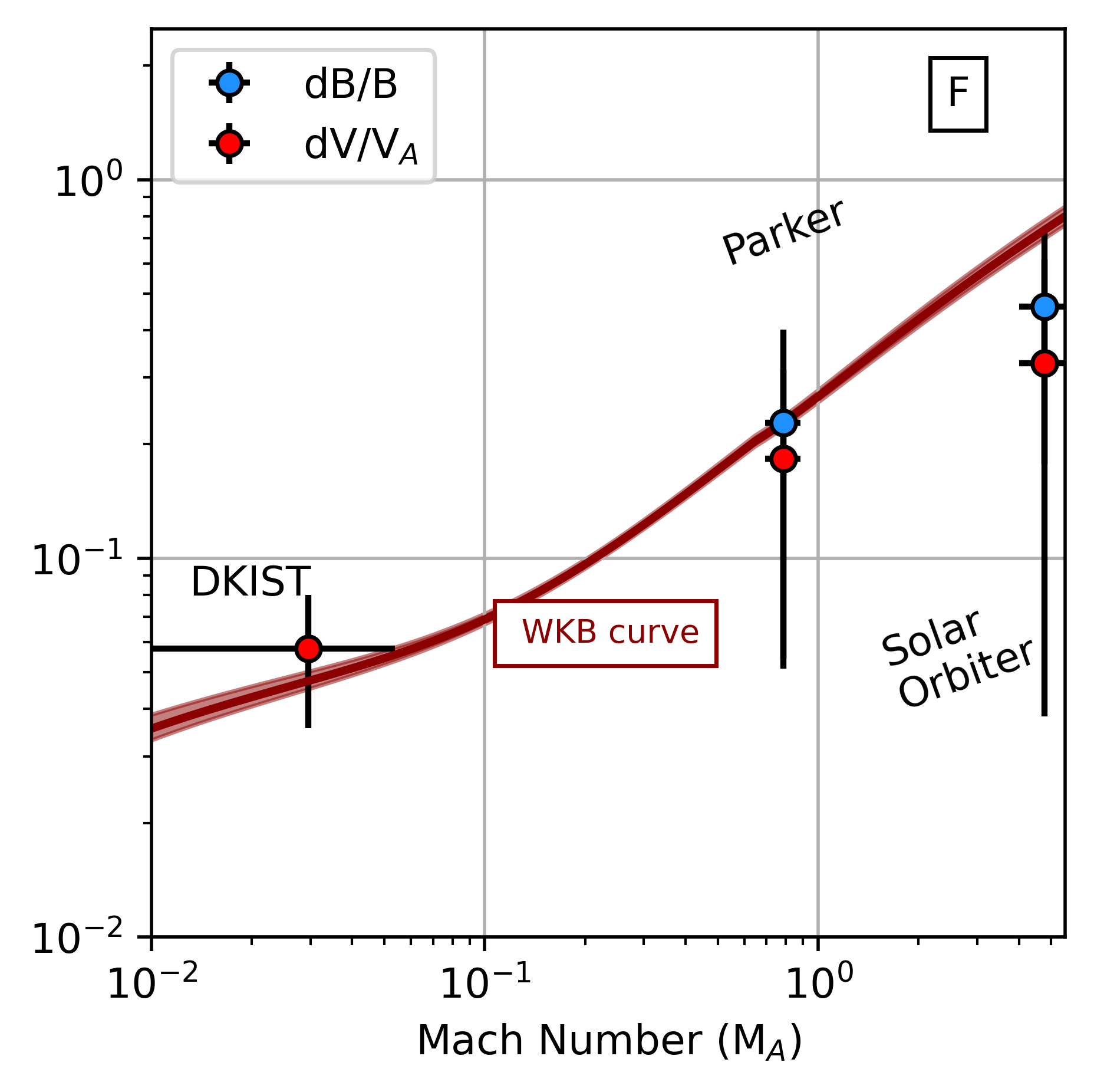}
\caption{The combined coronal hole properties with distance from the Sun and Mach number. Panel~A: Density, B:~electron (lime), proton (green) temperature, C:~bulk speed (shades of blue), Alfv\'en speed (magenta), D:~magnetic field strength (coral/red), E:~normalized magnetic field (blue) and velocity (red) fluctuations, F:~magnetic and velocity fluctuations against Mach number. Panels A, B, D have the average parameter and standard deviation at 1.05~$R_{\odot}$. The background gray shaded region in panels A and C is the solution to the iso-poly + forcing model that captures the full range of statistical solar wind observations, see Appendix Section~\ref{sec:methods:isopoly}. The dark grey shaded region are the selected solution that best `fit' the full set of observations. Panel~B contains proton and electron temperature profiles corresponding to the selected dark grey solutions. Panels C and D contain dotted lines of the computed Alfv\'en speed and a simple fit to the UCoMP, \emph{Parker}, and Solar Orbiter magnetic field strength, respectively.}
\label{fig4_radialplots}
\end{figure}

Figure~\ref{fig4_radialplots} shows the combined remotely derived quantities and \emph{in situ} observations with distance from the Sun. The radial profiles provide a comprehensive set of observational constraints of a moderately fast solar wind stream which reaches a mean speed of 450~km~s$^{-1}$ by 72.06~$R_{\odot}$. All the quantities are plotted as altitude from the surface of the Sun, as $r/R_{\odot}-1$ in log space. The figure panels show the proton/electron density~(A), temperature (electron/lime and proton/dark green in~B), speed (bulk flow in blue and Alfv\'en speed in purple, in~C), magnetic field magnitude~(D),  velocity normalized to the Alfv\'en speed and the magnetic fluctuations normalized to the magnetic field magnitude~(E), and same as panel~E except against Mach number, M$_{A}$. All measurements below $\sim$10~$R_\odot$ are assembled from the patchwork of remote sensing observations described in Section~\ref{subsec:remote} and Appendix~\ref{sec:methods}. At \emph{Parker} and Solar Orbiter distances, we include average bulk properties from the purple shaded regions of Figure~\ref{fig3_insitu} as described in Section~\ref{subsec:in-situ}. The shaded continuous profiles in the density, speed, and temperature profiles are a range of simulated curves from radially expanding ``iso-poly + forcing'' Parker solar wind solutions \citep{Dakeyo2022, shi2022_polyt,Rivera2024} that span the full range of statistical heliospheric parameters, where specific implementation details can be found in Appendix Section~\ref{sec:methods:isopoly} and \citet{Badman2025}. 


\subsubsection{Density}
The coronal hole's electron density was probed using an ensemble of overlapping remote observations (Figure~\ref{fig4_radialplots}A). While results vary below 1~$R_{\odot}$ across different techniques, the standard deviation of these profiles provides the uncertainty for energy flux calculations discussed in Appendix Section~\ref{sec:methods:energy_flux_calc}. Between 1.3--4~$R_{\odot}$, observed densities exceed what is expected from mass flux conservation (dark grey band), which otherwise matches extremely well the combined coronal and \emph{in situ} data. These higher densities reported by K-Cor, Metis, and LASCO-C2 likely stem from a pB overestimation. Because it is difficult to cleanly isolate the dim coronal hole along the LOS, the pB likely includes some foreground and background contributions from surrounding brighter structures. Although there is some variability, overall the coronal hole density in the lower corona (below $\sim$1.3~$R_\odot$) among DKIST, SDO/AIA, and UCoMP align well despite the derived diagnostics utilizing different wavelength ranges (EUV and near-infrared) and methodology, see details in Appendix Section~\ref{sec:methods:density}.

\subsubsection{Temperature}
In Figure~\ref{fig4_radialplots}B, from remote observations, we are limited to only electron temperature estimates, at \emph{Parker} both proton and electron temperatures of the solar wind are available, with only proton temperature measurements available at Solar Orbiter for this interval. At the Sun, the electron temperature exhibits significant heating from the chromosphere (SPICE) across the transition region into the coronal base (AIA) as the temperature rises to above 1~MK. The derived electron temperature at the base, 1.23~MK, is typical of coronal holes at these altitudes (see details in Appendix Section~\ref{sec:methods:temperature}). Beyond the low corona, the modeled proton temperature profiles match well the evolution from this $\sim$1~MK state to the cooling observed between the \emph{in situ} measurements and is plausibly explained via the ``iso-poly'' profiles. However, the lack of available estimates \citep{Rivera2025_IAU} of temperature beyond the low corona and below \emph{Parker}'s orbit limit our knowledge of the stream's true radial profile at these altitudes. The iso-poly profile parameters for the shaded solutions are: $T_{p0} = 0.9-1.1$~(MK) and $\gamma_{p}=1.29-1.31$, and  $T_{e0} = 1$~(MK) and $\gamma_e=1.20$.

\subsubsection{Bulk Speed and Alfv\'en Speed}
The iso-poly + forcing acceleration profiles match relatively well to the observed radial speed derived from SPICE, EIS, Metis, and SOHO/LASCO-C2 as seen in Figure~\ref{fig4_radialplots}C. However, some discrepancy is found in the derived speed of Metis and SOHO/LASCO-C2 at lower altitudes ($\sim$1~$R_{\odot}$) which later converges by 3~$R_{\odot}$ with the LASCO measurements hewing closer to expected mass conservation. In the heliosphere, the modeled speed curves intercept both \emph{Parker} and Solar Orbiter speeds, indicating the acceleration driven by the combination of the observed thermal and empirical Alfv\'en wave pressure used in the iso-poly profiles are appropriate throughout.

The Alfv\'en speed at the Sun, \emph{Parker}, and Solar Orbiter distances are computed or derived from observations and presented in the same figure. The shaded magenta curve is the derived Alfv\'en speed based on the fit from the magnetic field, in Figure~\ref{fig4_radialplots}D, discussed in the next section, and the dark grey density profile from Figur~\ref{fig4_radialplots}A. Direct measurements of the Alfv\'en speed in the low corona (magenta dots from Fig.~ure\ref{fig:EDF8_UCoMP_b}C) are obtained from measurements of the phase speed of fluctuations in the coronal hole stream observed above the limb, derived using the coronal seismology technique described in \cite{Yang2020a, Yang2024} and Appendix Section~\ref{sec:methods:magnetic_field}. The plot also includes a mean value with standard deviation at 1.05~$R_{\odot}$ (i.e., approximately the same heights as DKIST and AIA measurements of the other key quantities) that is used for the energy flux calculations, Section~\ref{subsec:energy-fluxes}. Although we find large variability in the phase speed at the Sun, especially with increasing altitude as the signal to noise of those observations degrades, the shaded Alfv\'en speed profiles show good compatibility with the mean $V_A$ at 1.05~$R_\odot$, \emph{Parker}, and Solar Orbiter indicating self-consistent solar wind and magnetic radial profiles.

\subsubsection{Magnetic Field}
Figure~\ref{fig4_radialplots}D compares plane-of-sky (POS) magnetic field measurements from UCoMP (coral) from Figure~\ref{fig:EDF1_UCoMP_den_b}A. A fit (dashed black) connects UCoMP and \emph{in situ} observations at larger distances (red) with a sum of three power laws (see annotation in figure) with a 1/$r^2$ term dominating at solar wind distances. A power-law different than $r^{-2}$ close to the Sun suggests non-radial expansion in the low corona while near spherical expansion farther from the Sun. This non-radial expansion low down in the corona is consistent with the expansion factor calculation discussed in the energy flux calculation in Section~\ref{subsec:energy-fluxes} and Appendix Section~\ref{sec:methods:energy_flux_calc}. Given the relatively large spread in the UCoMP measurements, the power law parameters are varied to produce a minimum and maximum curve which captures this spread, and is used along with uncertainty in the density profile to produce the range of Alfv\'en speed profiles shaded in purple in Figure~\ref{fig4_radialplots}D. The functional form of the minimum and maximum curve is given as, $\vert B\vert = Ar^{-20}+Br^{-7}+Cr^{-2}$, where for the minimum: $A = 0.1$ , $B = 2$, and $C = 1.3$, and the maximum, $A = 12$ , $B = 2$ , and $C = 1.3$.  

\subsubsection{Magnetic and Velocity Fluctuations}
Lastly, Figure~\ref{fig4_radialplots}E and~F presents the measurements of $dB/B$ and $dV/V_A$ with distance from the Sun and against mach number~(M$_A$), respectively. Panel~F also includes the predicted Wentzel–Kramers–Brillouin~(WKB) approximation amplitude scaling for dissipation-free fluctuations (dark red), normalized to match the \emph{Parker} observations, described in Appendix Section~\ref{sec:methods:WKB_curves}. Observed line-of-sight fluctuation amplitudes in the corona are traverse to the magnetic field geometry suggesting they are Alfv\'enic in nature. Additionally, we find no significant \ion{Fe}{13} intensity variability within the FOV of the coronal hole on the plane of sky, suggesting relatively small density fluctuations and small compressibility (panel~A, Figure~\ref{fig:EDF6_DKIST_NTW}), and that therefore Alfv\'enic (non-compressive) fluctuations dominate. A detailed analysis of compressibility would include the comparison of the line amplitude and Doppler velocities in time, however the observations taken during the present period are limited to a single raster \citep{Morton2025}. 

A comparison of the fluctuations with distance from the Sun (Panel~E) indicates an increase in these fluctuation with heliocentric distance and a growing deviation between the velocity and magnetic fluctuations at \emph{Parker} and Solar Orbiter, as also indicated by a slightly more negative residual energy (Figure~\ref{fig3_insitu}B, H), noted before for increasing Mach number \citep{Sioulas2025}. In panel~F, plotting the same quantities against Mach number and thereby minimizing biases from the specific radial profiles (see Appendix Section~\ref{sec:methods:WKB_curves}), we find that the fluctuations follow WKB extrapolations within measurement uncertainties between the low corona out to \emph{Parker}'s orbit while deviating beyond \emph{Parker}'s orbit. This result would be consistent with recent statistical observations from \emph{Parker} below and above the Alfv\'en surface \citep{Sioulas2025}.

\begin{figure}[h]
\centering
\includegraphics[width=0.37\linewidth]{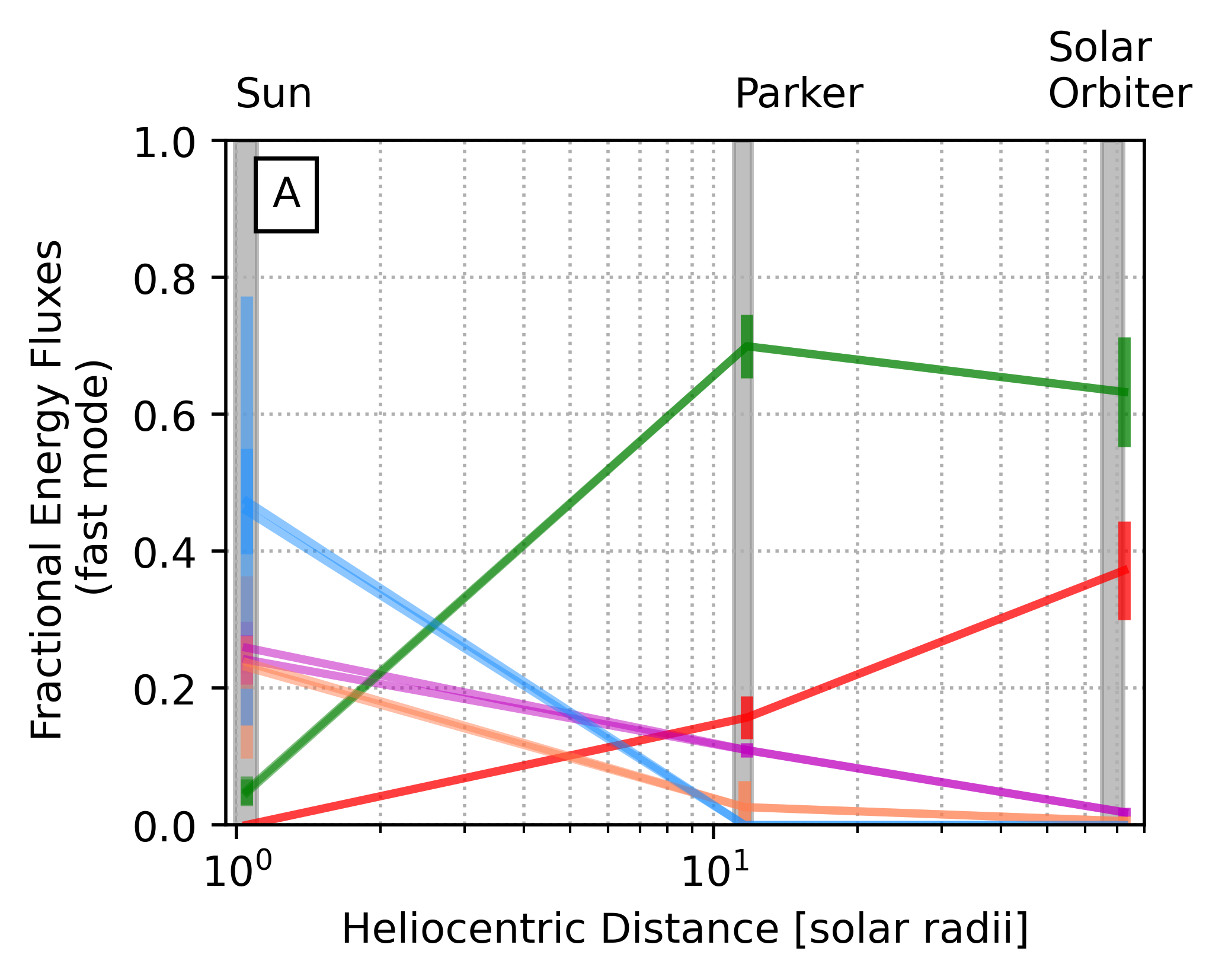}
\includegraphics[width=0.46\linewidth]{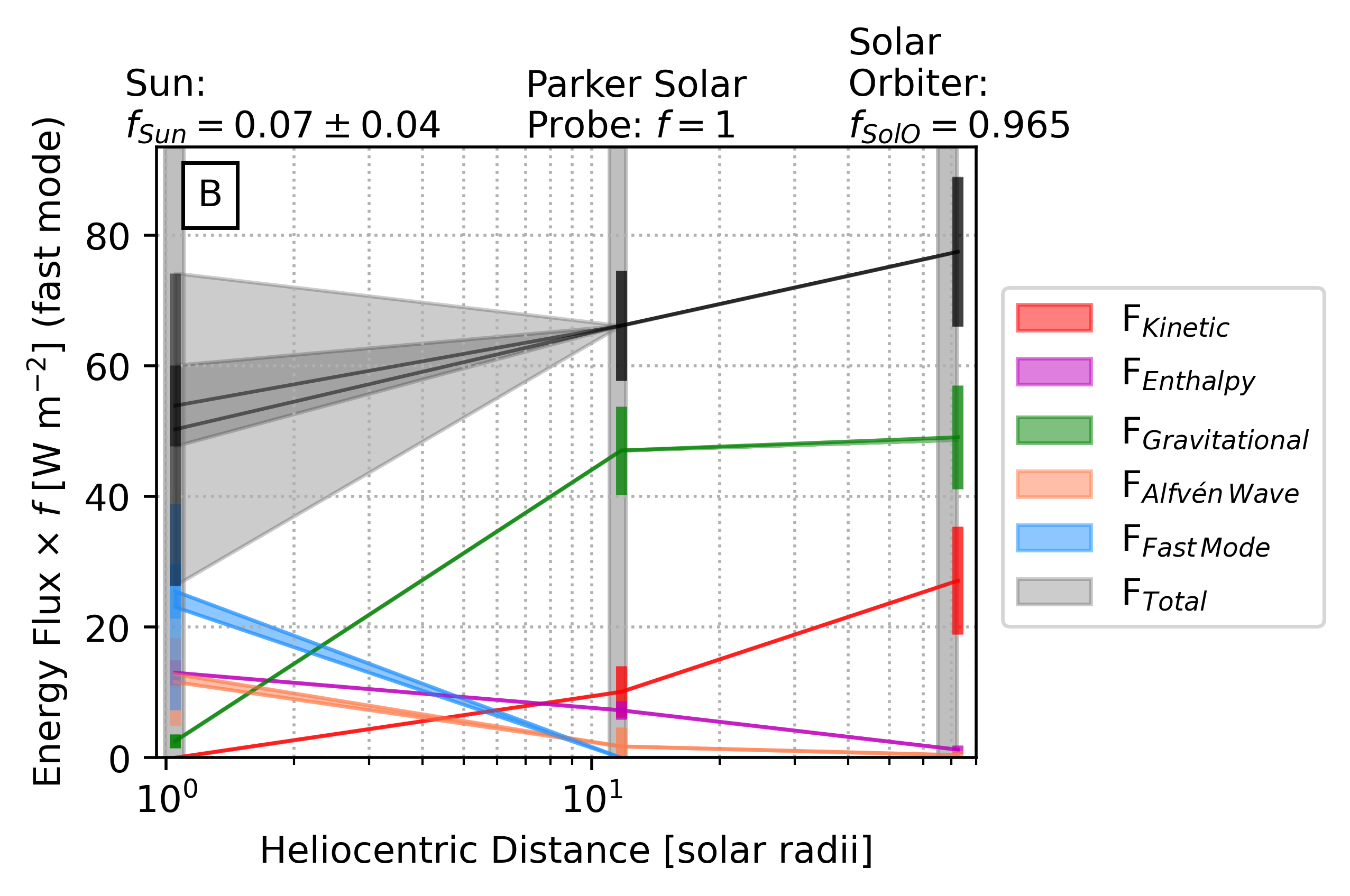}
\includegraphics[width=0.37\linewidth]{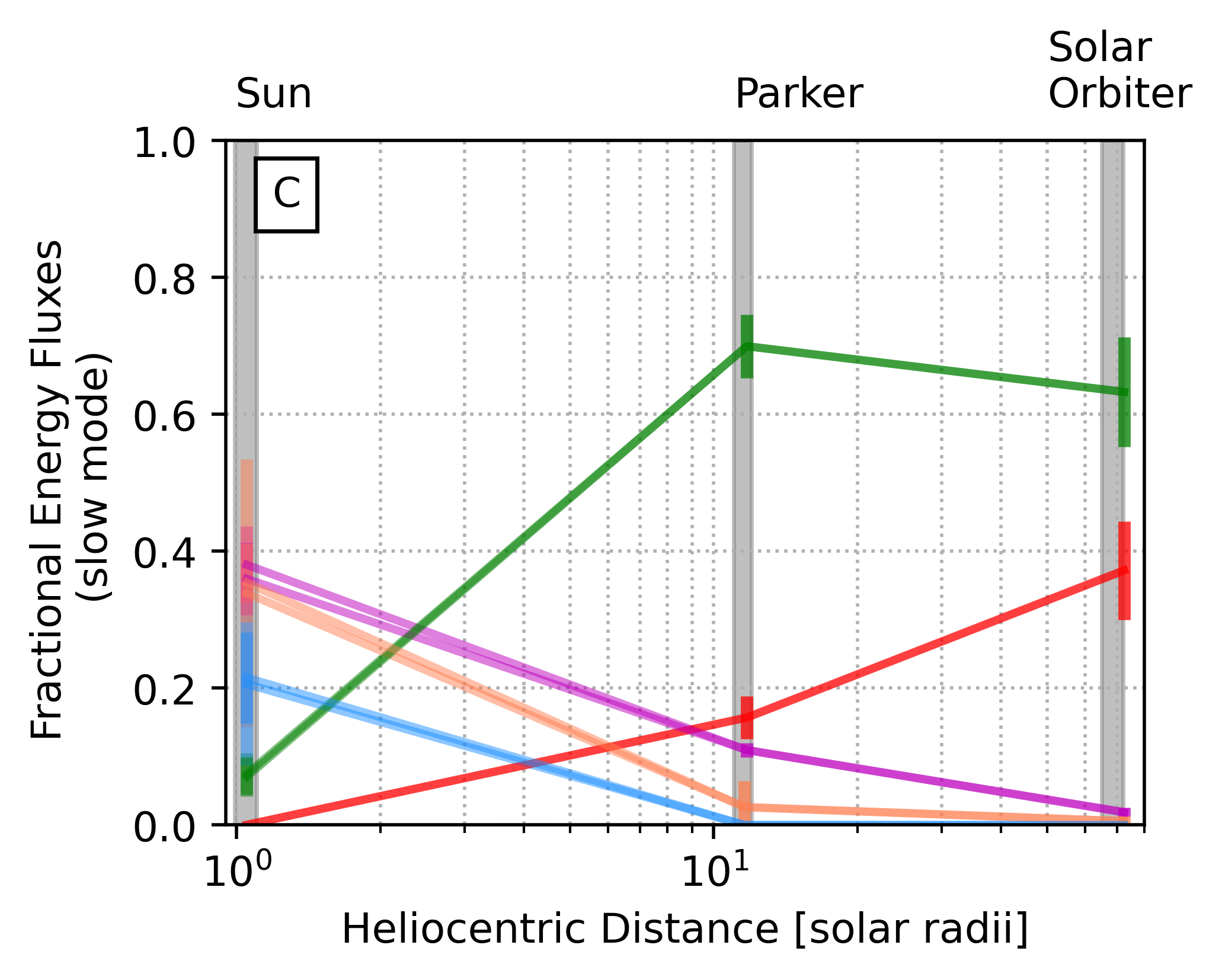}
\includegraphics[width=0.46\linewidth]{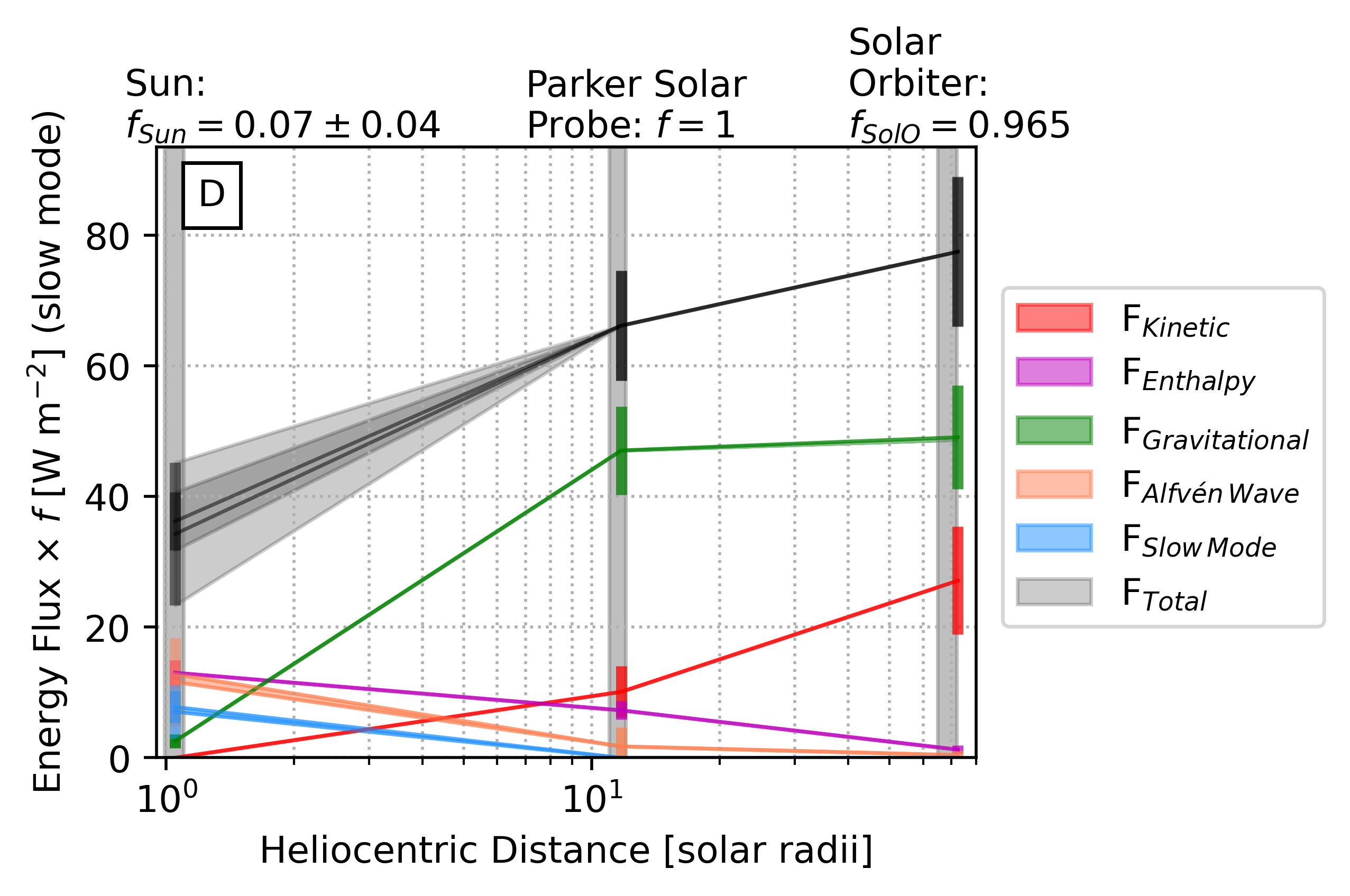}
\caption{Fractional energy fluxes~(A) and total energy fluxes~(B) computed at three heliocentric locations, at 1.05, 11.60, and 72.06~$R_{
\odot}$ within the corresponding coronal hole stream. Details of each energy flux component can be found in Appendix Section~\ref{sec:methods:energy_flux_calc} and a summary of the values are found in Table~\ref{tab:energyflux_3}.}
\label{fig5_energetics}
\end{figure}

\subsection{Energy Fluxes}\label{subsec:energy-fluxes}

To understand how energy is exchanged within the coronal hole solar wind between the low corona and throughout its propagation to the inner heliosphere, we compute the Alfv\'en wave, Slow or Fast wave mode, Kinetic, Enthalpy, and work done against gravity energy fluxes. Figure~\ref{fig5_energetics} presents the fractional energy fluxes~(A, C) and total energy fluxes~(B, D) with increasing heliocentric distance from the Sun, for scenarios where different compressive modes (slow or fast) are contributing the the total wave energy flux. Tables~ \ref{tab:swprops_2} and \ref{tab:energyflux_3} summarize the mean quantities used for the energy computation and the individual energy flux terms at each location, respectively.  

The potential fast and slow mode wave energy flux contribution are included, in addition to the terms typically observed to dominate in the solar wind \citep{Liu2021,Halekas2023,Rivera2024,Rivera2025}, as we find a large deficit in the coronal energy budget without them. We calculate the slow/fast mode contributions from LOS derived non-thermal broadening when the coronal hole was at disk center, see Appendix Figure~\ref{fig:EDF_EISNTW}. For completeness, we include their in situ contributions, confirming that they are negligible in the solar wind.  We interpret these non-thermal widths as being parallel or slightly oblique to the magnetic field geometry and therefore compressive in nature \citep{Fla1984_fast_mode_CHs, Hahn2013_tani, Banerjee2011_MHDwavesCH_review}. One caveat of these observations is that we are unable to identify which mode or combination of modes is responsible for the broadening. With slow mode being fully parallel to the magnetic field while the fast mode having a parallel component, for the case where the fast mode is only slightly oblique as to be a largely separate contribution from its Alfv\'en counterpart. Therefore, we include a contribution from either the fast (top row) or slow (bottom row) compressible modes in Figure~\ref{fig5_energetics} and estimate their potential impact of energy conservation in this stream. However, a detailed examination of these compressive modes in coronal holes is needed to further refine their energy contribution. This would require a timeseries of density fluctuations and Doppler shifts on-disk and off-limb from high-cadence and high-spectral resolution instruments with those capabilities, \cite[such as future MUSE or Solar-C][]{DePontieu2022, Shimizu2024}, to determine the compressive fluctuations' group velocity and confirm their presence through intensity fluctuations in addition to non-thermal widths.

Energy fluxes were calculated only where co-spatial data for density, temperature, magnetic field, fluctuations ($dV$, $dB$), and radial speed were available (see Appendix Section~\ref{sec:methods:energy_flux_calc}). Contributions from protons, helium, and electrons were accounted for wherever available. At the Sun, only electron information is available, at \emph{Parker} we include all populations, and Solar Orbiter includes only protons. We assume a constant helium abundance across all locations. These requirements restricted calculations to 1.05~$R_{\odot}$, $\sim$11.60~$R_{\odot}$, 72.06~$R_{\odot}$. At the Sun and Solar Orbiter, we normalize the energy fluxes using the mass flux ratio or expansion factor between that location and \emph{Parker} to account for super-radial expansion \citep{Rivera2024, Rivera2025}. From the Sun to \emph{Parker}, it was found to be  $0.07\pm0.04$ and was compared with MHD field line extrapolations, $f_{MAS}={0.1}$, from \citet{Downs2025_eclipse}. The total energy flux at the Sun is highly sensitive to the derived expansion factor, which in turn depends strongly on the density and outflow speed observed at the Sun (to derive the mass flux). These quantities can vary significantly across photospheric footpoints, as shown in Table~\ref{tab:swprops_2}. Given the range of expansion factors associated with this coronal hole, it becomes a leading contribution to the uncertainties, shown in Figure~\ref{fig5_energetics}. To constrain the range of possibilities at the Sun, we use the MHD-informed constraint on the mass flux of $f_{sun}=0.1$. This has the effect of anti-correlating the errors in the velocity and density measurement at the Sun and restricts the range of energy flux from the fainter shading to the darker shading in Figure~\ref{fig5_energetics}, and the numerical values are reported in the $f_{sun}=0.1$ column in Table~\ref{tab:energyflux_3}.

Once scaled for stream expansion, the total energy flux budget is consistent within error bars across these distances only for the case of the fast mode contribution (Figure~\ref{fig5_energetics}B), both with and without the imposition of the mass flux constraint. We note that, without the additional constraint the energy fluxes at the Sun exhibit a large standard deviation and range, with the total on the order of 50\% (for the fast mode case), with substantial contributions from uncertainty in the non-thermal widths, mass flux and Alfv\'en speed at the Sun used in the calculation. With the additional constraint this uncertainty is reduced to $10\%$ and the fast mode case remains consistent with the observed energy flux at \emph{Parker}. As discussed later in Section~\ref{futuredirection}, this error budget permits other terms to be contributing to the total energy budget and highlights the need for improved coverage and novel measurement techniques for coronal observations.

The contribution of the energy sources in the low corona in descending order are quantified as Fast wave (blue, top row), Alfv\'en wave (coral), Enthalpy (magenta), Slow wave (blue, bottom row), Gravitational (green), Kinetic (red) energy fluxes. As shown in Figure~\ref{fig5_energetics}B and D, the work done against gravity and the kinetic energy fluxes increase with distance, while the enthalpy, and all wave energy fluxes decrease rapidly. Therefore, its expected that the thermal and wave pressure gradients drive much of the acceleration starting at the low corona, and could largely balance the increase to the kinetic energy flux and work done against gravity, in the case of the Fast wave mode. The energy flux components at each location shows that the majority of energy is exchanged within the reported terms with the inclusion of fast magnetosonic energy flux, and that the majority of this transfer has taken place by \emph{Parker}'s orbit, below the Alfv\'en surface. Notably, without the inclusion of compressive energy, our results are not consistent with energy conservation between the low corona and solar wind.

\section{Discussion} \label{discussion}

\subsection{Implications of Radial Profile Constraints}

As shown in Figure~\ref{fig4_radialplots}, overall the observations taken of the stream at several stages of its evolution are broadly consistent with the mass-conserving, iso-poly model solutions (dark grey). This agreement indicates good compatibility with the statistical in situ trends used to derive the model solutions \citep{Badman2025}. Despite large variation in density and Alfv\'en speed in the corona, physically consistent values throughout largely validates the underlying assumptions across a large heliocentric range, suggesting the dominant physics is captured correctly across the sub-sonic to super-Alfv\'enic regimes. These iso-poly models assumed radial expansion in the low corona and the large expansion factor in the corona, $f_{Sun}=1/f_{exp}\simeq0.07-0.1$ indicates large non-radial expansion of the small coronal hole which should be incorporated in the low coronal velocity profiles in future detailed modeling of this coronal hole \citep{Dakeyo2024b}. However, we note that the energy flux calculations presented in this work are not dependent on the low coronal velocity profiles, but instead are scaled using the expansion factor which is empirically determined.

A quasi-isothermal temperature regime extending to near 10~$R_{\odot}$ is suggested to be maintained out to almost 10~$R_\odot$ as indicated by the measured coronal electron temperature of 1.23~MK and in situ proton temperature of 0.826~MK at \emph{Parker}, which is readily explained with the temperature profile of the iso-poly models: (proton/green and electron/lime in Figure~\ref{fig4_radialplots}B). Although weak non-isothermal behavior which differs between electrons and protons is possible in this distance range, as suggested by UVCS measurements \citep{Cranmer2020}, at the coronal heights focused on in this work, thermal equilibrium between protons, electrons (and presumably heavier species) is likely \citep{Cranmer2020b, Rivera2025b}. Therefore, although only electron temperature estimates are available in the low corona, they can be taken as good proxies for the proton temperatures too. This suggest the upper range of the proton temperature profiles used in Figure~\ref{fig4_radialplots}B is a good approximation at all distances, while the isothermal electron temperature is a slight underestimate in the low corona and over-estimate at \emph{Parker}. This could be resolved either with a weakly sub-isothermal electron temperature profile, or a transition to polytropic cooling closer in that the protons \citep{Dakeyo2025}.

Additionally, although simplistic, a quasi-isothermal temperature profile in the 1--10~$R_\odot$ region would be compatible with that simulated in reflection-driven Alfv\'en wave turbulence models where the heating experienced remains balanced by radiative cooling, conductive losses, and expansion of the solar wind of the solar wind out to $\sim$10~$R_{\odot}$ \citep{Chandran2021, Chandran2025}. Therefore, diagnostics of temperature with increasing coronal altitude beyond just single locations, as has been done in this work, are of considerable interest. 

Beyond \emph{Parker}, the median measured proton temperature is warmer than expected statistically, but would be consistent within the standard deviation with the modeled polytropic cooling observed. 

Our results show Alfv\'enic fluctuations follow close to the WKB approximation below \emph{Parker}'s orbit and deviate from it beyond \emph{Parker}'s orbit, consistent with \citet{Sioulas2025}. In this case, we trade continuous statistical measurements over a continuous range of $M_A$ for a 2-point constraint which probes down to much lower $M_A$ than is accessible with in situ measurements (Figure~\ref{fig4_radialplots}F). The WKB approximation suggests Alfv\'en waves propagate conservatively through a slowly varying background with no reflection, no dissipation, and no nonlinear coupling to other modes. Therefore, wave action is strictly conserved and the waves are expected to only interact with the stream through mechanical work. Although largely WKB-like in the corona, we find the extrapolated profile (which is normalized to match the fluctuation amplitude at \emph{Parker}) slightly undershoots the mean amplitude ($dV$/V$_A$) measured at 1.05~$R_\odot$. Although with the measurement error bar, this could reflect a small amount of turbulent or resonant collisionless damping \citep[e.g., ion-cyclotron;][]{Cranmer1999, Hollweg2002} or reflection of the outward-propagating waves from the steep Alfv\'en speed gradient, redistributing energy into an inward-propagating population \citep{Velli1993, Verdini2009}. Even a small discrepancy could correspond to a meaningful degree of heating due to the fluctuation energy constituting the second largest term in the energy budget at these altitudes. Independently testing this would require measurements of either the heating profile or the inward-propagating wave energy component, which should be explored in future work.

In summation, we infer from the match between our constraints and the statistical iso-poly profiles, specifically the flat coronal temperature profile, the large gradient in the Alfv\'en speed and the behavior of the velocity fluctuations, that our results largely resemble the expected large-scale trends produced in reflection-driven Alf\'ven wave turbulence dissipation models \citep[e.g.,][]{Verdini_Velli2007, Verdini2009, Chandran2021, Chandran2025}. However, we also note that this does not exclude other processes from taking place. Slow MHD waves are often excited alongside reflected Alfv\'en waves \citep{Suzuki2005, Suzuki2006JGR}. Because slow MHD waves are compressible, they are more readily damped and dissipated, complementing Alfv\'en wave dissipation. Our estimates of $F_{\text{Slow Mode}}$ suggest they could be comparable to the Alfv\'en waves in the low corona, in line with previous studies \cite[e.g.,][]{DeForest1998_slowMHDwaves_CH_flux}. Density fluctuations have been detected with radio occultation observations in the corona and solar wind \citep{Chiba2025}, suggesting that physical processes involving compressible waves/fluctuations could play a non-negligible role. As discussed in the next section, there is a general increase in the total energy flux from the coronal determination to the solar wind, and such compressive MHD modes do contribute substantially.

\subsection{Implications of the Energy Budget}
After considering a large uncertainty in the flux tube expansion factor, the upper range of the low coronal energy budget \citep[which is supported by MHD modeling of the coronal hole using the models from][]{Downs2025_eclipse} is consistent with heliospheric values, suggesting energy conservation from the Sun out to \emph{Parker} and Solar Orbiter within one standard deviation if and only if compressive energy flux is included and fully attributed to fast magnetosonic waves. Under this inclusion, we  find that the dominant terms are accounted for. Nevertheless, the mean coronal energy fluxes remains below the energy fluxes derived from \emph{Parker} near the Alfv\'en surface in all cases. This motivates further investigation into the detailed contributions of the different wave modes. 

We also note that at the Sun there is a large total energy flux and standard deviation, particularly stemming from the coronal density, Alfv\'en speed and non-thermal width measurements as well as the range of expansion factors based on MAS and mass flux ratio. These factors impart significant uncertainty on the energy budget at the Sun which may therefore permit other smaller terms not measured here to also be important contributors. We were able to restrict the uncertainty somewhat using mass flux ratios from modeling of the coronal hole. This highlights a trade-off in work such as this between the inclusion of modeling information, and maintaining fully data-derived constraints.

Nevertheless, at the Sun and across expansion factors, the mean enthalpy and wave energy fluxes (including compressive and non-compressive MHD modes) dominate the coronal energy flux budget and remain consistent in fractional contributions. While the thermal pressure and Alfv\'enic wave pressure are substantial and approximately equal, the compressive modes either add a similar size term (slow mode), or a significantly larger term (fast mode), with this latter case being necessary to close the energy budget. The enthalpy and wave energy fluxes decrease dramatically to just 11\% and 2.6\% of the energy budget near the Alfv\'en surface. Beyond the Alfv\'en surface at Solar Orbiter, the enthalpy (1.7\%) and wave (0.4\%) energies are significantly smaller and comparable with each other, and by 72.06~$R_{\odot}$ they are negligible, as reported previously in \cite{Rivera2024, Rivera2025}. 

\section{Conclusions} \label{conclusions}
Through a multi-mission coordinated effort, we examine near-contemporaneous and multi-wavelength observations of the corona to derive detailed plasma conditions and magnetic field properties used to compute an empirically-derived energy budget of an equatorial coronal hole solar wind stream which reaches a mean speed of 450~km~s$^{-1}$ by 70~$R_{\odot}$. The remote properties of the nascent coronal hole wind are connected to its heliospheric counterpart sampled by Parker Solar Probe and Solar Orbiter during a fortuitous spacecraft alignment. Together, we traced the Alfv\'en wave, enthalpy, kinetic, and gravitational energy fluxes of this single solar wind stream from deep in the corona (subsonic regime), across the Alfv\'en surface and beyond. 

Overall, the solar wind stream qualitatively follows mass-flux conserving, thermal pressure-driven wind with additional momentum from Alfv\'enic fluctuations supplied empirically, well described with ``iso-poly + forcing'' models \citep{Dakeyo2022,shi2022_polyt,Rivera2024,Rivera2025} where the proton and electron temperatures are near isothermal out to approximately 10~$R_\odot$, before starting to cool sub-adiabatically with distance. Alfv\'enic fluctuations below \emph{Parker}'s orbit are close to consistent with a dissipation-free, WKB approximation from the low corona out to the Alfv\'en surface, with a small (but potentially meaningful given the large energy flux content) amount of heating compatible with observations. Larger deviations with more substantial dissipation is observed from \emph{Parker}, well beyond the Alfv\'en surface. These key large-scale trends match well with the salient features developed in recent reflection-driven turbulence models \citep[e.g.,][]{Chandran2025} and therefore offer some support to the underlying physical mechanisms in those models, although we do not rule out other mechanisms  \citep[e.g.,][]{Suzuki2005}, especially given the apparent importance of compressive modes in the corona as shown by the energy flux analysis discussed next.

The empirical constraints are then used to quantify the energy fluxes in this coronal hole stream to identify principal contributions to the overall energy budget. The enthalpy (13~W~m$^{-2}$) and total wave energy (Alfv\'en wave, 12~W~m$^{-2}$, and fast/slow wave modes, 23/7~W~m$^{-2}$) fluxes dominate the coronal energy flux budget at 1.05~$R_\odot$ with the rest of the energy attributed to the work done against gravity from 1.0 to 1.05~$R_\odot$ (2.55~W~m$^{-2}$) and kinetic ($<$0.01~W~m$^{-2}$) energy flux terms. These quoted percentages can be contextualized with the reference total energy budget measured at \emph{Parker} of $\sim$66 W m$^{-2}$. Of the terms we were able to observe, the total (compressive + Alfv\'enic) wave energy flux is the most significant driver, although the enthalpy flux is also substantial. We note that this balance could evolve sharply over small differences in altitude in this low coronal regime. However, it is important to contextualize that the present stream as relatively slow solar wind far from the Sun (reaching $\sim$450~km~s$^{-1}$ at 70~$R_\odot$) and sourced from a very narrow coronal hole where the unscaled Alfv\'en wave flux (128~W~m$^{-2}$) is lower than that reported in larger coronal holes \citep{Withbroe1982}. Therefore, it is likely that the balance towards higher Alfv\'en wave energy flux is larger in asymptotically faster wind speeds, and the importance of compressive modes potentially reduced. In any case, at Solar Orbiter, far beyond the Alfv\'en surface, the wave and enthalpy fluxes dominating the coronal energy flux are now negligible.

The large total errors in the coronal energy budget highlights the need for more continuous coverage and novel diagnostic capability for remote sensing of the low corona. A particular identified weakness with current measurement capabilities lies in the lack of distributed ion temperature measurements, and the low signal-to-noise in coronal hole streams of the current state-of-the-art diagnostics for coronal magnetometry and density inversion. Lifting the restriction of building the energy budget at a single height in the corona would mean greatly strengthened constraints could be developed.

Finally, more generally, the observations derived from this single stream of solar wind represent a robust set of empirical constraints that must be satisfied by any viable theoretical framework. For example, the Poynting flux quantified here may be used directly for initial conditions in MHD solar wind simulations \citep[e.g.,][]{Mikic1999, Sokolov2013}, and more broadly for stellar wind models \citep[e.g.,][]{Cohen2014, Garraffo2018}.

\subsection{Study Limitations and Future Prospects} \label{futuredirection}

In this section we discuss the main limitations of this study, which fall into two categories: coordination and instrument coverage constraints, and observational uncertainties and variability. To the best of our abilities and within current instrumentation capabilities, we extracted all necessary plasma and magnetic field properties of the stream in order to assess the solar wind energy budget. Nonetheless, several roadblocks limit the completeness and precision of our results.

\subsubsection{Coordination and Instrumentation Coverage Constraints}

In an idealized version of the investigation we perform in this work, all quantities would be measured synchronously and radially distributed to provide continuous constraints over the range of heliocentric distances. However, for this real-world example several compromises were necessary to obtain our ``complete'' energy budget computation:

Due to the sparseness of \emph{in situ} measurements, we are limited to a two-point measurement of the stream, where all required diagnostics are measured simultaneously at each individual measurement point. For the coronal constraints obtained across a patchwork of different instruments, due to the variable field of view, scheduling constraints between instruments, and decreasing signal to noise with altitude (especially for intrinsically dim coronal holes), we were only able to derive a robust set of constraints (i.e., $n$, $V$, $T$, $V_A$, $B$, $\delta v$) at a single height in the corona. One of the most notable constraints is that there are no available proton or electron temperature diagnostics through active missions with routine observations and full coverage of the corona \citep{Vourlidas2023, West2023, Rivera2025_IAU}. This restricts us to evaluating the solar wind enthalpy flux at very low altitudes where it is reasonable to assume electrons and protons are in thermal equilibrium. While we make reasonable assumptions for comparing the radial profiles at a less restricted set of altitudes in Figure~\ref{fig4_radialplots} \citep[for example using historical UVCS observations to justify near-flat coronal temperature profiles;][]{Cranmer2020a}, this necessarily limits our measurements of the stream's dominant energy budget contributions to a three-point constraint. 

Additionally, some of these measurements were approximately a week apart due to the necessity of combining on-disk and off-limb observations, leading to some inevitable time evolution being averaged in our final measurements. We are able to mitigate this latter concern by verifying in EUV images that the morphology of the source region does not vary strongly between when it is observed on disk with SDO/AIA, and 7~days later when it is viewed on disk from Solar Orbiter. 

\subsection{Observational Uncertainties}

Even after restricting the coronal measurement to a single height, we still obtain a significant error budget on the individual and total energy fluxes in the solar corona. This is mainly driven by our best estimate of the density and the Alfv\'en speed at the coronal location. For the density, we obtained several independent measures with different observational methods (see Appendix Section~\ref{sec:methods:density}) which yielded some variability. This variability is examined in detail in a companion study \citep{Paraschiv2026} that analyzes the same observations presented here.

The Alfv\'en speed is estimated from measurements of fluctuation phase speeds in the coronal stream \citep{Yang2020a,Yang2024} and is observed to have large variability likely due to a sensitivity to overlying structures in the FOV and potentially out-of-plane projection effects given the non-radial field lines from the coronal hole's intrinsic expansion. These same issues propagate to our determination of the POS magnetic field strength, which is derived from these phase speed measurements. The viability of  inferences of the LOS component of the magnetic field has been demonstrated \citep{Schad2024}, where such measurements can in the future complement the Alfv\'enic derived determinations. These developments underscore the continued community need for full 3D coronal magnetometry \citep{paraschiv&judge2022,Paraschiv2026_bfield}.

\subsection{Unresolved Contributions to the Error Budget}

Lastly, as mentioned above, we mitigate for the observational uncertainties through propagating them to an error budget in our final determined energy flux terms in Figure~\ref{fig5_energetics} which yields a total error in the corona of around 47$\%$, reducing to $10\%$ by supplying an external mass flux constraint. Although, with the inclusion of fast mode energy flux, our results are consistent with energy conservation within one standard deviation, the mean solar total and \emph{Parker} total still remain of order of 10~W~m$^{-2}$ apart. We therefore cannot rule out the contributions of other terms of this magnitude.

\begin{longrotatetable}
\begin{deluxetable}{llccc}
\tablecaption{Observational Diagnostics and Methodology: Summary of remote observations utilized in the study. The table includes the diagnostics, associated mission or instrument, observation dates, altitude of observations, and methodology for deriving corresponding diagnoses.
\label{tab:remote_obs_diag_1}}
\tablewidth{0pt}
\tablehead{
  \colhead{Diagnostic} &
  \colhead{Instrument} &
  \colhead{Observation Date (UT)} &
  \colhead{Altitude} &
  \colhead{Methodology}
}
\startdata
\sidehead{Density}
 & SDO/AIA          & 2024 Apr 01 -- 2024 Apr 15  & 1--1.25~$R_\odot$  & DEMT \\
 &&&& \citep{frazin_2009, lloveras_2022} \\
 & DKIST Cryo-NIRSP & 2024 Apr 09 18:00           & 1.05~$R_\odot$        & \ion{Fe}{13} line ratio \\
 & MLSO/UCoMP       & 2024 Apr 09 18:00           & 1.05--1.95~$R_\odot$ & \ion{Fe}{13} line ratio \\
 & MLSO/K-Cor       & 2024 Apr 09 18:00           & 1.05--1.95~$R_\odot$ & WL pB \citep{vandehulst1950} \\
 & Solar Orbiter/Metis & 2024 Mar 29 05:15        & 2--4~$R_\odot$      & WL pB \citep{vandehulst1950} \\
 & SOHO/LASCO-C2    & 2024 Apr 01 -- 2024 Apr 15  & 2.5--6~$R_\odot$   & WL pB SRT \\
 &&&&\citep{frazin_2002, Vasquez2024} \\
\sidehead{Electron Temperature}
 & SDO/AIA          & 2024 Apr 04 18:30--19:30    & On disk & DEM \citep{Cheung2015} \\
 & Solar Orbiter/SPICE & 2024 Apr 04 06:20        & On disk & DEM \citep{Plowman_EMToolKit_A_Standardized} \\
\sidehead{Velocity}
 & Hinode/EIS       & 2024 Apr 01 -- 2024 Apr 03 (daily) & On disk & Doppler velocities \\
 & Solar Orbiter/Metis & 2024 Mar 29 05:15        & 2--4~$R_\odot$      & Ly\,$\alpha$ Doppler dimming \\
 & SOHO/LASCO-C2    & 2024 Apr 01 -- 2024 Apr 15  & 2.5--6~$R_\odot$   & Mass conservation and expansion factor$^{*}$ \\
\sidehead{Magnetic Field}
 & MLSO/UCoMP       & 2024 Apr 09 18:00           & 1.05--1.95~$R_\odot$ & Coronal seismology \citep{Yang2020a} \\
\sidehead{Non-thermal Widths}
 & DKIST Cryo-NIRSP & 2024 Apr 08 18:00           & 1.05~$R_\odot$        & Spectral line fitting$^{**}$ \\
 & Hinode/EIS       & 2024 Apr 01 & On disk & Spectral line fitting$^{**}$ \\
\enddata
\tablecomments{$^{*}$Velocity derived from mass conservation and the expansion factor, see Appendix Section~\ref{sec:methods:outflow_speed}. \\
$^{**}$Non-thermal widths obtained from profile fitting of the spectral line, see Appendix Section~\ref{sec:methods:nonthermal_widths}.}
\end{deluxetable}
\end{longrotatetable}

\begin{deluxetable*}{lccc}
\tablecaption{Properties of the Solar Wind at Three Heliocentric Distances: Summary of all the mean values used to compute the energy budget at the Sun, Parker Solar Probe, and Solar Orbiter. At the Sun, we take the median values between 1.04--1.06~$R_{\odot}$. At Parker Solar Probe and Solar Orbiter, we use the average value across the Carrington longitudes 85--121 and 70--105, respectively. For $\delta \textbf{b}$ and $\delta \textbf{v}$ in situ, we record the 30~min averages.  \label{tab:swprops_2}}
\tablewidth{0pt}
\tablehead{
  \colhead{Property} &
  \colhead{Sun (1.05~$R_\odot$)} &
  \colhead{Parker Solar Probe (11.60~$,R_\odot$)} &
  \colhead{Solar Orbiter (72.06~$R_\odot$)}
}
\startdata
$n_{p,e}$~(cm$^{-3}$)
  & $2.29\pm0.71\times10^{8}$
  & $3.7\pm0.48\times10^{3}$
  & $57.85\pm19.20$ \\
$v_p$~(km~s$^{-1}$) 
  & $8.09\pm3.36$ (EIS), $9.98\pm10.60$ (SPICE, 1.01~$R_\odot$)
  & $266\pm28.37$
  & $443.03\pm79.49$ \\
$T_p$~(MK)
  & $1.23\pm0.04$
  & $0.762\pm0.186$
  & $0.230\pm0.126$ \\
$V_A$~(km~s$^{-1}$)
  & $387.90\pm148.60$ 
  & $348.97\pm28.01$
  & $94.42\pm24.01$ \\
$C_s$~(km~s$^{-1}$)
  & $175.70\pm17$ 
  & --
  & -- \\
$B_R$~(G, nT)
  & $-$
  & $-928.07\pm100.67$
  & $-19.19\pm6.96$ \\
$|B|$~(G, nT)
  & $4.30\pm1.42,\ 4.30\pm1.42\times10^{5}$
  & $9.81\pm0.2\times10^{-3},\ 981.28\pm24.90$
  & $2.5\pm0.1\times10^{-4},\ 24.50\pm0.96$ \\
$n_\alpha/n_p$
  & $0.04$
  & $0.04\pm0.01$
  & $0.04$ \\
$v_\alpha$~(km~s$^{-1}$)
  & $8.09\pm3.36$
  & $377.61\pm49.01$
  & $-$ \\
$T_\alpha$~(MK)
  & $1.23\pm0.04$
  & $6.26\pm1.04$
  & $-$ \\
$T_e$~(MK)
  & $1.23\pm0.04$
  & $0.671\pm0.031$
  & $-$ \\
$\delta \textbf{B}$~(nT)
  & --
  & $286.01 \pm 182.38$
  & $16.41 \pm 10.79$ \\
$\delta \textbf{b}$~(km~s$^{-1}$)
  & --
  & $91.88\pm 63.85$
  & $48.02 \pm 30.79 $ \\ 
$\delta \textbf{v}$~(km~s$^{-1}$)
  & $24.1\pm1.60$ ($\zeta_\perp$), $32.6\pm10.63$ ($\zeta_\parallel$)
  & $93.92 \pm 57.56$
  & $46.03\pm35.94$ \\
$\delta \textbf{n}$~(cm$^{-3}$)
  & -- 
  & $198.32\pm0.12$
  & $4.13\pm0.0006$ \\
  $|\sigma_C|$
  & --
  & $0.92\pm0.14$
  & $0.43\pm0.43$ \\
  $\sigma_R$
  & --
  & $-0.07\pm0.22$
  & $-0.12\pm0.48$ \\
\enddata
\end{deluxetable*}

\begin{deluxetable*}{lccc}
\tablecaption{Energy Flux Budget at Three Heliocentric Distances: List of all the energy flux budget terms at the Sun, \emph{Parker}, and Solar Orbiter. Each term at the Sun and Solar Orbiter has been scaled to Parker Solar Probe by $f_\text{Sun} = 0.07$ and $f_\text{Sun} = 0.1$ and $f_\text{SolO}=0.965$, respectively, to account for non-spherical expansion. \label{tab:energyflux_3}}
\tablewidth{0pt}
\tablehead{
  \colhead{Energy Flux (W~m$^{-2}$)} &
  \colhead{Sun (1.05~$R_\odot$) $\times f_{Sun}$} &
  \colhead{Parker Solar Probe (11.60~$R_\odot$)} &
  \colhead{Solar Orbiter (72.06~$R_\odot$) $\times f_\mathrm{SolO}$}
}
\startdata
& $f_{Sun} = 0.07\pm0.04$ $\vert$ $f_{Sun} = 0.1$  &   &   \\ 
\hline
Kinetic Flux      & $0.0092\pm0.0056$ $\vert$ $0.0075\pm0.0019$   & $10.09\pm3.85$  & $27.11\pm8.25$  \\ 
Enthalpy Flux      & $13.02\pm1.92$ $\vert$ unchanged & $7.27\pm1.47$   & $1.25\pm0.68$   \\ 
Alfv\'{e}n Wave Flux & $11.60\pm6.71$ $\vert$ $12.81\pm2.08$ & $1.71\pm1.61$   & $0.41\pm0.67$   \\ 
Fast Mode Wave Flux & $23.08\pm15.76$ $\vert$ $25.50\pm4.15$  & $0.06\pm0.82$  & $0.002\pm0.004$  \\ 
Slow Mode Wave Flux & $7.05\pm4.07$ $\vert$ $7.79\pm2.41$  & --  & --  \\ 
Gravitational Energy Flux & $2.55\pm1.04$ $\vert$ unchanged & $47.04\pm7.94$ & $48.696\pm8.57$ \\ 
\hline
Total Energy Flux (slow)  & $34.24\pm10.92$ $\vert$ $36.19\pm4.46$ & $66.20\pm9.10$  & $77.46\pm11.90$ \\
Total Energy Flux (fast)  & $50.27\pm23.92$ $\vert$ $53.89\pm6.19$ & $66.20\pm9.10$  & $77.46\pm11.90$ \\
Total Energy Flux (Alfv\'en only)  & $27.18\pm6.05$ $\vert$ $28.39\pm2.08$ & $66.14\pm9.10$  & $77.46\pm11.90$ \\
\enddata
\end{deluxetable*}
\begin{acknowledgments}

Parker Solar Probe was designed, built, and is now operated by the Johns Hopkins Applied Physics Laboratory as part of NASA’s Living with a Star (LWS) program (contract NNN06AA01C). Support from the LWS management and technical team has played a critical role in the success of the Parker Solar Probe mission. Solar Orbiter is a mission of international cooperation between ESA and NASA, operated by ESA. Solar Orbiter SWA data were derived from scientific sensors that were designed and created and are operated under funding provided by numerous contracts from UKSA, STFC, the Italian Space Agency, CNES, the French National Centre for Scientiﬁc Research, the Czech contribution to the ESA PRODEX program, and NASA.  Solar Orbiter SWA work at the UCL/Mullard Space Science Laboratory is currently funded by STFC (grant Nos. ST/W001004/1 and ST/X/002152/1).

The SWA team at INAF/IAPS is currently funded under ASI grant 2018-30-HH.1-2022.

The development of SPICE has been funded by ESA member states and ESA. It was built and is operated by a multinational consortium of research institutes supported by their respective funding agencies: IAS (Centre National d'Etudes Spatiales (CNES), operations lead), STFC RAL (the UK Space Agency (UKSA), hardware lead), GSFC (NASA), MPS (Deutsches Zentrum für Luft- und Raumfahrt (DLR)), PMOD/WRC (Swiss Space Office (SSO)), SwRI (NASA), UiO (Norwegian Space Agency). The efforts at SwRI for Solar Orbiter
SPICE are supported by NASA under GSFC subcontract 80GSFC20C0053 to
Southwest Research Institute.

The March--April 2024 instance of the Fast Wind SOOP was a Solar Orbiter observing campaign led by L. Franci in collaboration with S. Mzerguat and C. Froment, with scientific support from V. Reville, R. Pinto, and M. Janvier, and technical support from E. Buchlin, A. De Groof, D. Williams, C. Nelson, A. Walsh, and all the Solar Orbiter Instruments PIs/teams, and coordinated with the teams of the Hinode mission and of the Interface Region Imaging Spectrograph (IRIS).

L.F. is supported by the Royal Society University Research Fellowship No. URF/R1/231710 and at the time of the Fast Wind SOOP 2024 was supported by the Science and Technology Facilities Council (STFC) grant ST/W001071/1.

SDB acknowledges support from the Royal Society Wolfson Visiting Fellowship program.

{\em Hinode} is a Japanese mission developed and launched by ISAS/JAXA, with NAOJ as domestic partner and NASA and STFC (UK) as international partners. It is operated by these agencies in co-operation with ESA and NSC (Norway).

The research reported herein is based on data collected with the U.S. National Science Foundation (NSF) Daniel K. Inouye Solar Telescope (DKIST), built and operated by the NSF National Solar Observatory (NSO). NSO is managed by the Association of Universities for Research in Astronomy, Inc. (AURA), and funded under a cooperative agreement with the NSF. DKIST is located on land of spiritual and cultural significance to Native Hawaiian people. The use of this important site to further scientific knowledge is done so with appreciation and respect.

UCoMP (DOI: \url{https://doi.org/10.26024/g8p7-wy42}) data  and  KCor data  (DOI: \url{https://doi.org/10.5065/d69g5jv8}) are courtesy of the Mauna Loa Solar Observatory, operated by the High Altitude Observatory, as part of the US NSF NCAR.

GdT was supported by the NSF National Center for Atmospheric Research, which is a major facility sponsored by the U.S. National Science Foundation under Cooperative Agreement No. 1852977.

Metis was built and operated with funding from the Italian Space Agency (ASI), under contracts to the National Institute of Astrophysics (INAF) and industrial partners. Metis was built with hardware contributions from Germany (Bundesministerium für Wirtschaft und Energie through DLR), from the Czech Republic (PRODEX) and from ESA. The Metis team and the SWA team at INAF/IAPS is currently funded under ASI grant 2018-30-HH.1-2022.
\end{acknowledgments}

\begin{contribution}

All authors contributed equally to the manuscript.

\end{contribution}

\appendix


\begin{figure}[h!]
    \centering
    \includegraphics[width= 0.35\linewidth]{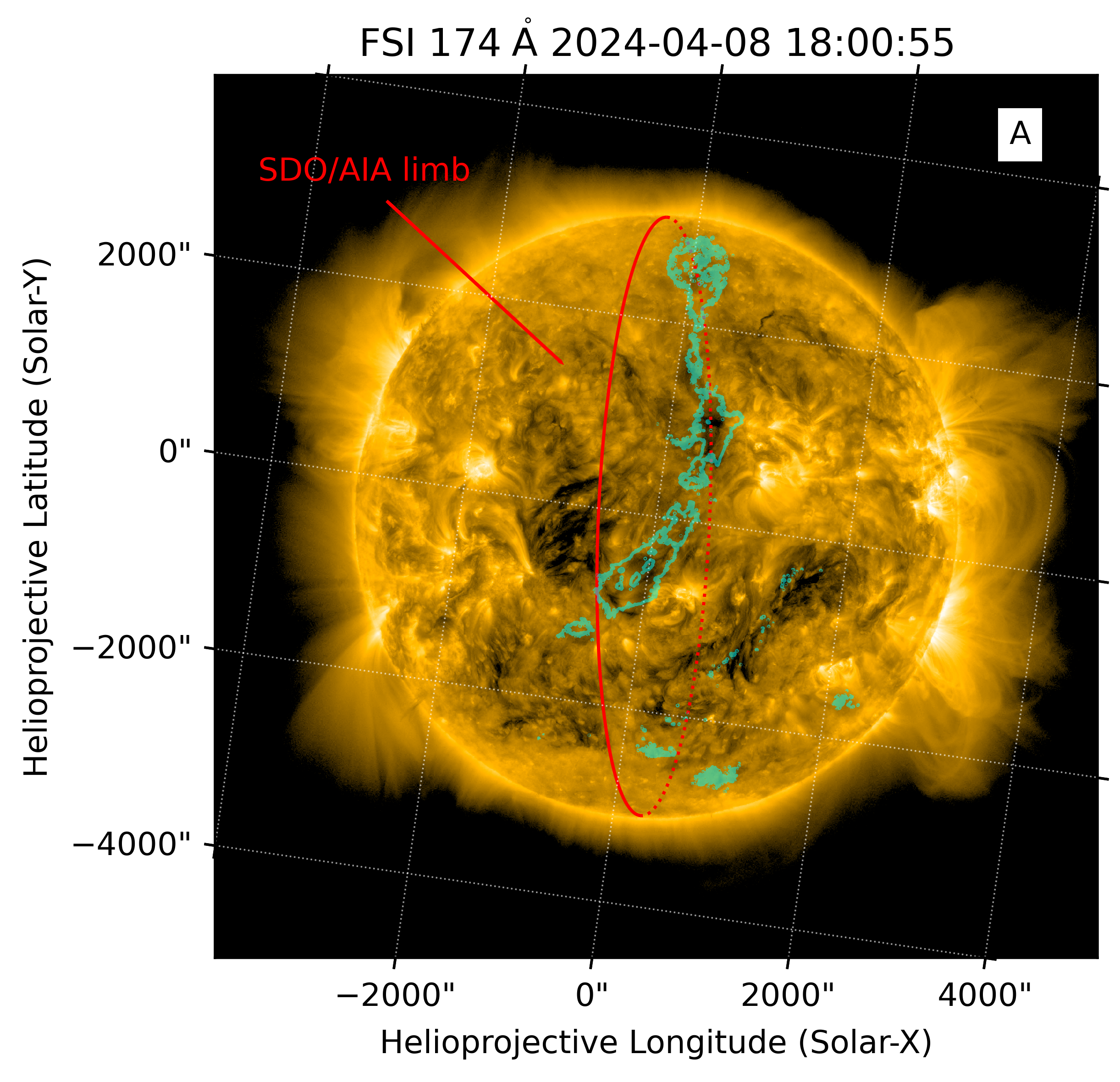}
    \includegraphics[width= 0.35\linewidth]{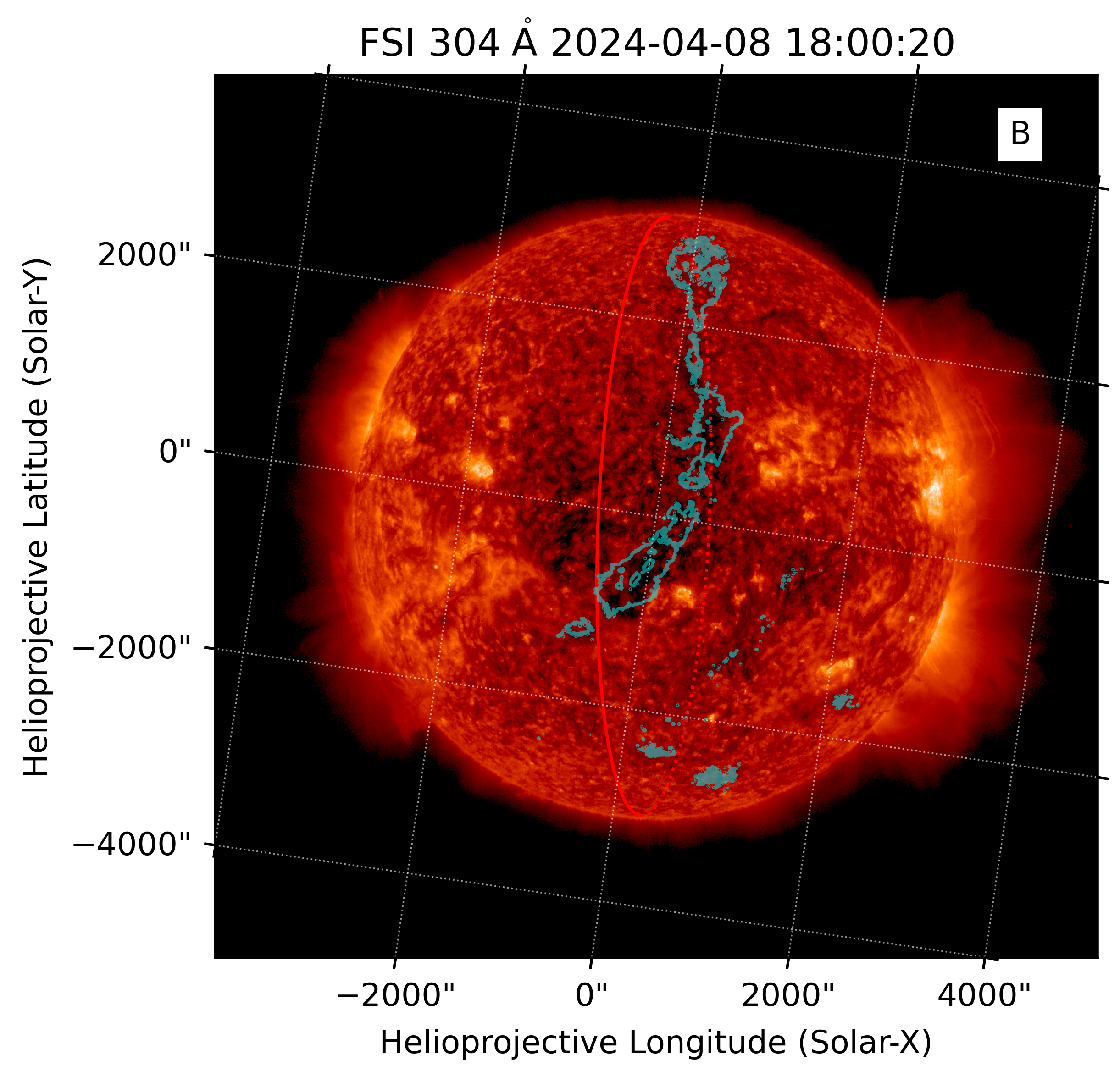}
    \caption{Solar Orbiter Full Sun Imager's 174 and 304~{\AA} images on 8~April around the total solar eclipse campaign. The coronal hole outline overlaid is derived from Figure~\ref{fig1_orbit_mapping}B on 1~April to indicate the position of the coronal hole on the disk from Solar Orbiter's perspective.}
    \label{fig:EDF7_fsi_ch_outline}
\end{figure}

\section{Coordination Overview and Derived Diagnostics}\label{sec:methods}

The coronal hole examined is sampled over a week of remote and \emph{in situ} observations between Solar Orbiter's Metis observations in quadrature with \emph{Parker} Solar Probe's 19th perihelion (29~March 2024 to 30~March 2024), Solar Orbiter's 8th perihelion pass (4~April 2024), and the solar eclipse (8~April 2024), that collectively span 29~March to 9~April 2024, up to a day after the solar eclipse.

\subsection{Remote Observations}
Direct observations of the coronal hole span roughly a week where the FOV of all instruments are summarized in Figure~\ref{fig2_coordination}A (1~April) and B (8--9~April). Figure~\ref{fig2_coordination}A shows the on-disk observation from SDO/AIA and pointing locations of Hinode/EIS and Solar Orbiter/SPICE. Figure~\ref{fig2_coordination}B shows a plot of the pointing locations of DKIST Cryo-NIRSP, SDO/AIA (for DEM diagnostics), K-Cor, and observations from UCoMP and SOHO/LASCO. Table~\ref{tab:remote_obs_diag_1} is a summary of all the coronal hole observations listing the derived diagnostic, observation date, approximate altitude coverage, and methodology used to derive each diagnostic. These techniques are discussed in detail in the following sections. Several ground and space-based instruments observe the coronal hole from the Sun-Earth line while it was on disk (near disk center at 1~April) and as it rotated to the west limb on 8--9 April, as shown in Figure~\ref{fig:EDF7_fsi_ch_outline}. There was also off Sun-Earth line remote observations taken by Solar Orbiter throughout its orbit covering the west limb during the entire timeframe of interest. 

\begin{figure}[h]
    \centering
    \includegraphics[width=0.35\linewidth]{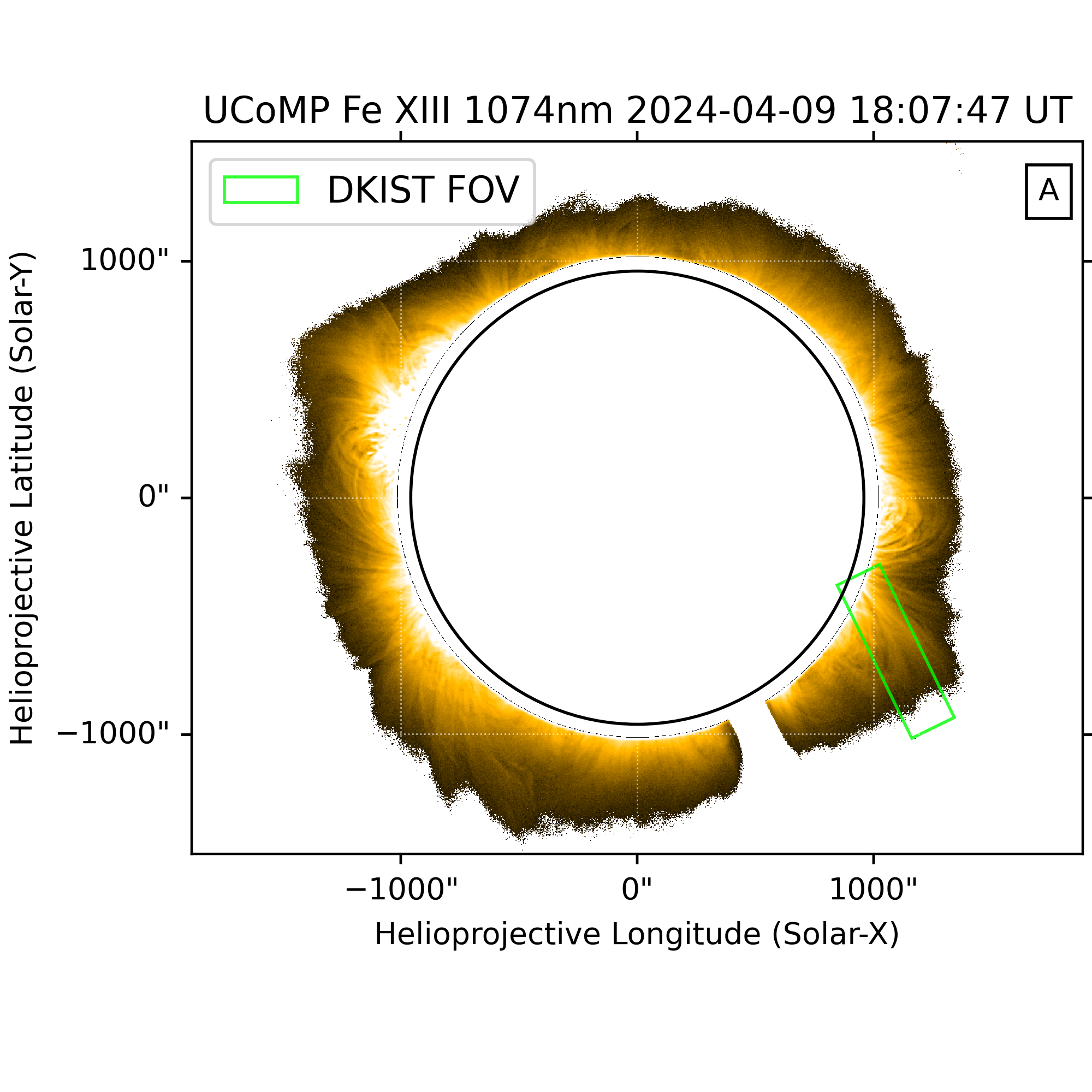}
    \includegraphics[width=0.35\linewidth]{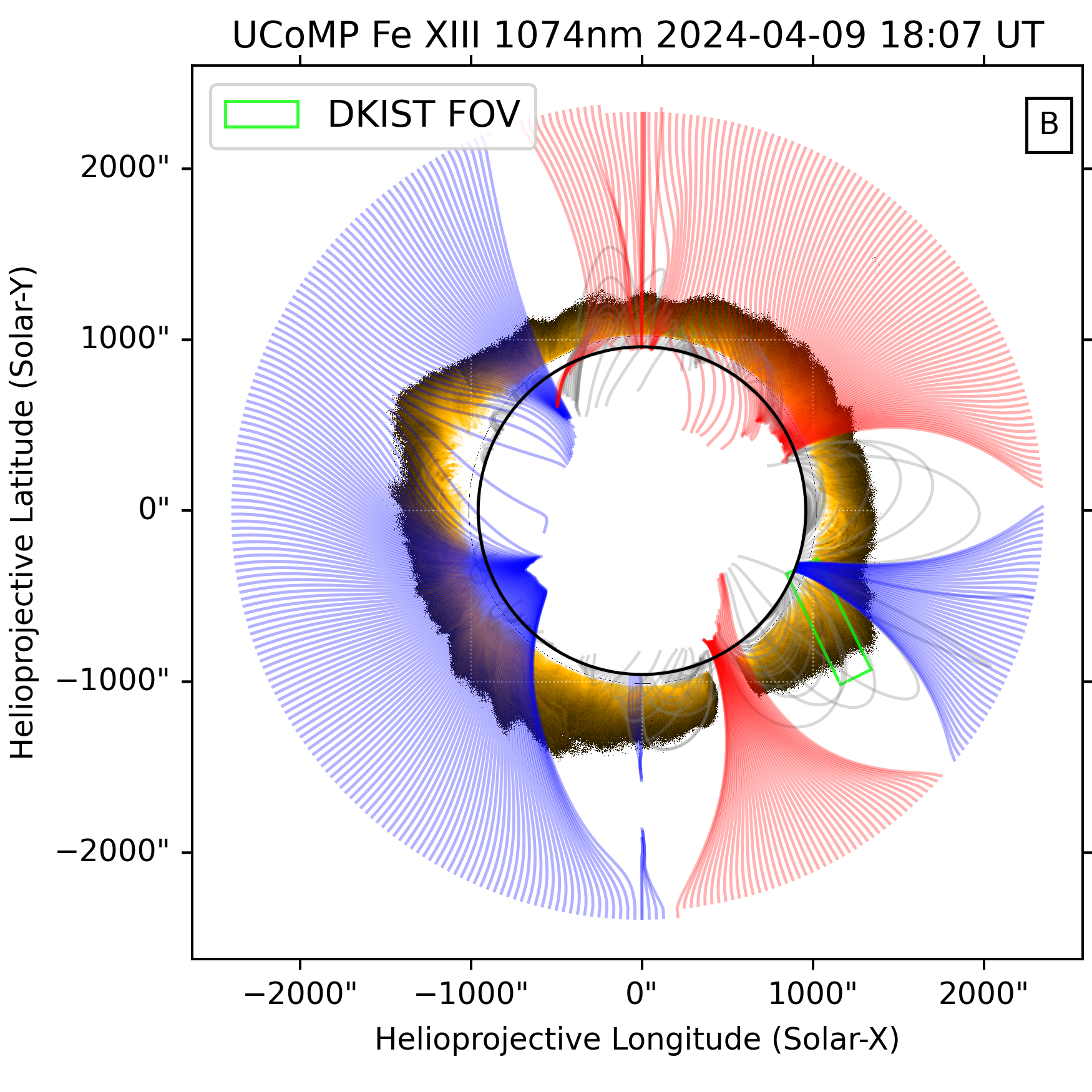}
    \caption{Identification of coronal hole off-limb on 9~April 2024. (A)~MLSO UCoMP 1074~nm (\ion{Fe}{13}) emission and DKIST Cryo-NIRSP FOV in green (also plotted in Figure~\ref{fig2_coordination}C). (B)~Same data as panel~(A) with PFSS field lines traced in the POS on 9~April with blue and red colors indicating open negative and positive polarity field lines respectively. Closed field lines are plotted in grey. A blue funnel of open field lines is co-located with a reduction in \ion{Fe}{13} emission at the expected location of the coronal hole on the west limb.}
    \label{fig:EDF0_UCoMP_CH_onlimb}
\end{figure}

\begin{figure}[h]
    \centering
    \includegraphics[width=0.4 \linewidth]{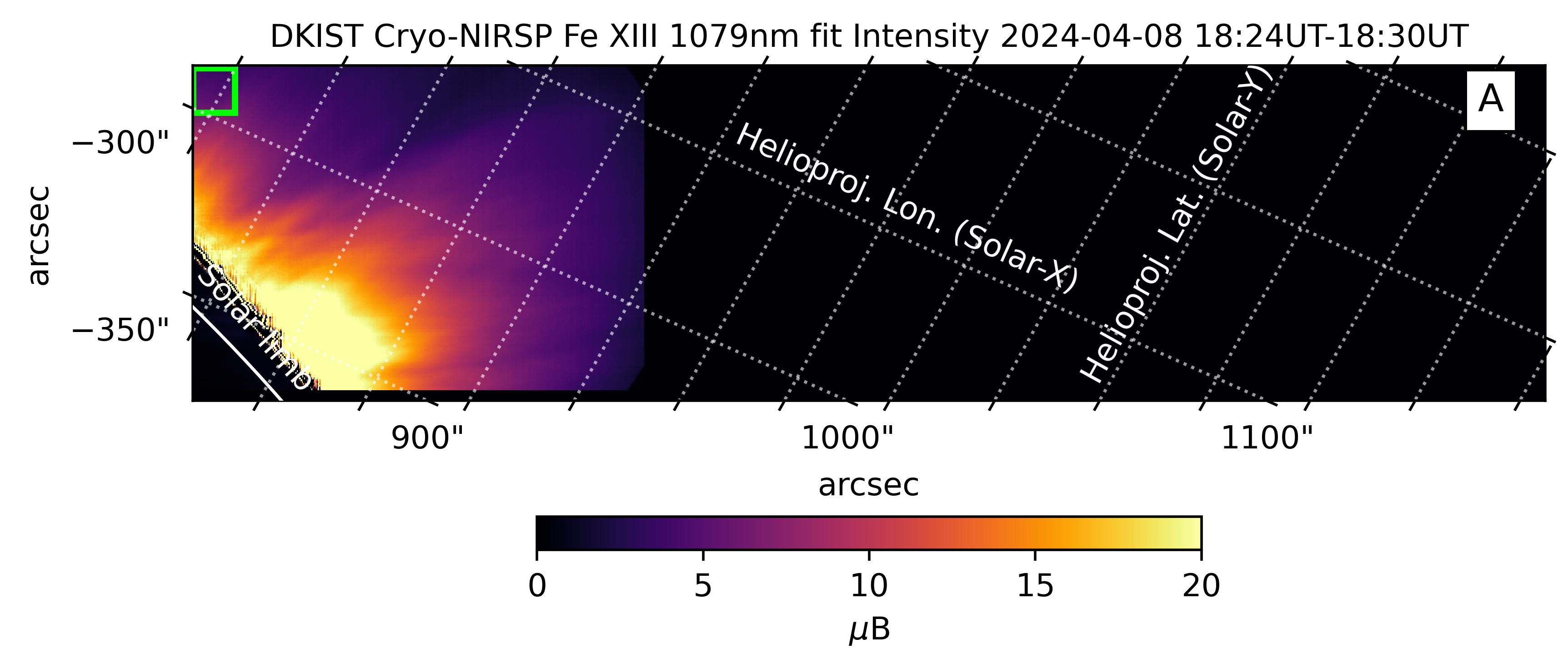}
    \includegraphics[width= 0.4\linewidth]{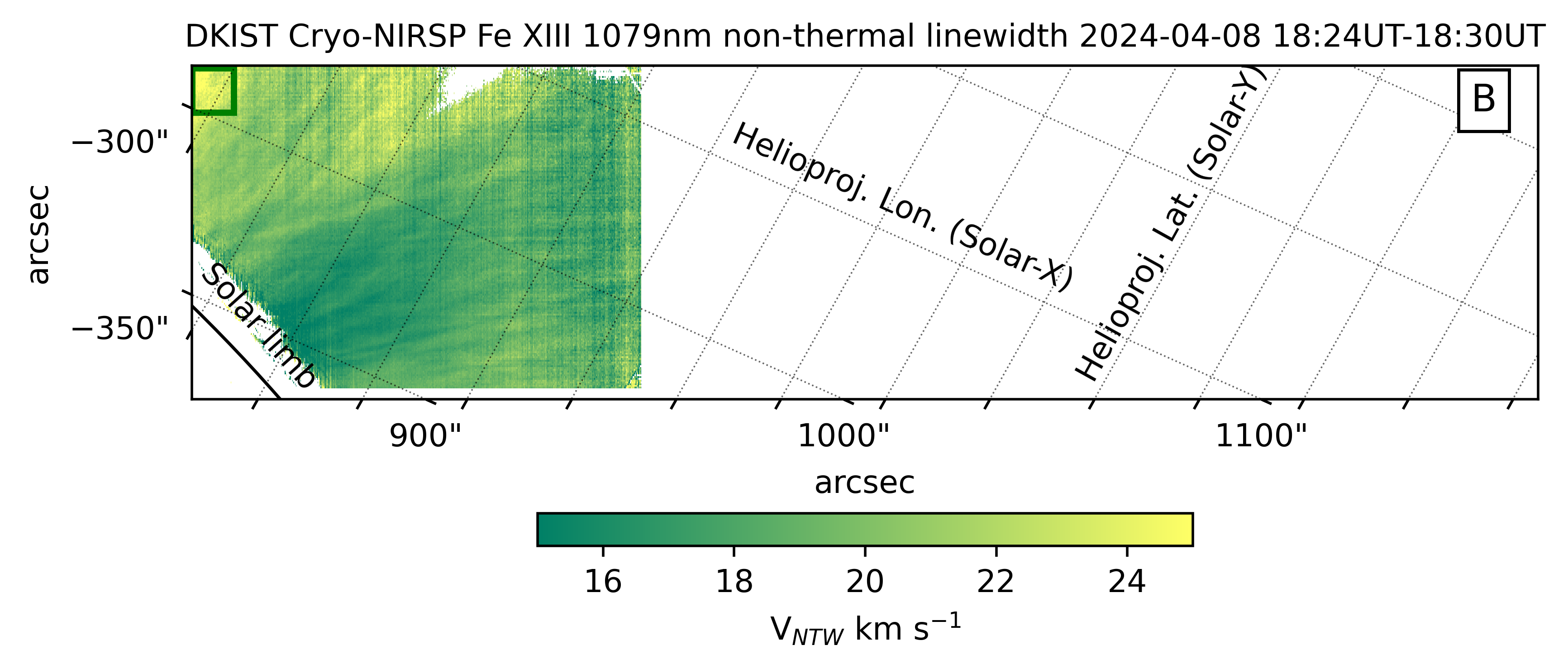}
    \caption{Fitted intensity~(A) and non-thermal linewidth in km~s$^{-1}$~(B) of the \ion{Fe}{13} 1079~nm spectral line. The green square is the location of the coronal hole sampled by DKIST Cryo-NIRSP on the west limb, as indicated in Figure~\ref{fig2_coordination}.}
    \label{fig:EDF6_DKIST_NTW}
\end{figure}

\begin{figure*}[h]
    \centering
    \includegraphics[width= 0.3
    \linewidth]{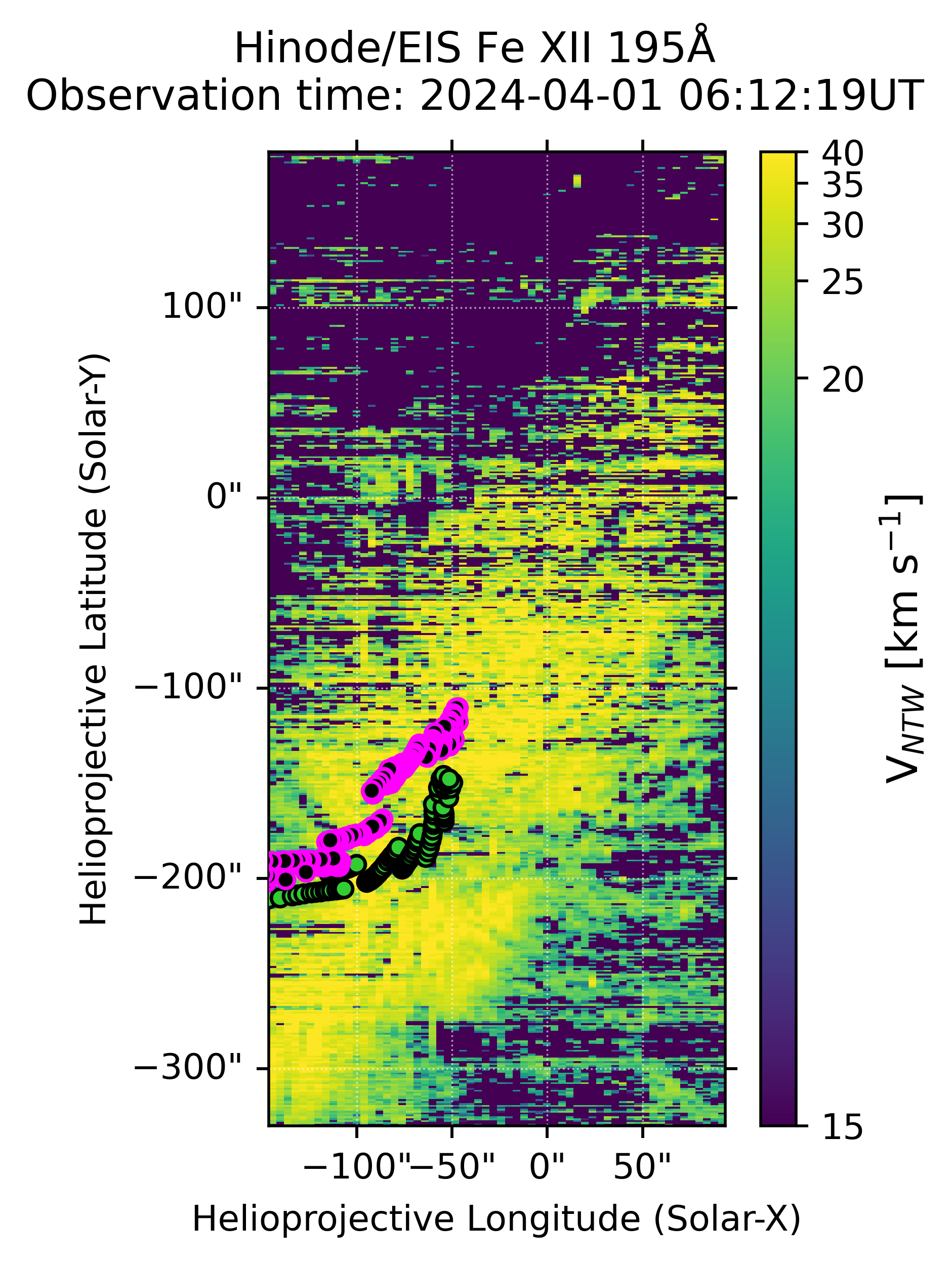}
    \caption{Non-thermal linewidth in km~s$^{-1}$ from Hinode/EIS \ion{Fe}{12} 195~{\AA} spectral line. The raster includes the \emph{Parker} and Solar Orbiter footpoints from Figure~\ref{fig2_coordination}.}
    \label{fig:EDF_EISNTW}
\end{figure*}

From the Sun-Earth line perspective during the TSE and day after, the coronal hole was identified on the west limb using the coronal structure observed with UCoMP and the co-located open magnetic field lines from the PFSS model, shown in Fig.~\ref{fig:EDF0_UCoMP_CH_onlimb}A,B. Fig.~\ref{fig:EDF0_UCoMP_CH_onlimb}A shows the Fe XIII 1074nm peak intensity along with the DKIST Cryo-NIRSP field of view on 9 April 2024. The coronal hole coincides with the dimming of the Fe XIII between two coronal loops structures located above/below a funnel of negative polarity open field lines, later sampled by \emph{Parker} and Solar Orbiter (Fig.~\ref{fig1_orbit_mapping}). This coronal location is used to sample the coronal hole properties on the limb in subsequent observations described later.

\begin{figure}[h]
    \centering
    \includegraphics[width= 0.45\linewidth]{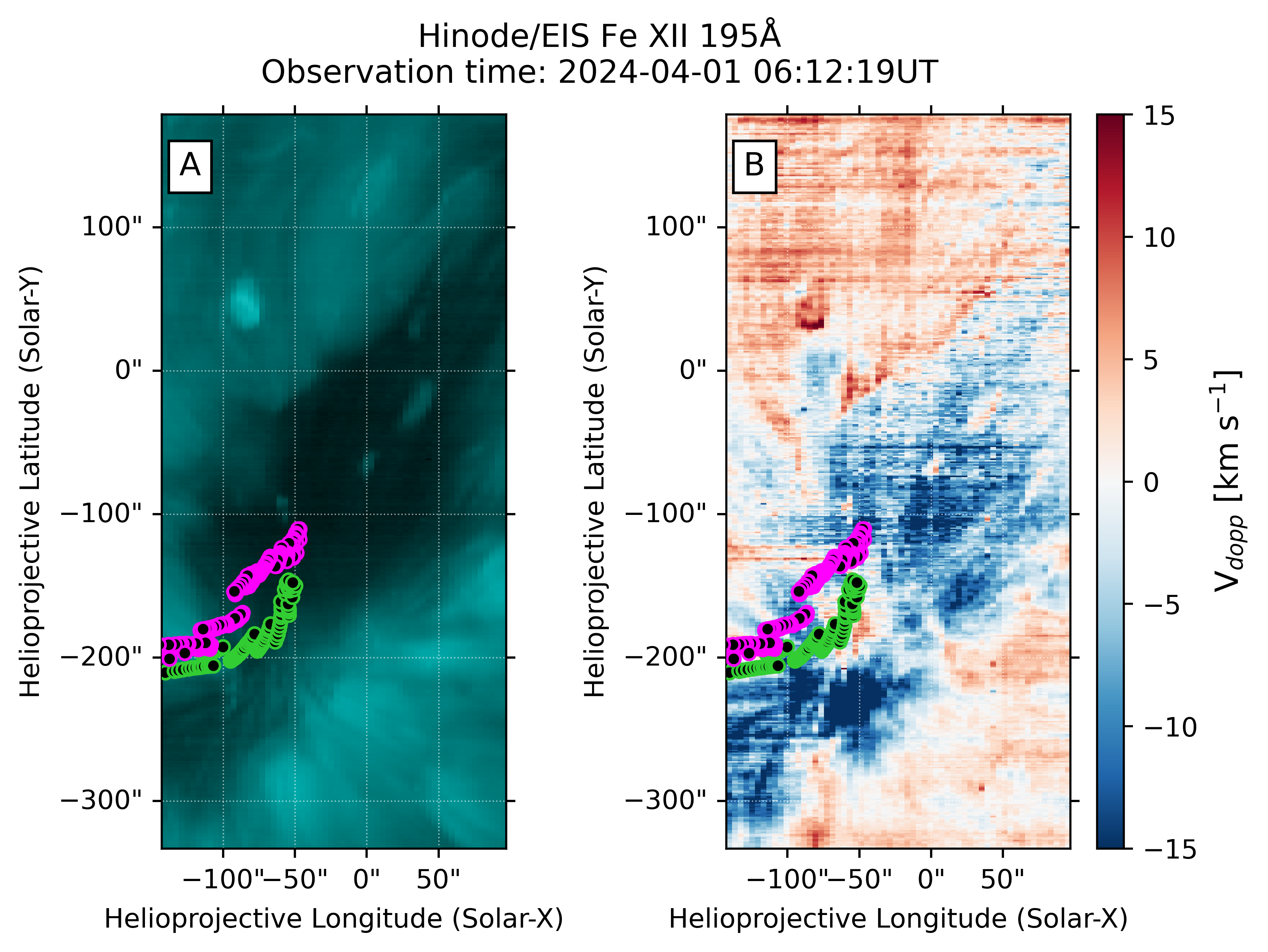}
    \includegraphics[width= 0.32\linewidth]{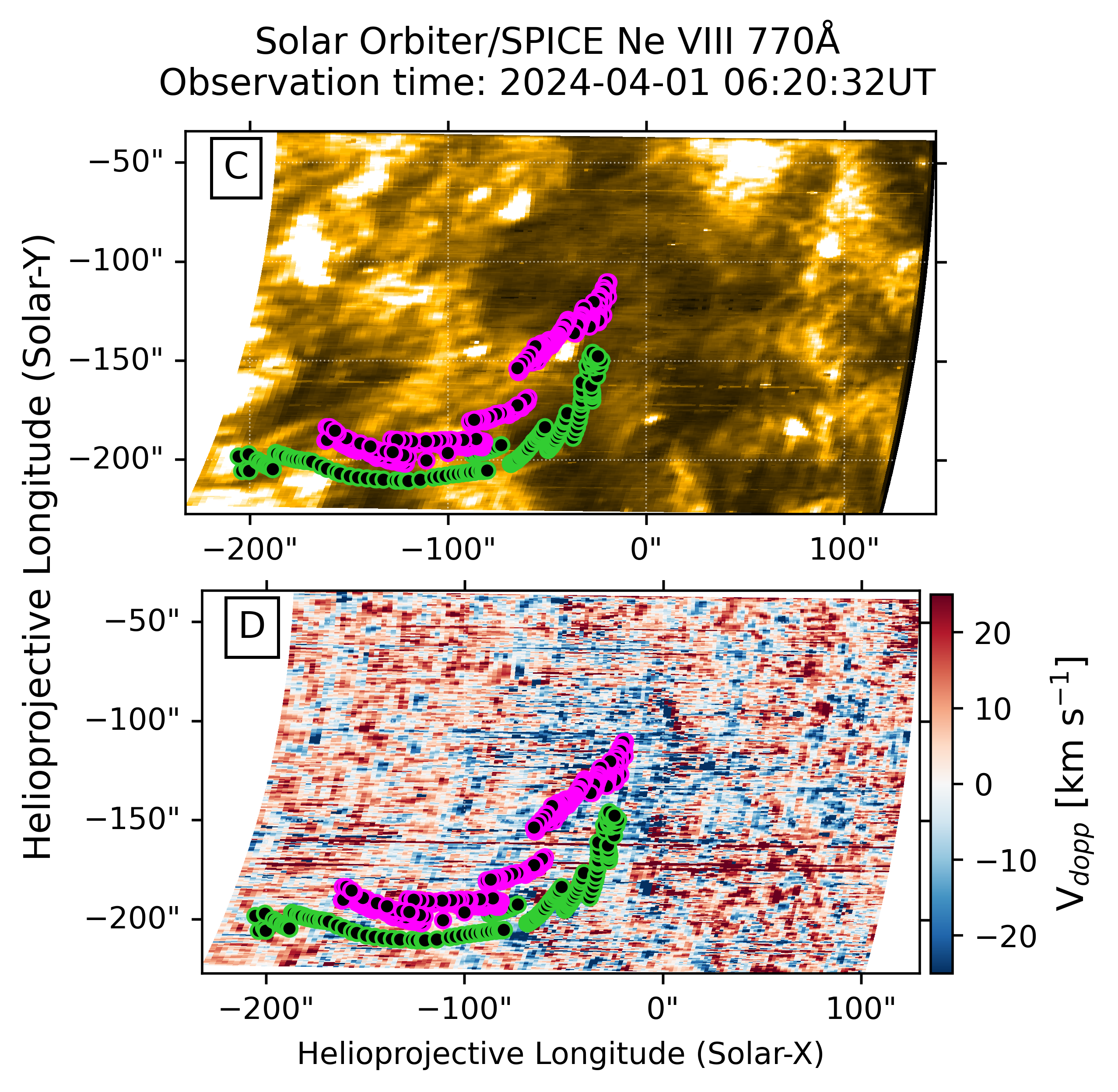}
    \caption{On-disk fitted intensities and Doppler velocities within the coronal hole boundary from Hinode/EIS \ion{Fe}{12} 195~{\AA}~(A, B) and Solar Orbiter/SPICE \ion{Ne}{8} 770~{\AA}~(C, D). All plots include the \emph{Parker} and Solar Orbiter footpoints from Figure~\ref{fig2_coordination}A,~B.}
    \label{fig:EDF5_EIS_SPICE_doppshifts}
\end{figure}

There are three main periods of interest: 1)~the timeframe leading up to the eclipse where SDO/AIA, Solar Orbiter/SPICE (1~April), and Hinode/EIS (1--3~April) observed the coronal hole, as shown for observations taken on 1~April in Figure~\ref{fig:EDF5_EIS_SPICE_doppshifts}A,~C. These remote observations were obtained as the result of a specific Solar Orbiter Observing Plan, the March--April~2024 instance of the ``Fast Wind SOOP'' during Solar Orbiter's Remote Sensing Window~14, aimed at investigating the release mechanisms of the fast solar wind.
2)~Between 29~March to 9~April, Solar Orbiter Full Sun Imager (FSI) captured the on disk evolution of the coronal hole up to the solar eclipse and covering a different perspective compared to ground-based observations, as shown for 8--9~April in Figure~\ref{fig:EDF7_fsi_ch_outline}. Also, while in quadrature with \emph{Parker} during its perihelion pass, the Metis coronagraph on Solar Orbiter imaged the coronal hole outflow (29~March 2024). 3)~On 8 and 9~April, ground-based observations from UCoMP \citep{ucomp} and K-Cor \citep{kcor} at the Mauna Loa Solar Observatory (MLSO) were taken of the full corona, shown in Figure~\ref{fig2_coordination}B. Also, DKIST's Cryo-NIRSP \citep{Fehlmann2023,Schad2023} located at Haleakal\={a} Observatory (Maui, HI) captured a small region of the southern part of the west limb of the Sun on 8 and 9~April, whose field of view is denoted in Figure~\ref{fig2_coordination}B. Although the main observations span 1~week, there were also SOHO/LASCO-C2 polarized brightness observations taken prior, during, and after the eclipse, totaling 14~days, utilized for tomographic reconstructions, as discussed in Section~\ref{sec:methods:tomography}.

\begin{figure}[h]
    \centering
    \includegraphics[width=0.4\linewidth]{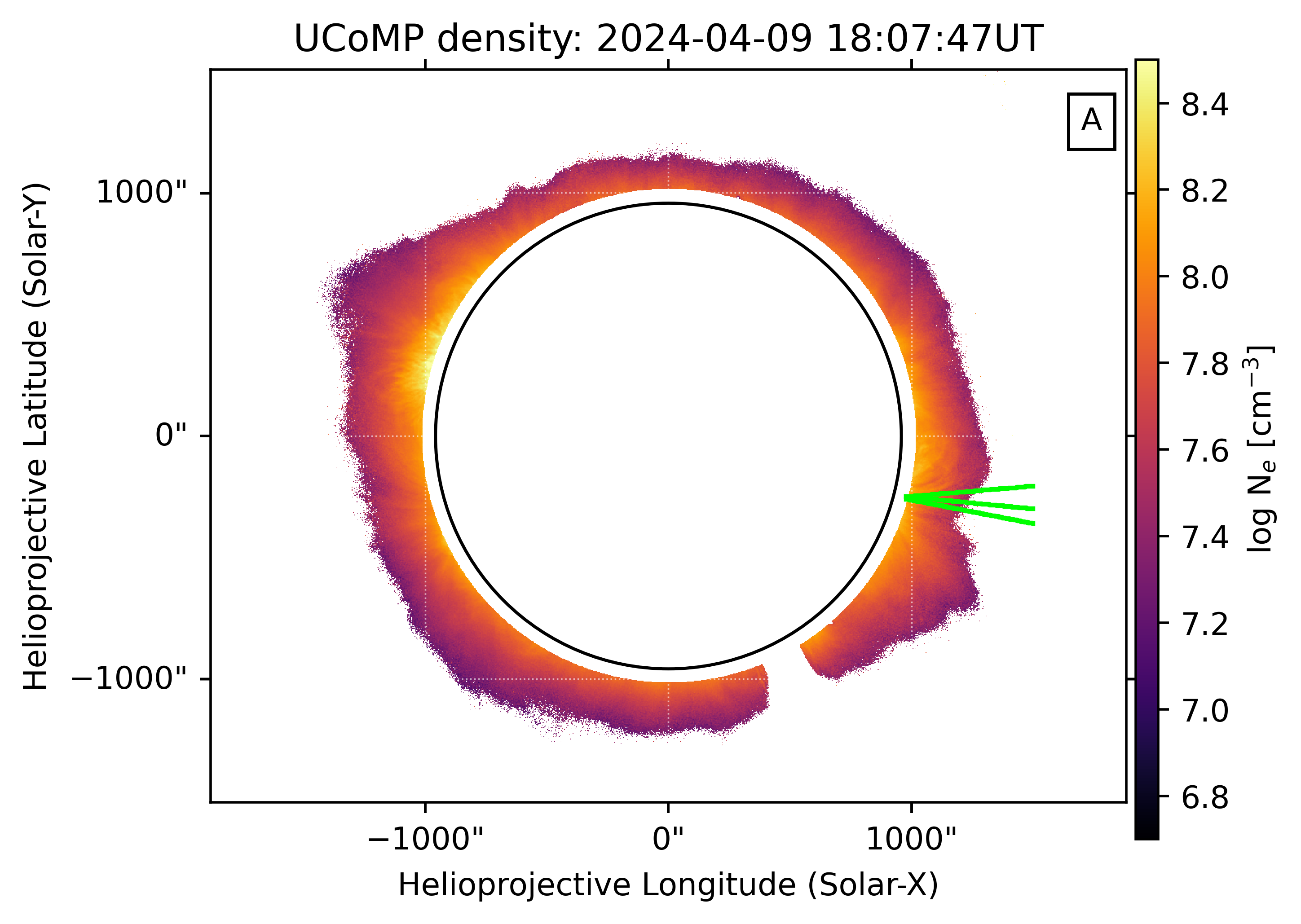}
    \includegraphics[width=0.4\linewidth]{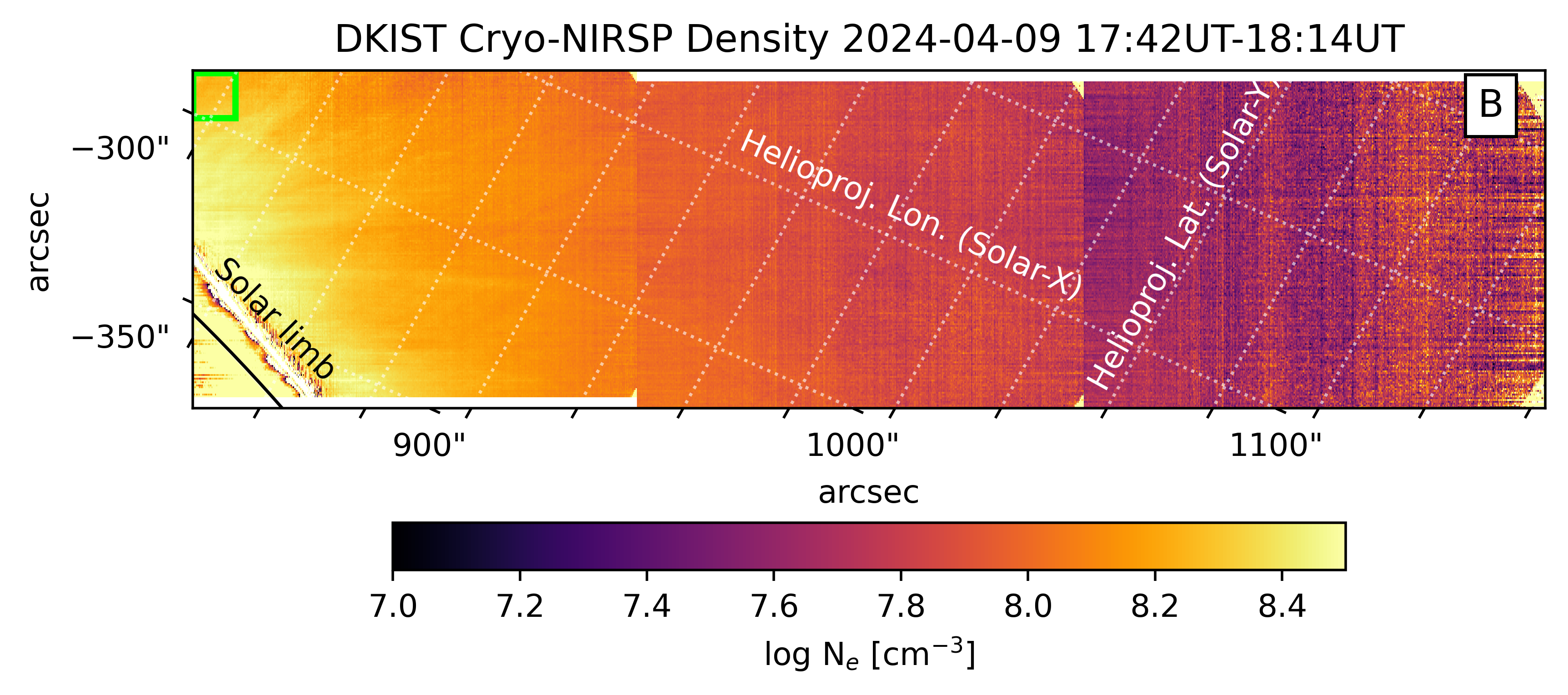}
    \caption{The electron density from MLSO UCoMP~(A) and within the DKIST FOV~(B). The green lines in~(A) are the locations sampled within the corona along the coronal hole. The green box in ~(B) is the location of the DKIST raster sampled along the corresponding coronal hole located on the west limb, as indicated in Figure~\ref{fig2_coordination}.}
    \label{fig:EDF1_UCoMP_den_b}
\end{figure}

\subsection{Density}
\label{sec:methods:density}

\begin{figure}[h]
    \centering
    \includegraphics[width=0.9\linewidth]{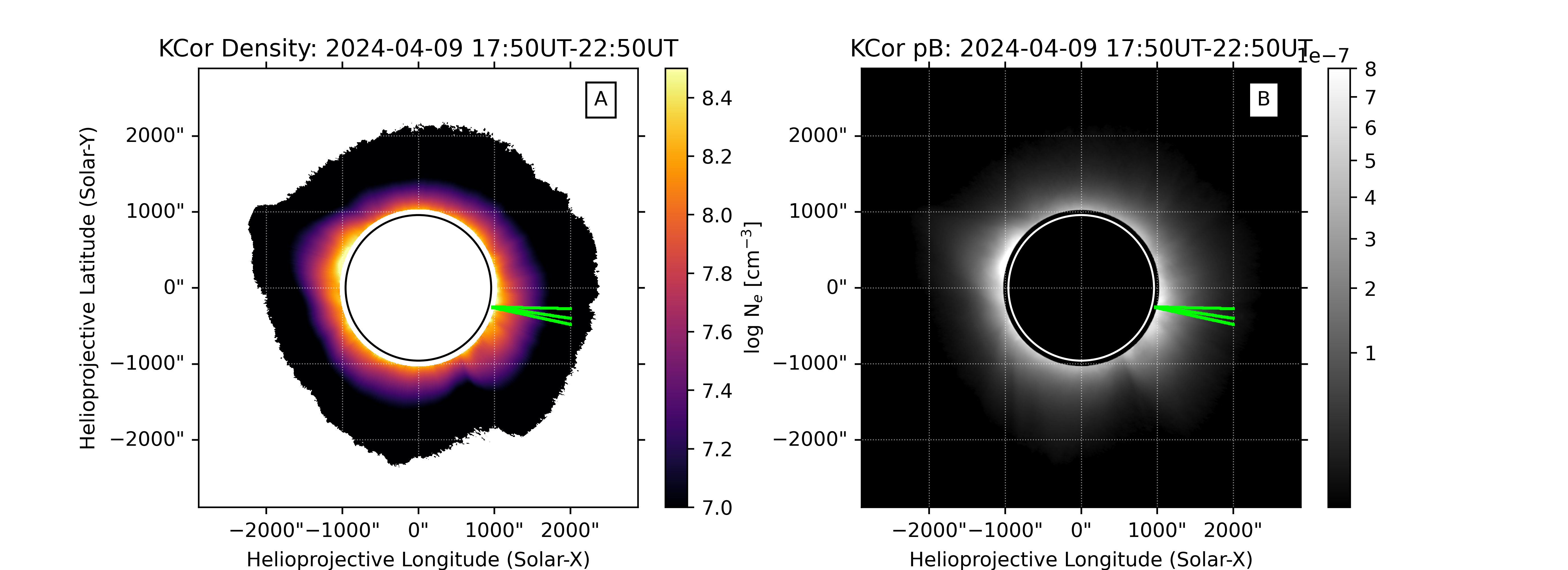}
    \caption{KCor derived density~(A) and pB~(B) on 9~April 2024. The green lines are the locations of the extended coronal hole sampled on the west limb, as indicated in Figure~\ref{fig2_coordination}.}
    \label{fig:EDF2_k_cor_dens}
\end{figure}

Figure~\ref{fig:EDF1_UCoMP_den_b} reveals the infrared (IR) low corona plasma electron density as observed by the DKIST Cryo-NIRSP and MLSO UCoMP instruments with only small temporal separation. We utilized DKIST Cryo-NIRSP L1 spectroscopic data, downloaded through the DKIST datacenter\footnote{\url{https://dkist.data.nso.edu/}}, and then further processed to line integrated quantities (peak and integrated line intensity, Doppler shifts, total line widths, etc.) through a customized analysis and line parameter fitting based on the public available Cryo-NIRSP Community code \citep{cryotutorial}. Three raster scans that are covering a combined radial height of $\sim$1.0--1.4~$R_{\odot}$ are observed subsequently in time. The rasters have been aligned as shown in Figures~\ref{fig:EDF1_UCoMP_den_b}B and~\ref{fig:EDF6_DKIST_NTW}. We have utilized L2 MLSO UCoMP Gaussian peak intensity data \citep{ucomp} obtained through the MLSO data archive\footnote{\url{https://www2.hao.ucar.edu/mlso/mlso-data-and-movies}}. Both instruments observe the low IR corona from the ground, using the highly ionized \ion{Fe}{13} 10746.8~{\AA} and 10798.0~{\AA} emission lines forming at temperatures of $\sim$1.5--2.0~MK. The spectral and line formation regimes along with the diagnostics potential of this IR \ion{Fe}{13} line pair have been discussed in the literature \citep{Judge1998, DelZanna2018B, Ali2022}. Figure~ \ref{fig:EDF2_k_cor_dens} shows an analogous derived density map computed using MLSO K-Cor L2 white-light polarized brightness data \citep{kcor}. This inference is based on the adaptation of the \citet{vandehulst1950} method, where we assumed a cylindrically-symmetric limb-darkened radiation field case, following the work of \citet{Schad2022}.

\begin{figure}[h]
    \centering
    \includegraphics[width= 0.4\linewidth]{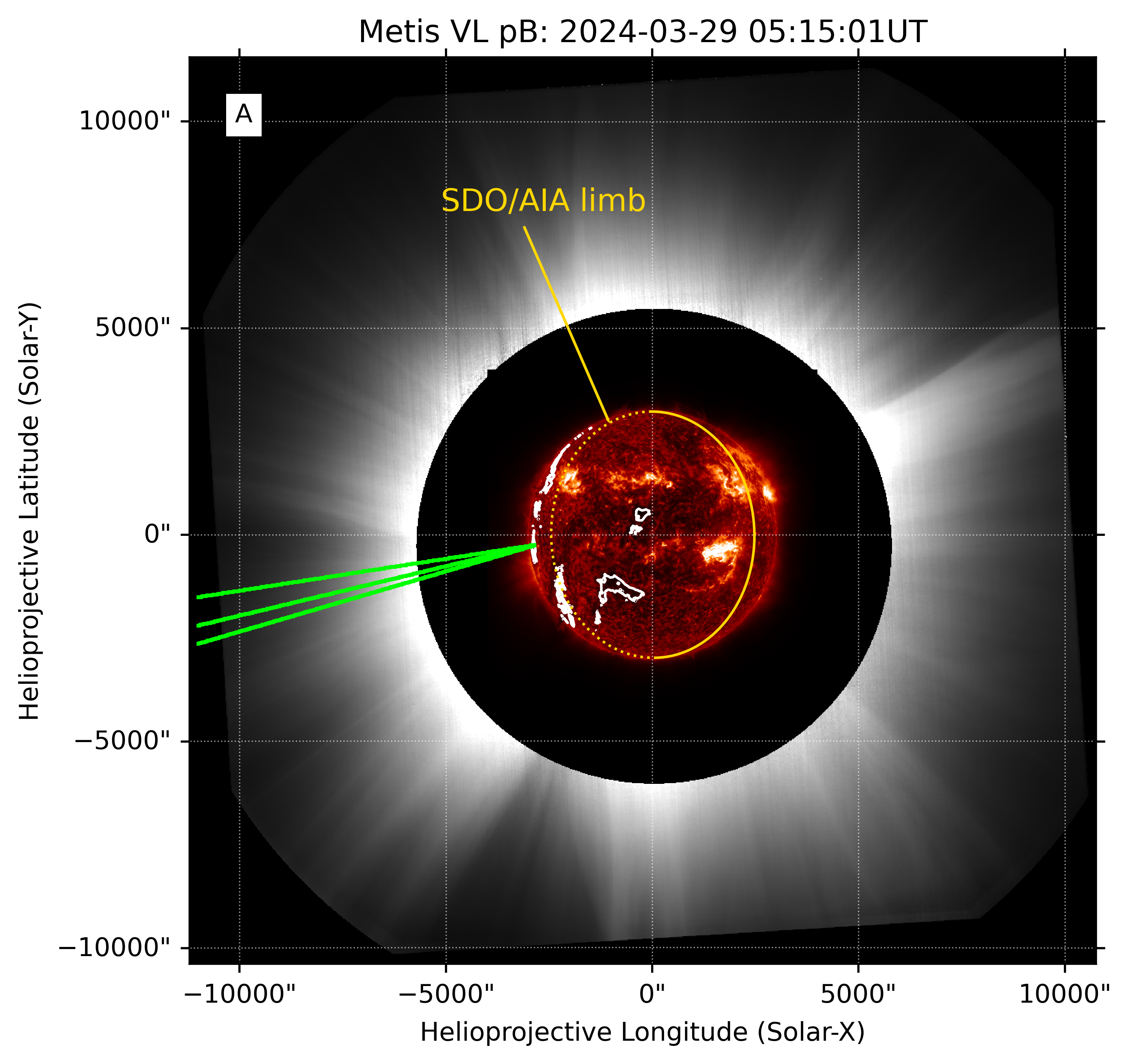}
    
    \includegraphics[width= 0.38\linewidth]{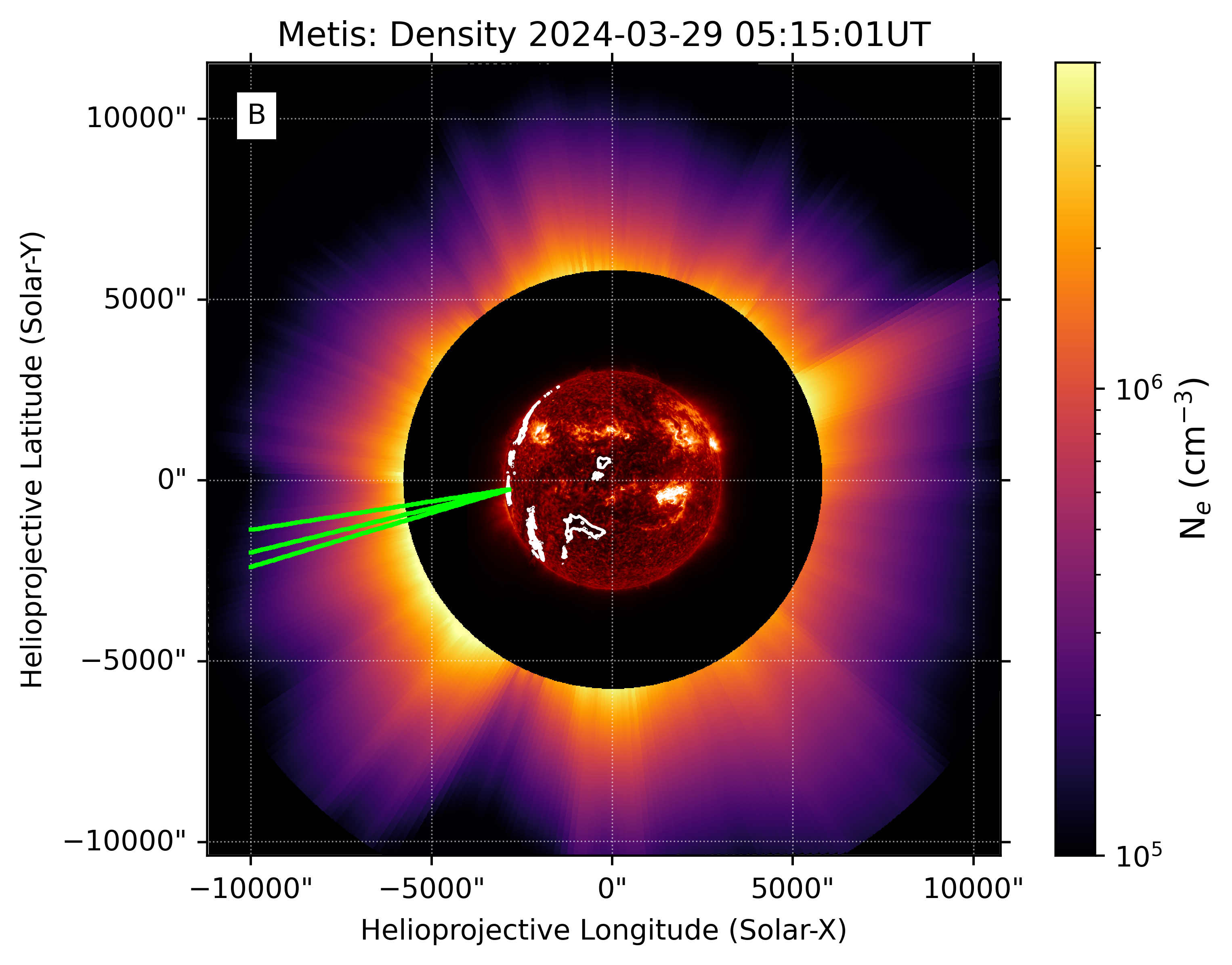}
    \includegraphics[width= 0.38\linewidth]{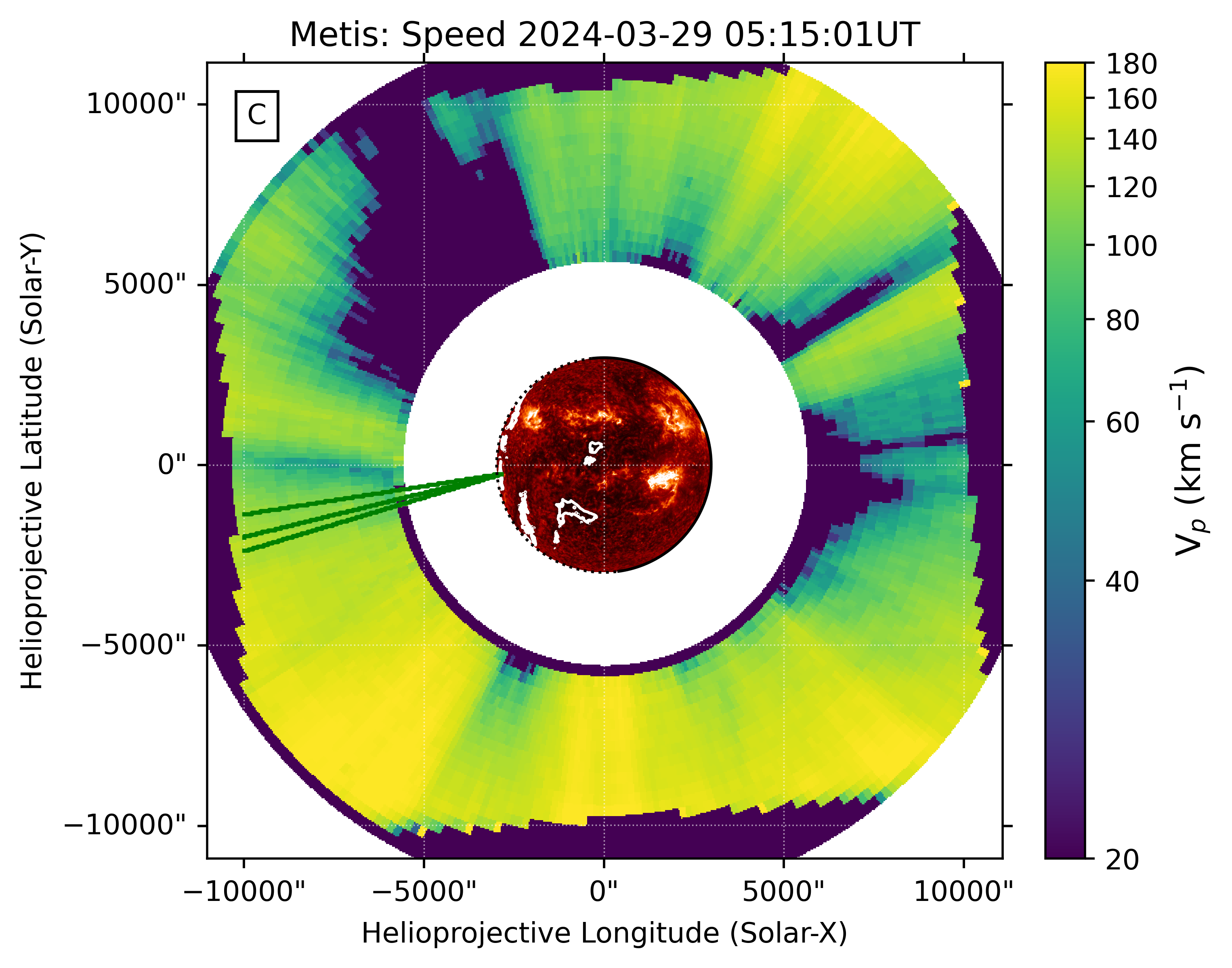}
    \caption{Visible Light (VL) pB observations taken by the Solar Orbiter/Metis coronagraph~(A) and derived diagnostics of electron density~(B) and outflow speed~(C) during 29~March 2024 at 05:15:01~UT. The white outline on the disk is the coronal hole outline from Figure~\ref{fig1_orbit_mapping}B showing the location on the limb. The limb of the AIA/SDO limb from the Sun-Earth line is shown in orange~(A). The green lines indicated the locations of the extended coronal hole sampled in the Metis observations. }
    \label{fig:EDF4_metis_dens_vels}
\end{figure}

A similar approach based on the inversion method of \citet{vandehulst1950} was also used to derive an electron density map from Metis white-light polarized brightness data acquired on 29~March, in the observed coronal region between 2--3~$R_{\odot}$ (see panels of Figure~\ref{fig:EDF4_metis_dens_vels}B).

For both DKIST Cryo-NIRSP and UCoMP, the electron density is derived using a community available \ion{Fe}{13} density calculator code \citep{fe13dens} through the use of the \ion{Fe}{13} 10747~{\AA}~/~10798~{\AA} line ratio, and by utilizing theoretical line formation calculations available through the CHIANTI~v11.0 \citep{Dere1997,delzanna2021} database. As discussed in \citet{DelZanna2018C}, careful consideration needs to be employed when interpreting such ratios, especially the IR \ion{Fe}{13} pair that is significantly influenced by photoexcitation. For this inference we have assumed that most of the observed coronal features form close to the POS and included collisional and radiative excitation effects, along with a limb-darkening effect, and a blackbody spectrum correction. A reasonable agreement between the DKIST Cryo-NIRSP, MLSO UCoMP, and MLSO K-Cor instruments has been reached, considering the different observing regimes, methods and assumptions, temporal offsets, alongside instrumental calibration, pointing, and resolution differences.

\subsection{Temperature}
\label{sec:methods:temperature}

\begin{figure}[h]
    \centering
    \includegraphics[width= 0.37\linewidth]{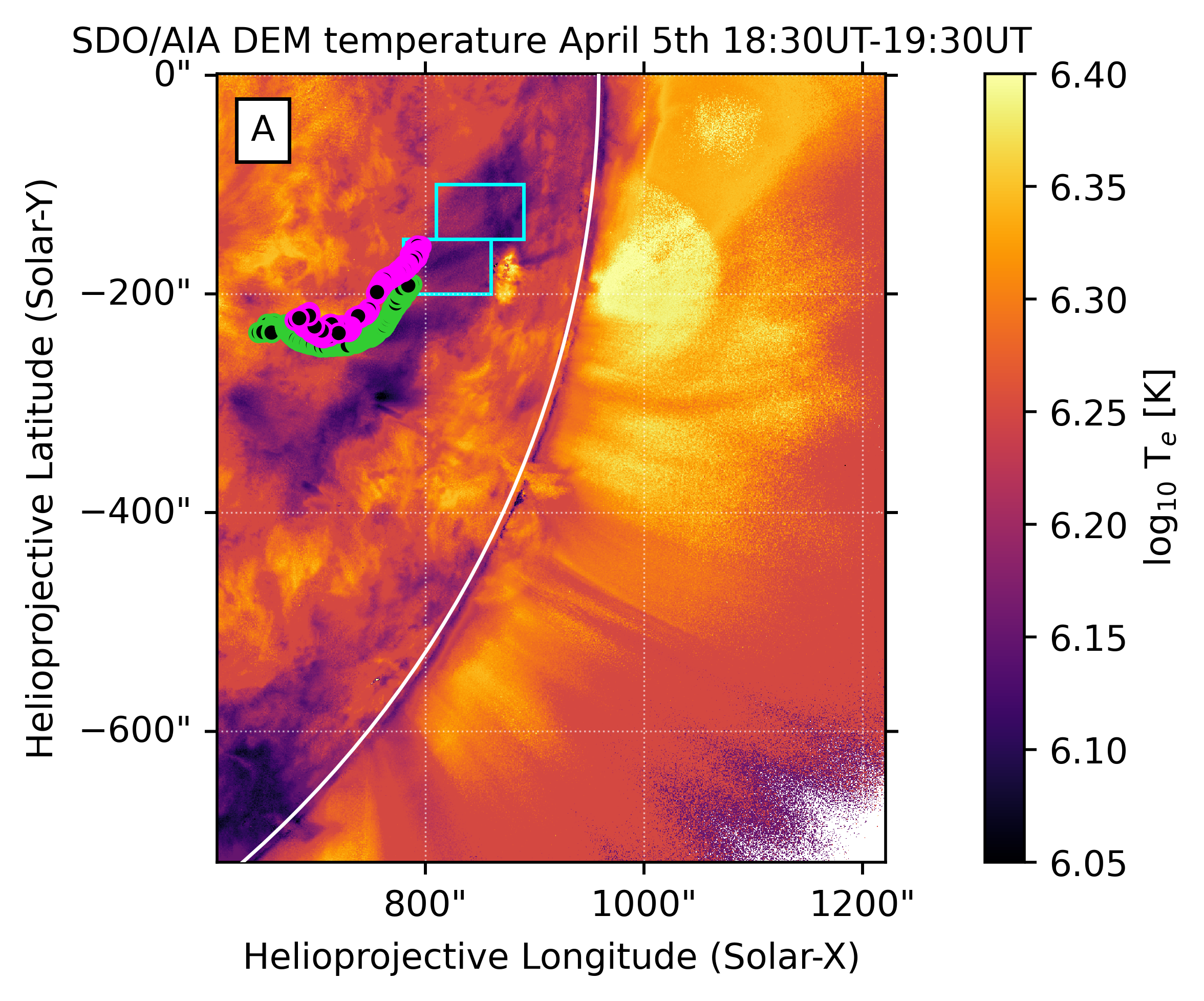}
    \includegraphics[width= 0.47\linewidth]{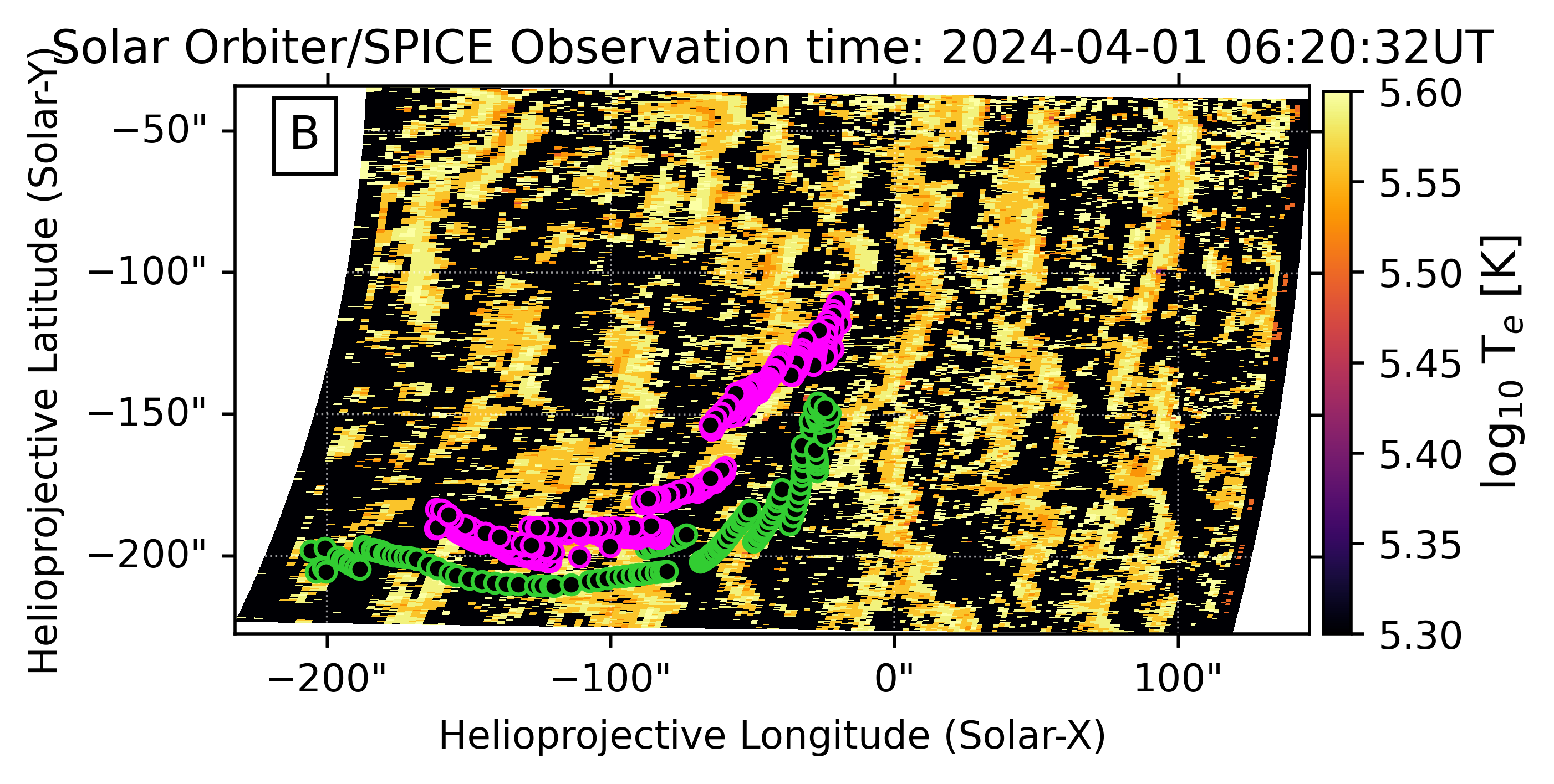}
    \caption{DEM showing the peak electron temperature of the corresponding coronal hole on the solar disk using SDO/AIA EUV channels~(A) and Solar Orbiter/SPICE UV wavelengths~(B). Both include the \emph{Parker} and Solar Orbiter footpoints from Figure~\ref{fig2_coordination}A,~B.}
    \label{fig:EDF3_spice_aia_temp}
\end{figure}

\textbf{SDO/AIA}: We measure the temperature of the coronal hole through the use of Differential Emission Measures (DEMs) of six optically thin AIA channels (94, 131, 171, 193, 211 and 335~{\AA}). To improve the signal-to-noise, we temporally bin the input data for each channel over one hour. We calculate the DEM using the sparse inversion DEM method presented in \citet{Cheung2015}. From the DEM, we calculate the EM-weighted mean temperature for each pixel, to generate maps of plasma temperature; see Figure~\ref{fig:EDF3_spice_aia_temp}A. Using the EM-weighted mean temperature as a measurement of temperature makes the assumption that plasma is close to isothermal in the LOS. From a manual inspection of pixels within our region of interest, we find typical single Gaussian-like profiles in the DEM, thus validating this assumption. 

While the coronal hole is on the limb (during the 8--9~April observation window), the DEM is dominated by hotter active region plasma near the coronal hole in the EUV channels, skewing the temperature measurement. Due to this, we instead resort to using on-disk measurements of the coronal hole from a few days prior on 4--5~April. During this period, we find an average coronal hole temperature of $\sim$1.1--$1.23\pm0.04$~MK. This is in the range of previously-published measurements of coronal hole temperatures, with values of 1.04~MK at solar maximum by \citet{Doyle2010}, $\sim$0.18--1.3~MK by \citet{Nistico2011}, 0.8--1.2~MK by \citet{Stansby2021} and $\sim$$0.94\pm0.1$~MK by \citet{Hegde2024}.

\textbf{Solar Orbiter/SPICE}: \citealt{SPICEConsortium_2020} was observing during the remote sensing windows in coordination with the rest of the Solar Orbiter instruments. SPICE observed the coronal hole connecting to the two spacecraft on 1--2~April 2024 during five different observations. SPICE observed many transition region lines with sensitivities between $\log(T~[K])=4$--6. We preprocessed the data with the PSF artifact removal algorithm developed by \citet{Plowman_2023,Plowman_2025}. Based on the result from this analysis, we were able to infer the spectral line parameters, such as the line intensity, Doppler shift and line width. Some of those parameters are shown in Figure~\ref{fig:EDF5_EIS_SPICE_doppshifts}D. In the following DEM inversion analysis based on \cite{Plowman_EMToolKit_A_Standardized}, we used only the lines forming above $\log(T) > 5$, so that we exclude all optically thick lines. In particular, we used the lines of \ion{S}{4} 748 \& 750~{\AA}, \ion{S}{5} 786~{\AA}, \ion{O}{5} 762~{\AA}, \ion{Ne}{8} 770~{\AA}, \ion{Mg}{8} 772~{\AA}, and \ion{Mg}{9} 706~{\AA}. The results from the SPICE observations provide a low temperature DEM constraint than available from other instruments and are shown in Figure~\ref{fig:EDF3_spice_aia_temp}B. 

\begin{figure}[h]
    \centering
    \includegraphics[width= 0.48\linewidth]{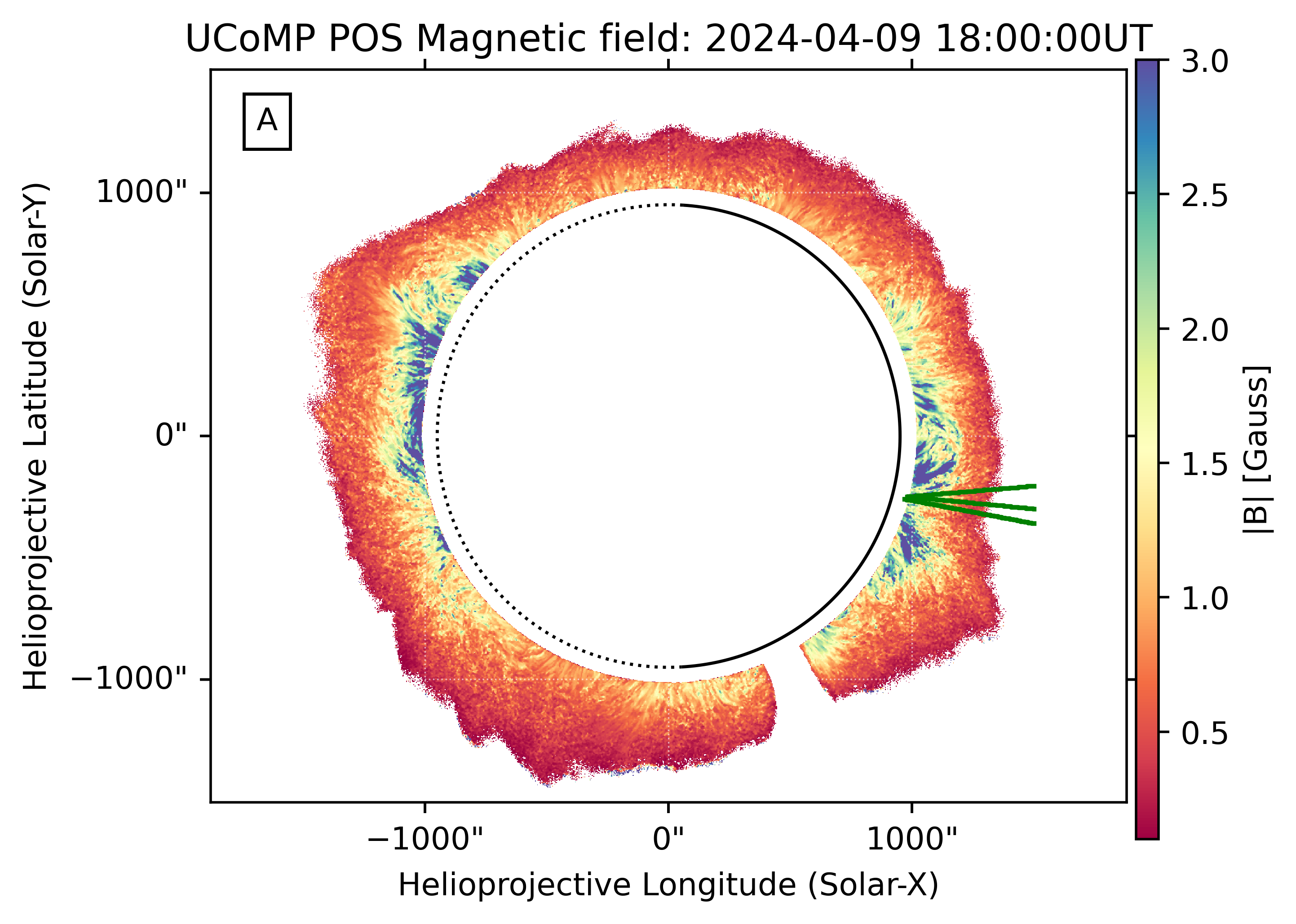}
    \includegraphics[width= 0.48\linewidth]{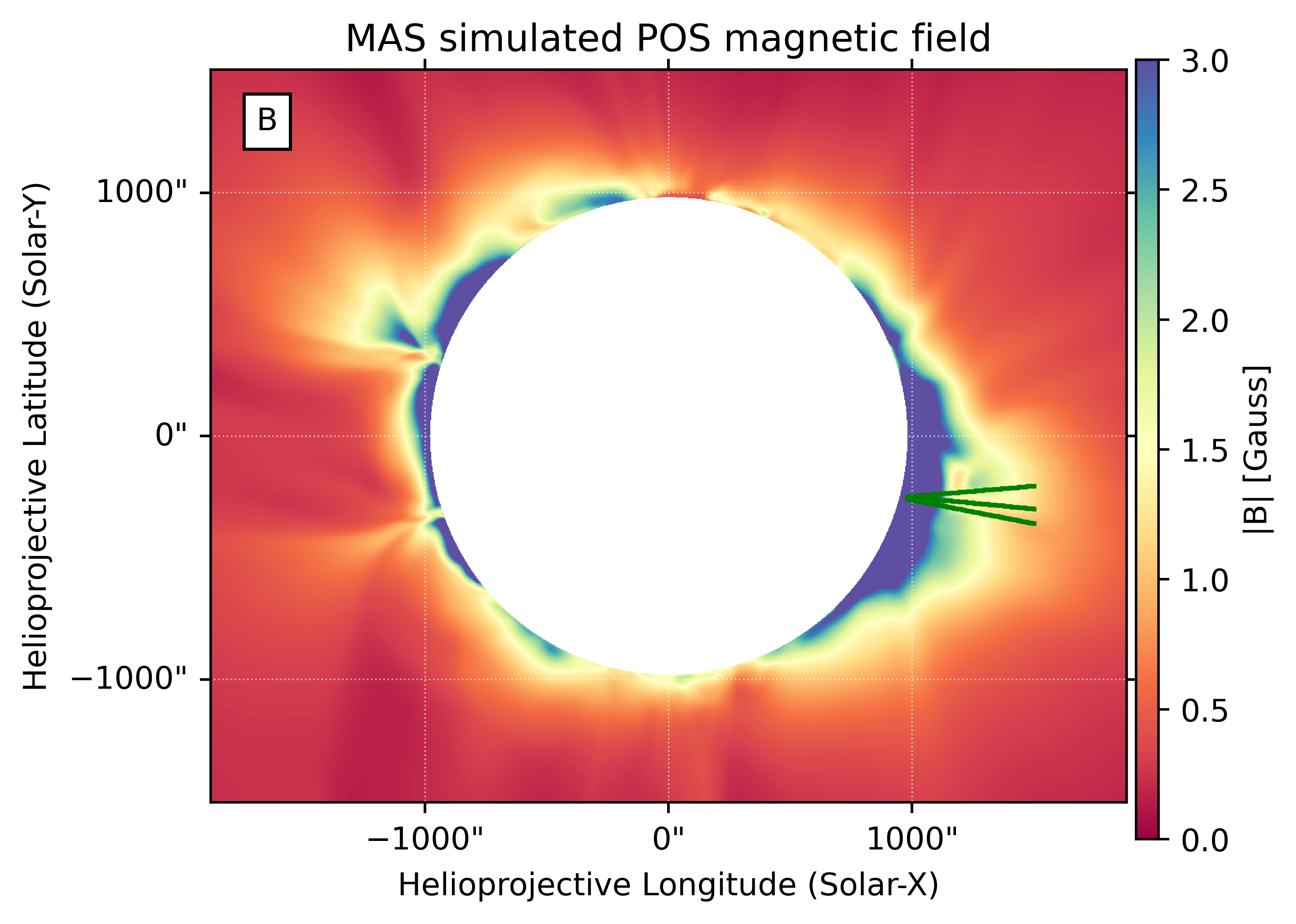}
    \includegraphics[width= 0.48\linewidth]{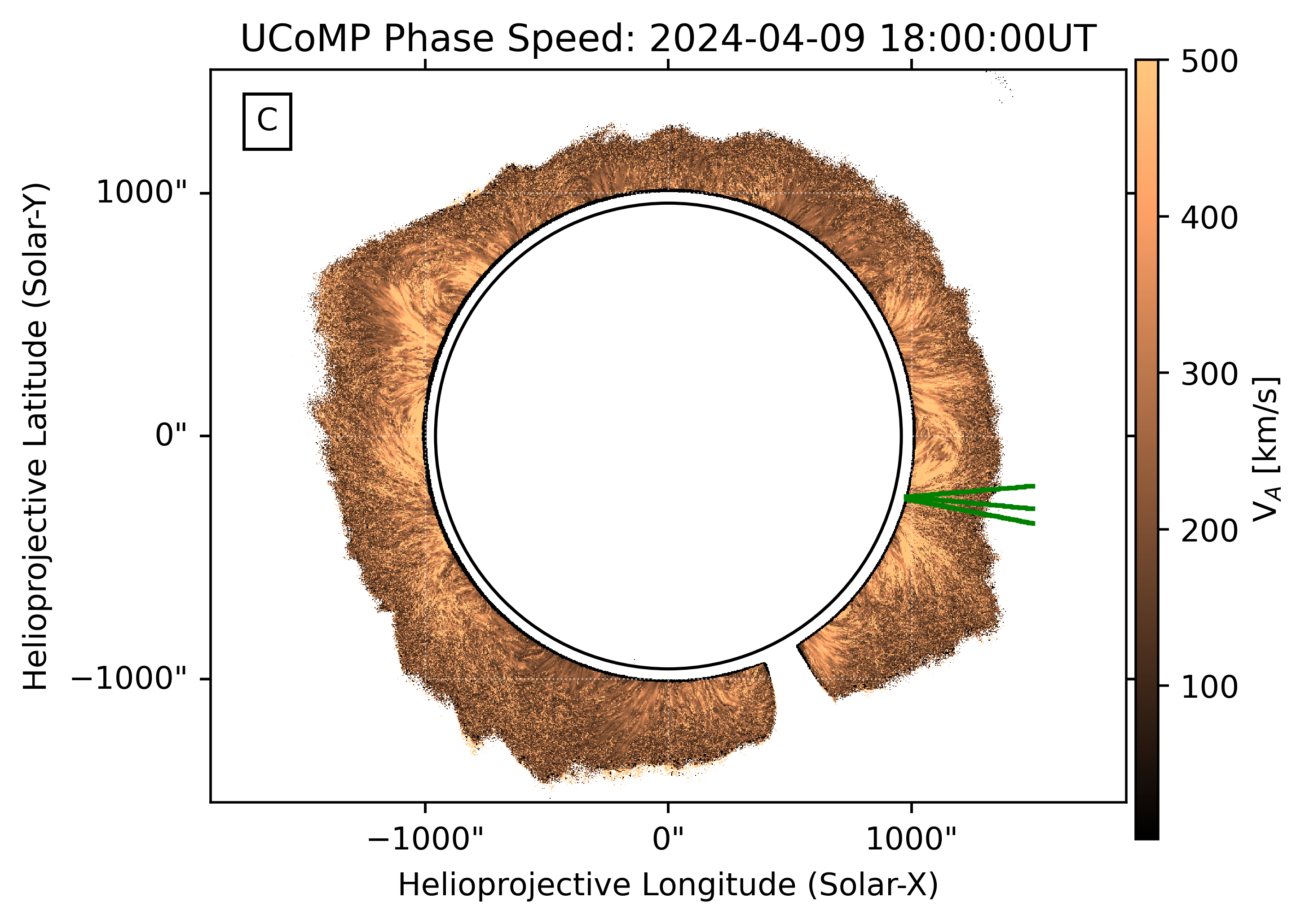}
    \caption{POS derived magnetic field magnitude for UCoMP~(A), equivalently computed for a MAS simulation~(B), and phase speed~(C). The green lines indicate the locations of POS magnetic field sampled within the corresponding coronal hole on the west limb, as indicated in Figure~\ref{fig2_coordination}.}
    \label{fig:EDF8_UCoMP_b}
\end{figure}

\subsection{Magnetic field}
\label{sec:methods:magnetic_field}

We used a two-dimensional coronal seismology technique to measure the global distribution of the coronal magnetic field strength based on UCoMP observations \citep{Yang2020a, Yang2020b, Yang2024}. The magnetic field results are shown in Figure~\ref{fig:EDF8_UCoMP_b}A. The phase speed of the ubiquitous transverse waves as observed by UCoMP is determined by the magnetic field strength and density in the corona. To calculate the phase speed of the transverse waves, we used a wave tracking method, see Figure~\ref{fig:EDF8_UCoMP_b}C. Since coronal Alfv\'{e}n waves are generally expected to exist only in the torsional mode, the waves observed by UCoMP are instead a type of kink wave that is nearly incompressible. However, given the spatial resolution of UCoMP and under the low plasma beta assumption in the low corona, the expression for the kink speed has the same form as that for the Alfv\'en speed. Therefore, in practice, the measured phase speed value is generally the same as the Alfv\'en speed, which we use from Figure~\ref{fig4_radialplots}C. This method involves two main steps: determining the wave propagation direction and computing the wave phase speed. For each pixel, the local propagation direction is derived by calculating coherence between the Doppler velocity time series of surrounding pixels and fitting a weighted linear function to regions with high coherence. Along the identified wave paths, the phase speed is obtained by cross-correlating the central time series with those along the path and applying a weighted linear fit to the resulting time lags \citep[e.g.,][]{Tomczyk2009, Morton2015, Yang2020a, Yang2024}. To diagnose the coronal electron density, we used the intensity ratio of the two \ion{Fe}{13} infrared lines observed by UCoMP, which is sensitive to electron density \citep[e.g.,][]{Dudik2021}; the resulting density is shown in Figure~\ref{fig:EDF1_UCoMP_den_b}A. Combining the wave phase speed and coronal density results, we could obtain the global distribution of the coronal magnetic field strength. Previous studies have shown that these magnetic field measurements primarily reflect the POS component, and are likely representative of the emissivity-weighted average of the POS magnetic field along the line-of-sight (LOS) \citep{Yang2024}.

Because coronal spectroscopic measurements are intrinsically sensitive to emission integrated along the LOS, we represent the model magnetic field using an emissivity-weighted LOS average following \cite{Yang2024}. In particular, the POS magnetic field strength relevant for comparison with observations is defined as,
\begin{equation}
    B_\text{POS}=\frac{\int{B_{\text{POS},k}\varepsilon_k}\text{d}h}{\int{\varepsilon_k}\text{d}h} \quad,
\end{equation}
where $B_{\text{POS},k}$~is the local POS magnetic field strength at position~$k$ along the LOS, $\varepsilon_k$~is the corresponding \ion{Fe}{13} 10747~{\AA} line emissivity, and d$h$~denotes the differential path length.

The required magnetic field and emissivity values were sampled at each grid point intersected by the LOS within the field of view, using a coronal MHD simulation generated with the magnetohydrodynamic algorithm outside a sphere (MAS) code based on the photospheric magnetogram derived from an observation on 9~April 2024. The resulting coronal magnetic field map from the MAS model is shown in Figure~\ref{fig:EDF8_UCoMP_b}B. The MAS simulation setup and description is detailed in \cite{Downs2025_eclipse}.

\subsection{Non-thermal widths} \label{sec:methods:nonthermal_widths}
In order to tighten our energy budget estimation, we have analyzed the non-thermal line widths ($V_{NTW}$) at the base of the low corona using the three previously discussed DKIST Cryo-NIRSP spectroscopic scans that partially cover the base of the coronal hole as shown by the green plots in Figure~\ref{fig2_coordination}B. To estimate the total line widths, we follow the primary data, calibration and analysis procedures as described in Section~\ref{sec:methods:density}. In order to estimate the non-thermal component, we have to disentangle the components contributing to the inferred line width. These are: the instrumental broadening, thermal broadening, and non-thermal broadening. DKIST Cryo-NIRSP is a high-efficiency spectrograph where the instrumental effects not corrected by the primary data reduction can be considered negligible with respect to coronal measurement uncertainties. A default analysis approach involves removing a constant thermal component from the line formation using the peak line formation temperature, which for \ion{Fe}{13} is $\sim$6.23~MK. In this work, we have adopted a more complex procedure aimed at extracting feature dependent non-thermal widths. Chapter~7 of \citet{DelZanna2018C} and applied to DKIST Cryo-NIRSP \ion{Si}{10} observations by \citet{Schad2024B}. This method involves combining emissivity maps derived from SDO/AIA as discussed in Section~\ref{sec:methods:temperature} with theoretical line formation forward calculations performed using the \citet[PyCELP]{pycelp2021} forward synthesis code. In both the theoretical PyCELP calculation and the SDO DEM inversions discussed in Section~\ref{sec:methods:temperature} we utilize the same ``chianti2021'' abundance estimation and optimize for the best observation matching effective formation temperatures. This method is described in more detail in the related eclipse observational analysis work \citep{Paraschiv2026}. The fitted intensity of \ion{Fe}{13} and non-thermal widths are shown in Figure~\ref{fig:EDF6_DKIST_NTW}A,~B.

For EIS, non-thermal widths from \ion{Fe}{12} 195~{\AA} are derived by fitting the observed profile with a double Gaussian to separate the 195.12~{\AA} line from its self-blend at 195.18~{\AA}, then decomposing the resulting FWHM $W$ into instrumental ($W_I$), thermal ($W_{\rm th}$), and non-thermal contributions using $4\ln 2\,(\lambda_0/c)^2\,V_{NTW}^2 = W_{NTW}^2 - W_I^2 - W_{\rm th}^2$, where $W_I$ is the instrumental broadening and $W_{\rm th}$ is computed from the \ion{Fe}{12} ion mass and an assumed formation temperature ($\log T \approx 6.2$); solving for $V_{NTW}$ then gives the non-thermal velocity \citep{Doschek2007, Baker2013, Brooks2016}.

\subsection{Scaling of WKB approximation} \label{sec:methods:WKB_curves}
The WKB approximation assumes that the background plasma properties (density, magnetic field strength) change slowly compared to the wavelength of the Alfv\'en waves. In this approximation, outward-propagating Alfv\'en waves from the Sun should propagate outward without reflecting and interacting with the medium, i.e., dissipation-free, with their amplitude scaling according to conservation of wave action \citep{Jacques1977}. 

In Figure~\ref{fig4_radialplots}E and F, we compare the amplitude of the observed fluctuating quantities to the WKB scaling for non-accelerating and accelerating solar wind to compare the progression of these quantities into the heliosphere. In the non-accelerating solar wind, the theoretical curve is computed for the WKB approximation that predicts that the magnetic fluctuation power should scale with Mach number.

We compute the energy density under the conservation of wave action as, \textbf{$E_D \propto V_A/(r^2(v_{p} + V_A)^2)$} \citep{Jacques1977}. $E_D \propto \langle\delta B^{2}\rangle/4\pi$, therefore for $M_A = v_p/V_A$, we plot $dB/B \propto \sqrt{4\pi E_D}/B$ as the dark red curve in panel~F with an upper and lower $V_A$ bound of Figure~\ref{fig4_radialplots}C. The curve is empirically anchored to the value of $dB/B$ derived from \emph{Parker} observations. For the computation of $E_D$, we use the curves of $V_A$ (Figure~\ref{fig4_radialplots}C), magnetic field, $B$, (Figure~\ref{fig4_radialplots}D), and $V_R$ curve in dark grey (Figure~\ref{fig4_radialplots}C).

\subsection{Outflow Speed} \label{sec:methods:outflow_speed}
\textbf{Solar Orbiter/Metis:} Simultaneous measurements of the coronal white-light polarized brightness and UV neutral-hydrogen Lyman-$\alpha$ intensity provide us with the possibility of inferring the solar wind outflow speed in the corona by exploiting the Doppler-dimming effect \citep[see][]{Withbroe1982,Noci1987} which affects the resonance scattering of the coronal Lyman-$\alpha$. 

The diagnostic method consists of a forward modeling of the Lyman-$\alpha$ intensity measured by Metis by adjusting the solar wind outflow velocity (treated as a free parameter) in the calculation of the Lyman-$\alpha$ emission expected on the basis of the coronal electron density derived from Metis polarized brightness and assuming suitable electron and neutral-hydrogen (proton) temperatures and the spectral profile of the exciting chromospheric Lyman-$\alpha$ line.
The full inversion technique is described in \citet{Dolei2018} and \citet{Giordano2025}, where assumptions and limitations are discussed in detail.

For the present work, we derived an outflow velocity map from Metis data by adopting the electron temperature radial profile of \citet{Gibson1999}, a uniform hydrogen kinetic temperature of 1.2~MK, and the analytical profile of \citet{Auchere2005} for the chromospheric Lyman-$\alpha$ line, scaled to match the average Lyman-$\alpha$ intensity measured from Earth on the observation date and available at the LASP Interactive Solar Irradiance Data Center. 
The \citet{Gibson1999} electron temperature profile is a good representative choice for both streamer and coronal-hole regions, as also shown by \citet{Giordano2025} in their investigation of the solar wind velocity distribution in the corona inferred with Metis during solar minimum in 2020--2021.

The assumption of a uniform hydrogen kinetic temperature is motivated by the recent analysis of SOHO/UVCS measurements of the \ion{H}{1} Lyman-$\alpha$ line intensity reported by \citet{Cranmer2020}, showing that the proton temperature does not exhibit a significant radial variation between heliocentric distances of 1.4 and 4$~R_{\odot}$ in polar coronal holes \citep[see also][]{Antonucci2023}.
The adopted value matches the $\sim$1.2~MK temperature value computed for the base of the coronal hole under investigation.
The resulting map is shown in the bottom panel of Figure~\ref{fig:EDF4_metis_dens_vels}C.

\textbf{Hinode/EIS \& Solar Orbiter/SPICE:} On-disk observation from Hinode/EIS provide LOS Doppler velocities while the coronal hole was on the disk on 1~April (fitted intensities are shown in Figure~\ref{fig:EDF5_EIS_SPICE_doppshifts}A and Doppler velocities are shown in Figure~\ref{fig:EDF5_EIS_SPICE_doppshifts}B). The Doppler shifts are computed as shifts from the theoretical peak wavelength of the \ion{Fe}{12} Gaussian fitted spectral profiles. Using the same approach for Solar Orbiter/SPICE, we utilize the \ion{Ne}{8} on 1~April 2024 (Figure~\ref{fig:EDF5_EIS_SPICE_doppshifts}C), where the Doppler shifts are shown in Figure~\ref{fig:EDF5_EIS_SPICE_doppshifts}D.  

\textbf{SOHO/LASCO-C2:} Additionally, we estimate outflow speeds with SOHO/LASCO-C2 using densities derived from SOHO/LASCO-C2 pB observations and mass conservation. and mass conservation. The speed is estimated by using the continuity equation as $(n_p v_p A)_\text{Parker} = (n_p  v_p A)_\text{LASCO}$, where $A$~is the cross-sectional area of the stream. The ratio $A_\text{LASCO} / A_\text{Parker}$ can be replaced by expansion factor, $f$, which measures the increase in the cross-sectional area of a magnetic flux tube beyond a spherical expansion. The ratio of the cross-sections $A_\text{LASCO} / A_\text{Parker}$ can be replaced by the inverse ratio of the magnetic magnitude, $B_\text{Parker} / B_\text{LASCO} \times (r_\text{LASCO}/r_\text{Parker})^2$, where $B$~is the magnetic field strength and $r$~is the heliocentric distance. We assume magnetic flux conservation. Substituting the expansion factor and rearranging the equation, 
\begin{equation}
    v_{LASCO} = \frac{(n_pv_p)_{Parker}}{(n_p)_{LASCO}}
    \frac{B_{LASCO}}{B_{Parker}} \left( \frac{r_{Parker}}{r_{LASCO}} \right)^2 \quad,
\end{equation}
where $B_{LASCO}$~is the magnetic field strength traced along the PFSS extrapolation (discussed in Section~\ref{sec:methods:PFSS_helio_mapping}) to \emph{Parker} within the SOHO/LASCO-C2 tomographic reconstructions. $B_{Parker}$~is the strength at \emph{Parker}'s orbit. The radial distances, $r_{LASCO}$ and $r_{Parker}$, are the heliocentric distances at each point along the SOHO/LASCO-C2 FOV and at \emph{Parker}'s location, respectively.

\subsection{Three Dimensional Density} \label{sec:methods:tomography}
In this study, solar rotational tomography \citep[SRT;][]{frazin_2009} was utilized to determine the three-dimensional (3D) distribution of the coronal electron density~($n_e$) and temperature~($T_e$). SRT relies on a time series of optically thin coronal images, each of which provides two-dimensional projected information of the underlying 3D coronal structure. This framework sets up a linear inversion problem that, when solved, allows for the estimation of a quantity related to emissivity. For white-light~(WL) coronagraph images, the emissivity is proportional to $n_e$; therefore, the inversion problem is solved for the 3D distribution of the electron density \citep{altschuler_1972, frazin_2002, vibert_2016, Vasquez2024}. For extreme ultraviolet~(EUV) images, SRT enables determination of the 3D distribution of the DEM in the corona. This technique, known as differential emission measure tomography~(DEMT), allows for the reconstruction of the 3D distributions of both the electron density \citep{frazin_2009, vasquez_2009, vasquez_2016}.

The series of images used in the tomography process is collected over a duration sufficient to provide a range of viewing angles for a complete reconstruction of the corona under the assumption that coronal evolution is small. For instance, data from the SOHO/LASCO-C2 or SDO/AIA instruments is gathered over approximately 14~days to encompass half of a synodic rotation. In this research, the time series of images utilized for tomography was centered around 8~April 2024, when the coronal hole of interest was positioned on the western limb of the Sun. SRT was applied to the SOHO/LASCO-C2 WL images to reconstruct the 3D distribution of $n_e$ in the middle corona, specifically within the heliocentric height range of 2.5--6.0~$R_{\odot}$. Meanwhile, DEMT was employed on AIA images in the 171, 193, and 211~{\AA} bands to determine the 3D distributions of $n_e$ in the inner corona (below 1.25~$R_{\odot}$). The various parameters used for both reconstructions—such as the time cadence of the image series and the tomographic grid size—mirror those used by \cite{lloveras_2022}, where further details can be found.

The SOHO/LASCO-C2 and SDO/AIA coronal hole density profiles (Figure~\ref{fig4_radialplots}A) were extracted from the 3D density cubes produced through the SRT method (WL and EUV tomography, respectively) by tracing PFSS open field lines via Carrington coordinates of those sampled by \emph{Parker}. These lines were assumed radial beyond the source surface (2.5~$R_{\odot}$), assuming Parker spiral effects negligible, providing the radial profile of the coronal hole density.

\subsection{Heliospheric Stream Mapping} \label{sec:methods:PFSS_helio_mapping}
Figure~\ref{fig1_orbit_mapping}A shows the \emph{Parker} and Solar Orbiter trajectories in the Carrington frame within the timeframe used in the analysis. Both trajectories are colored by the polarity of the local interplanetary magnetic field measurement where red is positive and blue is a negative polarity. The flow lines trace the Archimedean spiral from each spacecraft trajectory down to the source surface of the Sun whose path is determined by the local solar wind bulk speed and the Sun's rotation \citep{Nolte1973, Stansby2019, Badman2020}. In this framework, the speed is kept constant and taken to be the local speed. The flow lines are colored by the local speed of the plasma at \emph{Parker} and Solar Orbiter. The negative polarity outflow associated with the coronal hole examined in this study is traversed by \emph{Parker} in approximately 18~hours (on 30~March) while spanning over 6~days at Solar Orbiter (8--14~April).

The mapped longitude and latitude projections to the source surface (2.5~$R_{\odot}$) from \emph{Parker} and Solar Orbiter's orbits can be traced to the surface of the Sun using coronal magnetic field extrapolations. We utilize Potential-Field Source-Surface (PFSS; \citealt{altschuler1969_pfss, Schatten1969_pfss}) modeling to simulate the global coronal magnetic field where open field lines associated with spacecraft source surface projections are traced to their photospheric footpoint, as shown in Figure~\ref{fig1_orbit_mapping}B. The figure illustrates the projected spacecraft trajectories (at the source surface) along with the magnetic field extrapolations connected to the spacecraft footpoints with the corresponding color to indicate polarity as well as the heliospheric current sheet (HCS) in orange. The local changes in the polarity of the solar wind (at both spacecraft) indicate two HCS crossing that surround solar wind from a negative polarity coronal hole on the disk. The two HCS crossings are well reproduced by the simulated coronal magnetic field, providing a strong indication of good model performance. We find that the corresponding photospheric footpoints of both \emph{Parker} and Solar Orbiter trajectories map to the southern region of the latitudinally elongated and fragmented coronal hole, as indicated by the cyan outline. The spacecraft footpoints suggest that even with a small latitudinal offset (at its largest, $\Delta9^{\circ}$ in latitude) in their orbits, each spacecraft samples outflows from a similar region of the coronal hole. We further discuss correspondences among \emph{in situ} observations in Sections~\ref{sec:methods:in_situ_connection} and~\ref{sec:methods:energy_flux_calc}.

\subsection{In situ observations} \label{sec:methods:in_situ_connection}

For the \emph{in situ} properties of the solar wind stream, at \emph{Parker}, the SPAN-i instrument measures proton and alpha number densities, temperatures, and bulk velocities \citep{Livi2022}. Because the spacecraft's solar panels partially obstruct SPAN-i, certain regions of velocity space may be unobservable depending on the flow direction. To ensure data integrity, we only analyzed periods where both species remained within the instrument's field of view. We then validated the SPAN-i proton densities against electron densities derived from quasi-thermal noise (QTN), finding strong agreement throughout the study period, as shown between gold and black proton density curves in panel~A of Figure~\ref{fig3_insitu}. Additionally, we incorporated SPAN-e electron measurements \citep{Whittlesey2020} to determine core electron temperatures, panel~D (red curve). We also  use the 3D magnetic field data from the FIELDS fluxgate magnetometer (MAG) sampled at 4.58~vectors~s$^{-1}$ \citep{Bale2016}, panel~E.

At Solar Orbiter, measurements of the magnetic field are taken by another fluxgate magnetometer (MAG) at 8~vectors~s$^{-1}$ \citep{Horbury2020}. Observations of protons across this period were taken by the Proton-Alpha System (PAS) instrument as part of the Solar Wind Analyzer (SWA) suite \citep{Owen2020}. The proton density, temperature, and velocity were measured with the PAS instrument with 4~s full scan 3D particle distributions.

We compute the cross-helicity as $\sigma_C = 2\delta \textbf{v} \cdot \delta \textbf{b}/(|\delta \textbf{v}|^{2} + |\delta \textbf{b}^{2}|)$, where the $\delta \textbf{b} = \delta \textbf{B}/|\textbf{B}|\cdot V_{A}$ are in Alfv\'en units \citep{Barnes1974, Matthaeus1982, Roberts1987}. We compute the residual energy as, $\sigma_R= (|\delta \textbf{v}|^{2} - |\delta \textbf{b}|^{2})$/($|\delta \textbf{v}|^{2} + |\delta \textbf{b}|^{2})$. The cross-helicity parameter quantifies the balance between outward-propagating and inward-propagating Alfv\'enic fluctuations and indicates the degree of correlation between the magnetic and velocity fluctuations in the solar wind \citep{Perri2010, DAmicis2019}. The residual energy quantifies the balance or imbalance between kinetic and magnetic energies in the solar wind where a negative value indicates magnetic dominance. $\delta B$~is computed as the magnetic field deviation from the mean field, where $\delta \textbf{B} = \textbf{B}-\langle\textbf{B}\rangle_t$, and $\langle\textbf{B}\rangle_t$ is time-averaged over 15-, 30-, and 60-minute intervals, $t$, at \emph{Parker} and at Solar Orbiter, which is on the order of a few correlation timescales of $B$ and $v$ \citep{Chen2020, Parashar2020}. $\delta v$~is computed in the same manner. We note only small but increasing trend in the fluctuation averages with increasing averaging timescales as listed in Table~\ref{tab:dbdv_timescales} therefore use 30-minute timescales as our averages. 

\begin{table}[h]
\centering
\caption{Mean $\delta \textbf{B}$~(nT), $\delta \textbf{b}$~(km~s$^{-1}$) and $\delta \textbf{v}$~(km~s$^{-1}$) with the standard deviation for each quantity within the shaded purple region of Figure~\ref{fig3_insitu}, averaged over 15-min, 30-min, and 1-hour intervals.}

\begin{tabular}{ll ccc}
\hline
\hline

Spacecraft & & 15~min & 30~min & 1~hour \\
\hline
{\emph{Parker}} & $|\delta \textbf{B}|$ & $ 283.62 \pm 183.34 $ & $286.01 \pm 182.38$ & $ 286.71\pm 181.68$ \\

& $|\delta \textbf{b}|$ & $ 91.08 \pm 62.47$ & $ 91.88\pm 63.85$ & $ 91.91\pm63.69 $ \\

& $|\delta \textbf{v}|$ & $ 91.72 \pm 58.29$ & $93.92 \pm 57.56$   & $ 95.66\pm 57.46$   \\
\hline
{\emph{Solar Orbiter}} & $|\delta \textbf{B}|$ & $14.83 \pm 10.67$ & $16.41 \pm 10.79$ & $17.76 \pm 10.97$ \\

&$|\delta \textbf{b}|$ & $ 44.11\pm 30.50$ & $48.02\pm30.79 $ & $ 51.41\pm 31.19$ \\

& $|\delta \textbf{v}|$ & $42.22 \pm 35.34$ & $46.03 \pm 35.94$ & $48.50 \pm 36.59$ \\
                                
\hline
\hline
\end{tabular}

\label{tab:dbdv_timescales}
\end{table}

The coronal hole of interest is traversed by both spacecraft at an average heliocentric distance of 11.60~$R_{\odot}$ (\emph{Parker}) and 72.06~$R_{\odot}$ (Solar Orbiter). The coronal hole outflow is surrounded by two CSCs occurring near a spacecraft longitude of 80$^{\circ}$ to 130$^{\circ}$ at \emph{Parker} and 60$^{\circ}$ to 110$^{\circ}$ at Solar Orbiter, maintaining a similar longitude distance of roughly 50$^{\circ}$ and act as key signatures for solar wind heliospheric mapping. At \emph{Parker}, the helium abundances are on average 4\%, typical for fast solar wind \citep{Kasper2007} and inner heliosphere \citep{Johnson2024}. The cross-helicity (Figure~\ref{fig3_insitu}B in green) indicates strong Alfv\'enicity throughout this period while maintaining a slightly negative residual energy suggesting a small dominance in the magnetic fluctuation energy. The entire fast wind period is sub-Alfv\'enic, with an average $M_A$ of $\sim$0.76. Several coherent switchback patches (on the order of 5$^{\circ}$ in longitude) are embedded through the coronal hole stream in between the CSCs. The patches are identified as clusters of individual, highly correlated radial velocity and radial magnetic field fluctuations, where across patches the fluctuation amplitude at the center is maximum. The patches also indicate an uninterrupted electron strahl throughout. Key properties of magnetic switchback patches can be found in the following articles: \citealt{Fargette2021, Woolley2021, Shi2022, Huang2023, Rivera2024b}. 

At Solar Orbiter, the solar wind properties resemble those observed at \emph{Parker} well. In the top panel, the proton density indicate two peaks at the CSC locations analogous to that observed at \emph{Parker}. We examine the solar wind bounded by the CSCs to compare with properties at \emph{Parker}. We also note that $\sigma_C$ has shift away from 1 and --1, moving from a strongly imbalanced state to more balanced state \citep{Bavassano2000, Bavassano2000b, Matthaeus2004, Breech2005}.

\subsection{Iso-poly + Forcing Radial Wind Speed Profiles}\label{sec:methods:isopoly}

In Figure~\ref{fig4_radialplots}, remote sensing and \emph{in situ} determinations of the stream evolution are compared to a set of modeled radial profiles of the density, wind speed and temperature. These profiles are generated using the ``iso-poly'' framework \citep{Dakeyo2022} with an additional boost from Alfv\'enic fluctuation energy, described as the ``iso-poly + forcing'' model in the text \citep{shi2022_polyt,Rivera2024,Rivera2025}. These models solve the \citet{Parker1958} isothermal solution out to a certain ``coronal'' radius (here 10~$R_\odot$) before smoothly transitioning to a polytropically cooling solar wind in line with statistical observations. The specific profiles shown are those presented in \citet{Badman2025} in which the profiles were constrained to span the statistical range of density, velocity and proton temperature observed by \emph{Parker} and Solar Orbiter from 10--215~$R_{\odot}$ from 2018--2025  \citep[see full computational details in ][]{Badman2025}. The profiles may therefore be interpreted as tracking the average behavior of the bulk properties of the solar wind and treat it as a 1D, radially expanding flux tube. Nevertheless, for the stream studied in this work, the observations follow the average behavior of the solar wind for a moderately fast stream (dark, shaded profiles in Figure~\ref{fig4_radialplots}), as would be expected for a steadily open, equatorial coronal hole.

\subsection{Energy Flux Calculations} \label{sec:methods:energy_flux_calc}
All energy flux terms are summarized in Table~\ref{tab:energyflux_3}. At each heliospheric location, we compute the kinetic, enthalpy, Alfv\'en wave and gravitational energy flux terms as described below. 

For the \emph{in situ} measurements at \emph{Parker} and Solar Orbiter, we compute each term following the prescription in \cite{Rivera2024, Rivera2025}. For the kinetic and gravitational energy flux, we include the contribution from protons (with helium maintained at 4\%), at their respective densities. The enthalpy flux includes the contribution from protons and electrons, where we assume both populations have equal number densities and use their measured temperature. We estimate the expansion factor between \emph{Parker} and Solar Orbiter using a ratio of their mass flux at both locations, as $(n_p v_p r^2)_\text{Parker} /(n_p v_p r^2)_\text{Solar Orbiter} = f_\text{SolO} = 0.965\pm0.43$. The mass flux ratio is on the order of unity, indicating roughly a spherical expansion between 11.60~$R_{\odot}$ and 72.06~$R_{\odot}$. We note that we scale the Solar Orbiter energy flux terms by expansion factor, $f_\text{SolO}$. At the Sun and in the heliosphere, we use the electron enthalpy flux as an equivalency for the ambipolar potential \citep{Halekas2022}. 

To compute the energy flux terms at the Sun, we compute a mean value of all the variables between 1.035--1.065~$R_{\odot}$, shown in Figure~\ref{fig4_radialplots}, where each parameter is co-spatially available. We compute all the terms to derive a total energy flux calculation following \cite{Pucci2013_energyflux, Paraschiv2015}. We also include the contribution from the mechanical energy flux carried by the compressive acoustic (slow mode) and magnetoacoustic (fast mode) waves parallel to the magnetic field. We estimate their contribution in the low beta coronal environment from the general relation: $F = m_pn_p \zeta^2V_{g}$ where the group velocity, $V_{g} = C_s$ for slow mode, and $V_A$ for fast mode. The mean values used in the calculations are shown in Table~\ref{tab:swprops_2}. The energy flux calculations are as follows: 
\begin{equation}
\begin{split}
    F_\text{Kinetic} = \frac{1}{2} m_in_iv_{p}^3 \quad,\\
    F_\text{Enthalpy} = \frac{\gamma}{\gamma-1} n_ik_BT_iv_{p} \quad,\\
    F_\text{Alfvén~Wave} =\sqrt{\frac{m_in_i}{4\pi}}\zeta_\perp^2|B| \quad,\\
    F_\text{Fast/Slow mode} = m_in_i\zeta_\parallel^2V_{g} \quad,\\
    F_\text{Gravitational} = m_in_iGM_{\odot}/R_{\odot}(1-R_{\odot}/r)v_p \quad.\\
    \end{split}
\end{equation}

All terms are normalized by $r^2/R_{\odot}^2$ and use $\gamma=5/3$. $m_i$, $n_i$ and $T_i$ are the mass, number density, and temperature of each population, respectively. $v_p$~is the outflow speed. $\zeta_{\perp/\parallel}$~is the non-thermal width in derived speed units ($V_{NTW}$), measured perpendicular (DKIST, off-limb) and parallel (EIS, on-disk/along LOS) to the magnetic field, respectively. $|B|$~is the magnetic field magnitude. $G$ is the gravitational constant, and $M_{\odot}, R_{\odot}$ are the Sun's mass and radius, respectively. $k_B$~is the Boltzmann constant.

We scale the energy fluxes at the Sun to \emph{Parker} using the mass flux ratio and the expansion factor. In this case, the mass flux ratio between the Sun and \emph{Parker} is $f = 0.07\pm0.04$ which aligns somewhat well with a value of $f_{MAS}=0.1$. This value suggests a super-radial expansion between the corona and 11.60~$R_{\odot}$.


\bibliographystyle{aasjournalv7}

\end{document}